\documentclass[fleqn,usenatbib]{mnras}
\usepackage{color, colortbl}

\usepackage{xcolor}
\definecolor{DarkGreen}{RGB}{0,100,0}
\definecolor{lightOrange}{RGB}{255,229,204}
\usepackage{multirow}
\usepackage{natbib}
\usepackage{amsmath}
\usepackage{amssymb}
\usepackage{gensymb}
\usepackage{bm} 
\hypersetup{linkcolor=red,citecolor=orange,filecolor=cyan,urlcolor=magenta}
\usepackage{longtable}
\usepackage{orcidlink}
\usepackage{verbatim} 
\usepackage[T1]{fontenc} 

\newcommand{\hst}{\textit{HST}}
\newcommand{\jwst}{\textit{JWST}}
\newcommand{\rst}{\textit{Roman}}
\newcommand{\euclid}{\textit{Euclid}}
\newcommand{\spitzer}{\textit{Spitzer}}

\newcommand{\zgrism}{$z_\mathrm{grism}$}
\newcommand{\magphys}{\texttt{MAGPHYS}}

\newcommand{\halpha}{$\mathrm{H}\alpha$}
\newcommand{\hbeta}{$\mathrm{H}\beta$}
\newcommand{\hgamma}{H$\gamma$}
\newcommand{\esc}{erg s$^{-1}$ cm$^{-2}$}
\newcommand{\se}{\texttt{SExtractor}}

\title[WISP Photometric and Emission Line DR]{WFC3 Infrared Spectroscopic Parallel (WISP) Survey: Photometric and Emission Line Data Release}

\author[A. J. Battisti et al.]{A. J. Battisti\orcidlink{0000-0003-4569-2285}$^{1,2}$\thanks{Email:andrew.battisti@anu.edu.au},
M. B. Bagley\orcidlink{0000-0002-9921-9218}$^{3}$,
M. Rafelski\orcidlink{0000-0002-9946-4731}$^{4,5}$,
I. Baronchelli\orcidlink{0000-0003-0556-2929}$^{6,7}$,
Y.S. Dai\orcidlink{0000-0002-7928-416X}$^{8}$,
\newauthor 
A. L. Henry\orcidlink{0000-0002-6586-4446}$^{4,5}$,
H. Atek\orcidlink{0000-0002-7570-0824}$^{9}$,
J. Colbert$^{10}$,
M. A. Malkan\orcidlink{0000-0001-6919-1237}$^{11}$,
P. J. McCarthy$^{12}$,
C. Scarlata\orcidlink{0000-0002-9136-8876}$^{13}$,
\newauthor
B. Siana\orcidlink{0000-0002-4935-9511}$^{14}$,
H. I. Teplitz\orcidlink{0000-0002-7064-5424}$^{10}$,
A. Alavi\orcidlink{0000-0002-8630-6435}$^{10}$,
K. Boyett\orcidlink{0000-0003-4109-304X}$^{2, 15}$,
A. J. Bunker\orcidlink{0000-0002-8651-9879}$^{16}$,
J. P. Gardner\orcidlink{0000-0003-2098-9568}$^{17}$,
\newauthor
N. P. Hathi\orcidlink{0000-0001-6145-5090}$^{4}$,
D. Masters\orcidlink{0000-0001-5382-6138}$^{10}$,
V. Mehta\orcidlink{0000-0001-7166-6035}$^{10}$,
M. Rutkowski\orcidlink{0000-0001-7016-5220}$^{18}$,
K. Shahinyan\orcidlink{0000-0001-5128-4160}$^{13}$,
\newauthor
B. Sunnquist\orcidlink{0000-0003-3759-8707}$^{4}$,
X. Wang\orcidlink{0000-0002-9373-3865}$^{19,20,21}$
\\
$^{1}$Research School of Astronomy and Astrophysics, Australian National University, Cotter Road, Weston Creek, ACT 2611, Australia\\
$^{2}$ARC Centre of Excellence for All Sky Astrophysics in 3 Dimensions (ASTRO 3D), Australia\\
$^{3}$Department of Astronomy, The University of Texas at Austin, Austin, TX 78712, USA\\
$^{4}$Space Telescope Science Institute, 3700 San Martin Dr., Baltimore, MD 21218, USA\\
$^{5}$Department of Physics and Astronomy, Johns Hopkins University, Baltimore, MD 21218, USA\\
$^{6}$INAF-Istituto di Radioastronomia, Via Gobetti 101, 40129 Bologna, Italy\\
$^{7}$Italian ALMA Regional Centre, Via Gobetti 101, 40129 Bologna, Italy\\
$^{8}$Chinese Academy of Sciences South America Center for Astronomy (CASSACA), National Astronomical Observatories of China \\
(NAOC), 20A Datun Road, Beijing, 100012, China\\
$^{9}$Institut d'Astrophysique de Paris, CNRS, Sorbonne Universit\'e, 98bis Boulevard Arago, 75014, Paris, France\\
$^{10}$IPAC, California Institute of Technology, 1200 E. California Boulevard, Pasadena, CA 91125, USA\\
$^{11}$Department of Physics and Astronomy University of California, Los Angeles Los Angeles, CA 90095-1547, USA\\
$^{12}$National Optical-Infrared Astronomy Research Laboratory, Tucson, AZ 85719, USA\\
$^{13}$Minnesota Institute for Astrophysics, University of Minnesota, Minneapolis, MN 55455, USA\\
$^{14}$Department of Physics and Astronomy, University of California, Riverside, 900 University Avenue, Riverside, CA 92521, USA\\
$^{15}$School of Physics, University of Melbourne, Parkville 3010, VIC, Australia\\
$^{16}$Department of Physics, University of Oxford, Denys Wilkinson Building, Keble Road, Oxford OX13RH, UK\\
$^{17}$Astrophysics Science Division, NASA Goddard Space Flight Center, 8800 Greenbelt Rd, Greenbelt, MD 20771, USA\\
$^{18}$Minnesota State University-Mankato, Telescope Science Institute, TN141, Mankato, MN 56001, USA\\
$^{19}$School of Astronomy and Space Science, University of Chinese Academy of Sciences (UCAS), Beijing 100049, China\\
$^{20}$National Astronomical Observatories, Chinese Academy of Sciences, Beijing 100101, China\\
$^{21}$Institute for Frontiers in Astronomy and Astrophysics, Beijing Normal University,  Beijing 102206, China
}

\date{Accepted XXX. Received YYY; in original form ZZZ}

\pubyear{2024}

\begin{document}
\label{firstpage}
\pagerange{\pageref{firstpage}--\pageref{lastpage}}
\maketitle

\begin{abstract}
We present reduced images and catalogues of photometric and emission line data ($\sim$230,000 and $\sim$8,000 sources, respectively) for the WFC3 Infrared Spectroscopic Parallel (WISP) Survey. These data are made publicly available on the Mikulski Archive for Space Telescopes (MAST) and include reduced images from various facilities: ground-based $ugri$, \hst\ WFC3, and \spitzer\ IRAC (Infrared Array Camera). Coverage in at least one additional filter beyond the WFC3/IR data are available for roughly half of the fields (227 out of 483), with $\sim$20\% (86) having coverage in six or more filters from $u$-band to IRAC 3.6\micron\ (0.35-3.6\micron). For the lower spatial resolution (and shallower) ground-based and IRAC data, we perform PSF-matched, prior-based, deconfusion photometry (i.e., forced-photometry) using the \texttt{TPHOT} software to optimally extract measurements or upper limits. We present the methodology and software used for the WISP emission line detection and visual inspection. The former adopts a continuous wavelet transformation that significantly reduces the number of spurious sources as candidates before the visual inspection stage. 
We combine both WISP catalogues and perform SED fitting on galaxies with reliable spectroscopic redshifts and multi-band photometry to measure their stellar masses. We stack WISP spectra as functions of stellar mass and redshift and measure average emission line fluxes and ratios. 
We find that WISP emission line sources are typically `normal' star-forming galaxies based on the Mass-Excitation diagram ([OIII]/\hbeta\ vs. $M_\star$; $0.74<z_\mathrm{grism}<2.31$), the galaxy main sequence (SFR vs. $M_\star$; $0.30<z_\mathrm{grism}<1.45$), $S_{32}$ ratio vs. $M_\star$ ($0.30<z_\mathrm{grism}<0.73$), and $O_{32}$ and $R_{23}$ ratios vs. $M_\star$ ($1.27<z_\mathrm{grism}<1.45$).
\end{abstract}


\section{Introduction}
Slitless spectroscopy provides simultaneous spectra for every source in an observed field of view (FOV). Such observations have the notable advantage that they can be used to efficiently perform blind spectroscopic surveys over large areas of the sky for the study of galaxy evolution, relative to slits, fibers, or integral field spectroscopy (IFS) that cover much smaller FOVs. This gain in mapping area from slitless spectroscopy usually comes at the expense of reduced sensitivity due to a higher background, more source confusion, and low to moderate spectral resolution ($R\sim100-1000$) relative to slits or IFS ($R\gtrsim2000$). The latter limits dynamical analyses and may result in lines blending together (e.g., \halpha+[NII]). 

Blind surveys provide a unique and unbiased view of galaxy evolution by avoiding issues of cosmic variance or photometric pre-selection. The grisms on the \textit{Hubble Space Telescope} (\hst) have been widely utilised to perform slitless spectroscopic surveys of thousands of galaxies at intermediate redshifts ($0.5\lesssim z \lesssim 2.5$; e.g., \citealp[APPLES][]{Pasquali03}, \citealp[GRAPES][]{pirzkal04}, \citealp[WISP][]{atek10}, \citealp[3D-HST][]{brammer12}, \citealp[PEARS][]{pirzkal13}, \citealp[GLASS][]{treu15}, \citealp[FIGS][]{pirzkal17}, \citealp[MAMMOTH–Grism][]{wang22}, \citealp[CLEAR][]{simons23}, \citealp[MUDF][]{revalski23}).  
Next generation facilities, including the \textit{James Webb Space Telescope} (\jwst), \euclid, and the \textit{Nancy Grace Roman Space Telescope} (\rst), will revolutionise this capability by providing slitless spectroscopy for millions of galaxies that span larger areas of the sky and a wider range in cosmic time \citep[e.g.,][]{bagley20}.

The WFC3 (Wide Field Camera 3) Infrared Spectroscopic Parallel \citep[WISP,][]{atek10} survey was the largest, multi-cycle \hst\ pure-parallel grism program. WISP obtained slitless spectroscopy for thousands of galaxies in 483 pointings using up to two near-IR grisms, G102 (0.80-1.15~\micron , $R\sim210$) and G141 (1.08-1.69~\micron , $R\sim130$). Pure-parallel surveys, such as WISP, have the benefit that the data are obtained `for-free' for numerous random fields that are independent and uncorrelated (i.e., unbiased). The versatility of large slitless surveys like WISP are demonstrated by the range of science cases that can be explored, such as: characterising the star forming galaxy main sequence \citep{atek14}, characterising dust attenuation \citep{dominguez13} and dust attenuation curves \citep{battisti22}, characterising the mass-metallicity relation \citep{henry13, henry21}, characterising massive, quenched galaxies \citep{bedregal13}, identifying galaxy pairs \citep{dai21}, identifying single spectral lines through machine learning \citep{baronchelli20, baronchelli21}, identifying Lyman-$\alpha$ emitters \citep{bagley17}, identifying bright, rare galaxies \citep{bagley24}, predicting emission line galaxy number counts for future surveys \citep{colbert13, mehta15, bagley20}, investigating the mass-size relation of passive galaxies \citep{zanella16}, crowdsourced analysis of slitless spectroscopic data \citep{dickinson18}, nitrogen enhancement of star-forming galaxies \citep{masters14}, and the discovery of very faint, distant (400pc) brown dwarfs \citep{masters12}.

A challenge for pure-parallel slitless surveys is that the random fields are sporadic in position and typically do not have deep ancillary photometric data available (e.g., compared to legacy deep fields). To remedy this, the WISP team has carried out several supplementary observing programs to obtain additional photometry with a variety of facilities, with priority given to the deepest $\sim$200 WISP fields. These include \hst\ WFC3/UVIS optical, ground-based optical, and \spitzer /IRAC near-IR data.

This paper describes the public data release of a self-consistent photometric catalogue, reduced images, and the emission line catalogue of WISP fields. These data are hosted at the WISP website\footnote{\url{https://archive.stsci.edu/prepds/wisp/}} on the Mikulski Archive for Space Telescopes (MAST). Currently $\sim$50\% of the WISP fields have additional photometric data beyond WFC3/IR. These data provide a valuable galaxy sample for legacy science and can serve as a useful reference for ongoing and future grism surveys with \jwst, \euclid, and \rst.

This paper is organised as follows: Section~\ref{data} describes the observational data, Section~\ref{photo_pipeline} describes the WISP photometric pipeline, Section~\ref{photo_catalog} presents the WISP photometric catalogue, Section~\ref{spec_pipeline} describes the WISP spectroscopic pipeline, Section~\ref{spec_catalog} presents the WISP emission line catalogue, Section~\ref{results} shows our results from combining both catalogues to study galaxies at $0.3<z_\mathrm{grism}<2.3$, and Section~\ref{conclusion} summarises our main conclusions. Throughout this work we adopt a $\Lambda$-CDM cosmological model, with $H_0=70$~km/s/Mpc, $\Omega_M=0.3$, and $\Omega_{\Lambda}=0.7$. All magnitudes are in the AB magnitude system \citet{oke&gunn83}.

\section{Observations and Data Reduction}\label{data}

\begin{figure}
\includegraphics[width=0.48\textwidth]{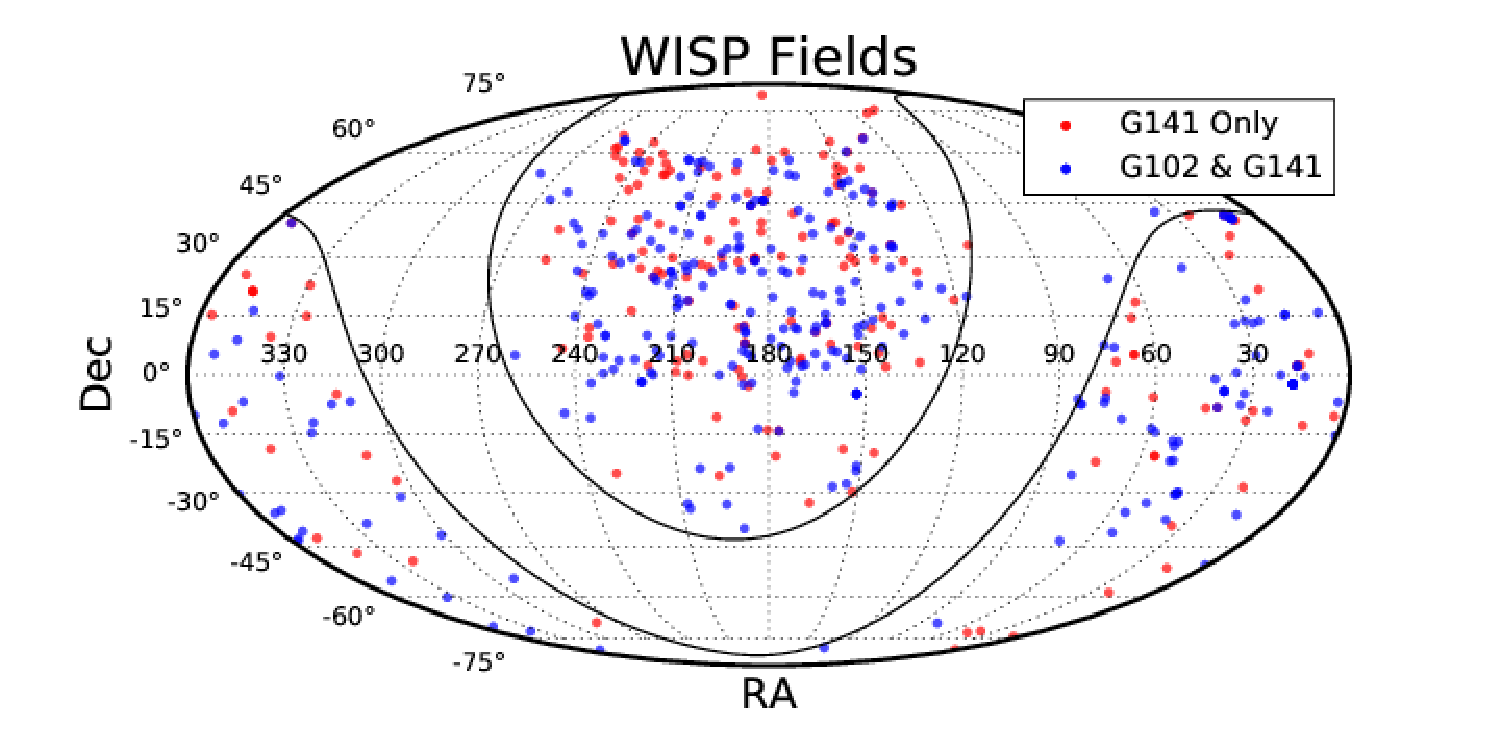}
\vspace{-4mm}
\caption{Locations of the 483 WISP fields. These fields are primarily located outside of the Galactic Plane. Most short parallel opportunities obtained WFC3/IR G141 grism data along with one imaging filter (F140W or F160W) and long opportunities obtained G102+G141 grisms and two imaging filters (F110W+F140W or F110W+F160W). 
 \label{fig:field_positions}}
\end{figure}

\begin{table*}
\caption{Summary of WISP \hst\ Survey (PI: M Malkan) \label{tab:wisp_obs_summary}}
\begin{center}
\begin{tabular}{lcccccc}
 \hline \hline
GO ID & Cycle & Observing Period$^a$ & WFC3/UVIS$^b$ & WFC3/IR  & WFC3/IR Grism & Orbits \\
 \hline
11696 & 17 & Nov 24, 2009 - May 1, 2011 & F475X,F600LP & F110W,F140W,F160W & G102,G141 & 250 \\
12283 & 18 & Oct 6, 2010 - Mar 11, 2012 & F606W,F814W$^c$ & F110W,F160W & G102,G141 & 280 \\
12568 & 19 & Oct 29, 2011 - May 19, 2013 & None & F140W & G102,G141 & 260 \\
12902 & 20 & Oct 16, 2012 - Mar 20, 2014 & F606W,F814W & F110W,F160W & G102,G141 & 260 \\
13352 & 21 & Oct 31, 2013 - Jun 14, 2015 & F606W,F814W & F110W,F160W & G102,G141 & 375 \\
13517 & 21 & Dec 8, 2013 - Feb 12, 2015 & F606W,F814W & F110W,F160W & G102,G141 & 200 \\
14178 & 23 & Nov 30, 2015 - May 1, 2017 & F606W,F814W & F110W,F140W,F160W & G102,G141 & 520 \\
 & & & & & Total: & 2145 \\
 \hline
 \end{tabular}
\end{center}
\textbf{Notes.} There are 184 short opportunity fields (one to three continuous orbits), 180 fields with G141 and either F140W or F160W, and 4 fields with G102 and F110W (see Section~\ref{wisp_wfc3ir}). There are 299 long opportunity fields (four or more continuous orbits); these obtained G102+G141 and either F110W+F140W or F110W+F160W, except for Cycle~19 which used F140W-only. \\
$^a$Observations often extended beyond the nominal \hst\ cycle period. \\
$^b$UVIS data were only obtained for a subset of the deep fields (155 fields; See Section~\ref{wisp_uvis}). \\
$^c$Most UVIS data in this cycle used 2x2 binning in the F475X and F600LP filters and those are not included in the current data release. 
\end{table*}

\subsection{WISP Survey - \hst\ WFC3/IR Grism Spectroscopy and Imaging}\label{wisp_wfc3ir}

\begin{figure}
\includegraphics[width=0.48\textwidth]{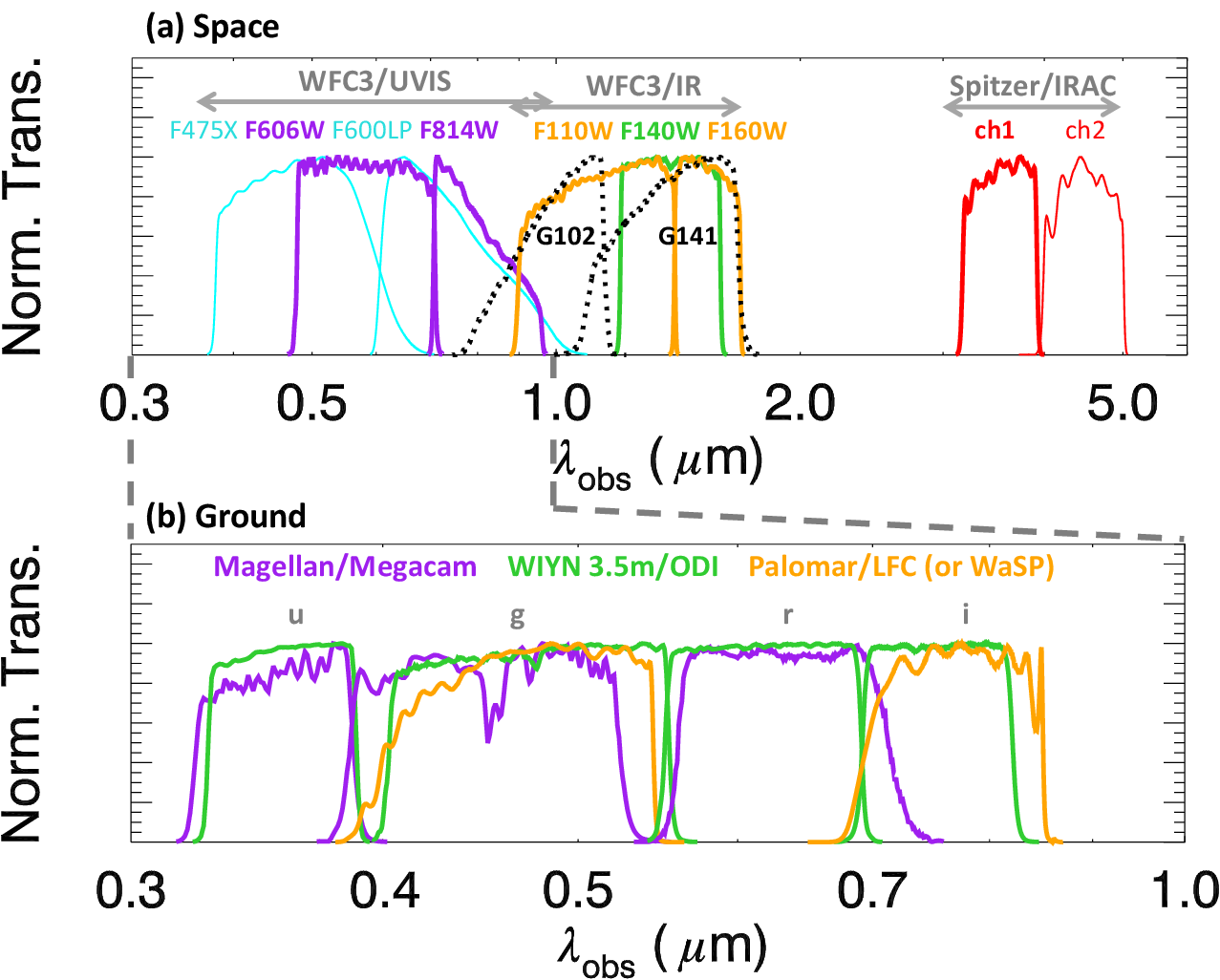}
\vspace{-4mm}
\caption{Filters used in the WISP survey and follow-up observations separated by \textbf{(a)} space-based and \textbf{(b)} ground-based facilities. The WISP survey obtained WFC3/IR imaging together with the grism spectroscopy (dotted black curves) and all WISP fields have either F140W or F160W data, with deep fields also having F110W imaging. Additional photometry is inhomogenous and comprises of WFC3/UVIS, Spitzer, and/or ground based optical data ($ugri$) from Magellan, WIYN 3.5m, and Palomar. The F475X, F600LP, and IRAC ch2 curves are shown as thinner lines to highlight that they are available for $<20$ fields (other filters are $>20$ fields), with a breakdown shown in Table~\ref{tab:filter_field_summary}.
 \label{fig:filters}}
\end{figure}

The WISP survey \citep[PI: M. Malkan;][]{atek10} is a multi-cycle (cycles 17-23) \hst\ pure-parallel program that obtained WFC3/IR observations for 483 pointings in random fields (i.e., location depended on the primary observing target and its position angle). Of the 483 parallel pointings (which we denote as `Par\#'; e.g., Par1, Par2, etc.), there are 40 fields with partial overlap
(i.e., there are 443 unique fields). The position of the WISP pointings are shown in Figure~\ref{fig:field_positions}. The program IDs and details for the different \hst\ cycles are summarised in Table~\ref{tab:wisp_obs_summary}. The effective grism area of each WISP pointing is $\sim$3.55~arcmin$^2$ in G102 and $\sim$3.8~arcmin$^2$ in G141, 
relative to the full 4.6~arcmin$^2$ WFC3/IR FOV, due to area on both the left and right side of each field being `lost'. On the left, this is because sources are not covered in the direct images necessary for source identification and wavelength calibration. On the right, this is because emission lines cannot be distinguished from contaminating zero order images \citep{bagley17}. In principle, high resolution ancillary data with a larger FOV, such as the WFC3/UVIS data available for a subset of fields, could be used to locate sources outside the WFC3/IR FOV. However, such an approach is not implemented for the current data release. Thus, the effective imaging and grism areas of the full WISP survey are $\sim$2200~arcmin$^2$ and $\sim$1600~arcmin$^2$, respectively (this also accounts for field overlap). An overview of the imaging and grism filters that are included in this data release are shown in Figure~\ref{fig:filters}.

\begin{figure}
\includegraphics[width=0.48\textwidth]{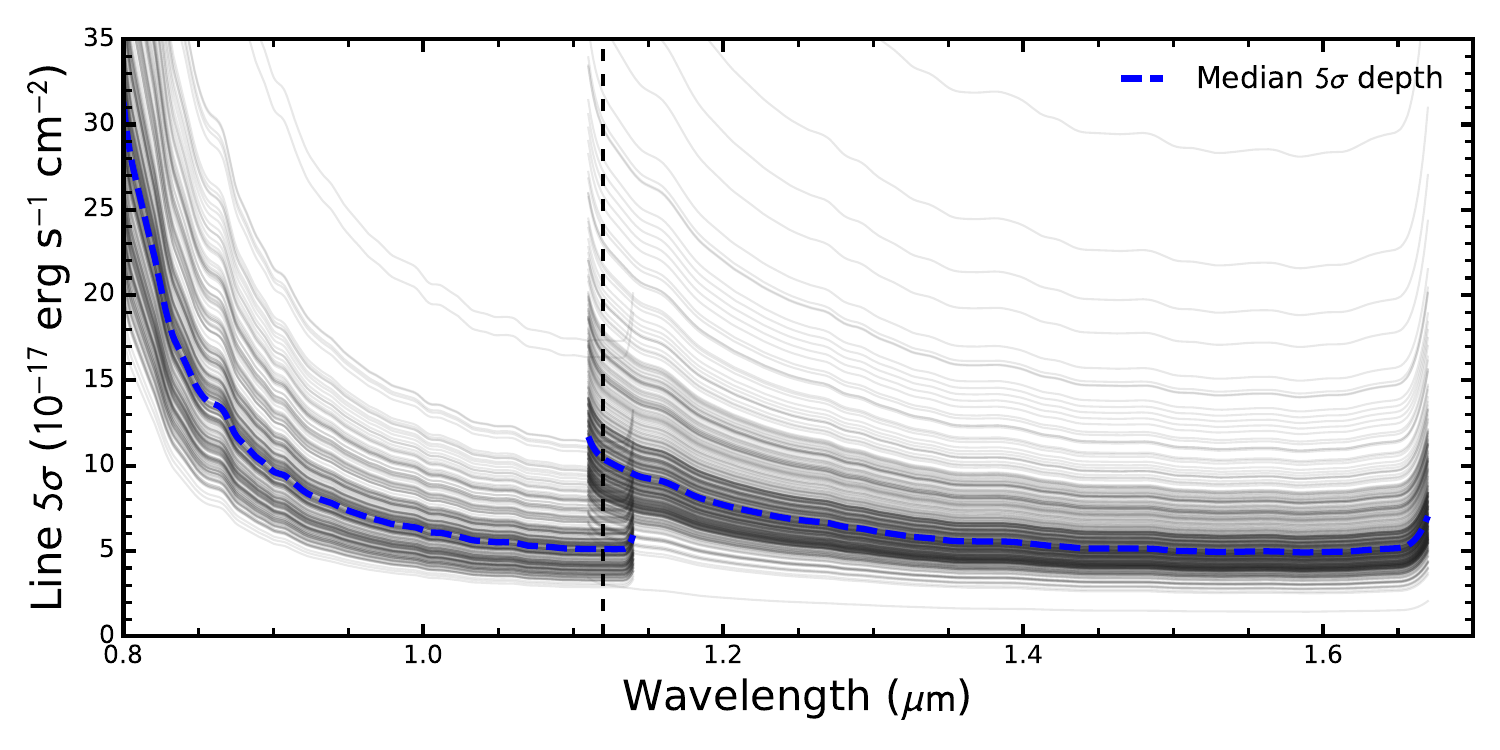}
\vspace{-6mm}
\caption{The $5\sigma$ emission line sensitivities in the individual WISP fields (light grey curves). The blue dashed line indicates the median $5\sigma$ field depth. The depth varies significantly field-to-field due to variations in the exposure time and background levels in each field.
\label{fig:line_depths}}
\end{figure}

Due to the nature of parallel observations, the integration times for each field was set by the primary target observations. We refer the reader to \citet{atek10} for a complete description of the observing strategy and data reduction. In brief, short opportunities (one to three continuous orbits; 180 fields) usually obtained G141 and one WFC3/IR filter, either F140W or F160W. However, a handful of short opportunities obtained G102 and F110W instead (4 fields). Long opportunities (four or more continuous orbits; 299 fields) obtained G102+G141 and two WFC3/IR filters, either F110W+F140W or F110W+F160W, except for Cycle~19 which used F140W-only. Cycle~20 and beyond are almost exclusively deep fields (G102+G141; See Table~\ref{tab:wisp_obs_summary}). In general, grism integration times are $\sim$6$\times$ those for the 
direct images. For the long opportunities, the integration times in the two grisms were set to achieve approximately uniform sensitivity for an emission line of a given flux across the full wavelength range ($\sim$5:2 for G102:G141). The median $5\sigma$ detection limit for emission lines fluxes (point source) in both grisms is $\sim5\times 10^{-17}$~erg~s$^{-1}$~cm$^{-2}$ \citep{atek10}, but we stress that this varies considerably with the length of the opportunity and variation in background levels (zodiacal light, Earth limb brightening, and telescope thermal emission). The line detection limits for each individual WISP fields are presented in Figure~\ref{fig:line_depths} and Table~\ref{tab:line_depths}.

\begin{table*}
\caption{WISP emission line depths for the 419 fields in the emission line catalogue \label{tab:line_depths}}
\begin{center}
\begin{tabular}{ccccc}
 \hline \hline
Par & RA & Dec & $1\sigma$ Depth G102$^a$ & $1\sigma$ Depth G141$^a$ \\
 &  &  & (\esc) & (\esc) \\
 \hline
1 & 01:06:35.29 & $+$15:08:53.8 & $9.67\times10^{-18}$ & $9.57\times10^{-18}$ \\
2 & 01:25:10.02 & $+$21:39:13.7 & ... & $1.17\times10^{-17}$ \\
5 & 14:27:06.64 & $+$57:51:36.2 & $7.72\times10^{-18}$ & $5.30\times10^{-18}$ \\
6 & 01:50:17.18 & $+$13:04:12.8 & $1.15\times10^{-17}$ & $8.76\times10^{-18}$ \\
7 & 14:27:05.90 & $+$57:53:33.7 & $1.54\times10^{-17}$ & $8.56\times10^{-18}$ \\
\multicolumn{5}{c}{ ... } \\[1ex] 
 \hline
\end{tabular}
\end{center}
\textbf{Notes.} A full ASCII version of this table is available online. \\
$^a$Grism flux limits depend on wavelength (see Figure~\ref{fig:line_depths}). Values presented here are at $\lambda=1.1\mu\mathrm{m}$ for G102 
and $1.5\mu\mathrm{m}$ for G141.
\end{table*}

Due to \hst\ only having a single dispersion direction for each grism, and position/orientation being tied to the Primary observing target, all grism data are single orientation. We note that multi-orientation pure-parallel observations are possible on \jwst\ due to the availability of multi-directional grisms, and that this is already being utilised (e.g., Cycle~1 PASSAGE survey with NIRISS, GO~1571; PI: M. Malkan, \citealt{malkan21}). Multi-orientation grism data can be used to construct spatially resolved emission line maps \citep[e.g.,][]{pirzkal18, wang22}.

All WISP data are reduced and calibrated with the WFC3 pipeline \texttt{calwf3}, together with custom scripts to improve the calibration and account for the specific challenges of undithered, pure-parallel observations.
The foundation of the WISP reduction pipeline is described in \citet{atek10}. Here we additionally implement a multiple component fit to the sky background in the grism images. While the ``input'' spectrum of the
sky can be considered uniform across the field-of-view, the sky emission is not uniform on the detector. Typically a single master sky image is scaled and fit to remove the background in each grism image, under the assumption that the relative strength of the background components remain the same exposure to exposure.
However, as noted by \citet{brammer12}, the structure in the sky of WFC3 grism images is highly variable, with zodiacal light and a helium emission line at 1.083 \micron\ contributing. While the zodiacal component should stay relatively constant throughout a full set of exposures taken on the same date, the helium emission comes from Earth's upper atmosphere and can vary on timescales shorter than the length of an exposure as HST moves closer or farther from Earth and the telescope pointing changes. We model and subtract the background in each grism exposure with a linear combination of the zodiacal light and helium sky images created by \citet{brammer15}, calculating the amplitudes of each sky component with two iterations of a least squares fit. 

We use the \texttt{AstroDrizzle} and \texttt{TWEAKREG} routines of \texttt{DrizzlePac} \citep{gonzaga12, hoffmann21} to combine the individual exposures, correcting for astrometric distortions and any alignment issues between exposures. The IR direct and grism images are drizzled onto a 0.08"/pixel scale. Object detection in the IR direct images (F110W, F140W, and F160W) is performed with Source Extractor \citep[version 2.5;][]{bertin&arnouts96}. For fields with imaging in two filters, we create a combined detection image and supplement the catalog with sources detected individually in only one of
the filters. We use the \texttt{aXe} software package \citep{kummel09} to extract and calibrate the spectra, using the Source Extractor catalogs as inputs.

\subsection{\hst\ WFC3/UVIS Imaging}\label{wisp_uvis}

For 155 of the long opportunity fields, WFC3/UVIS (FOV = 2.7\arcmin $\times$2.7\arcmin) imaging data were also obtained with some combination of F475X, F606W, F600LP, and/or F814W filters (only 2 of these at most for a single field). These observations were all carried out as part of the main parallel survey (PI: M. Malkan; see Table~\ref{tab:wisp_obs_summary} for cycles and GO IDs). The variety in UVIS photometric filters used is a result of a change in strategy between cycles. Specifically, in the first cycle of WISP observations (Cycle 17), the F475X and F600LP filters were utilised to maximise throughput. However, it was realised that the less-wide filters provided more information for spectral energy distribution (SED) fitting, and thus future observations switched to F606W and F814W. In Cycle 18, we switched to 2x2 binned observations for the UVIS imaging, although it was later realised that the combination of lack of binned calibrations and pixel based charge transfer efficiency corrections made the data less useful. In Cycle 19, we only obtained short parallel opportunities, and thus none included UVIS imaging. Starting with Cycle 20 and onwards, all UVIS observations were obtained in un-binned F606W and F814W data.  In total we have 70 and 43 fields with two and one un-binned UVIS filters, respectfully, for a total of 113 fields with G102+G141 grism spectroscopy and excellent UVIS imaging.

\subsubsection{\hst\ Data Reduction and Co-addition}\label{hst_reduction}

The UVIS images are also reduced through \texttt{calwf3}, along with additional custom calibrations described in \citet{rafelski15} including custom dark calibration files to address hot pixel masking, background gradients, and blotchy background patterns. These calibrations only apply to unbinned WFC3 data and, as a result, the binned UVIS data in Cycle~18 are not included in this data release. We first applied a pixel-based Charge Transfer Efficiency (CTE) correction on all individual raw dark files. When post-flashed dark files were available, we used those. The Cycle~17 UVIS data (F475X and F600LP) had high background and minimal CTE degradation. All UVIS data since Cycle~20 (F606W and F814W) use post-flash or have sufficient background on their own based on the recommendations at the time. We then used the dark files from a 3-5 day window to create a super dark to identify the hot pixels. These hot pixels use a lower threshold to identify hot pixels otherwise lost due to CTE degradation. We also create a master superdark that stacks all dark observations within the same anneal cycle as the observations, which we subtract from the data within \texttt{calwf3} to reduce the blotchy patterns and gradients. The details of all these steps are outlined in the Appendix of \citet{rafelski15}. This dark calibration methodology was partly incorporated as part of UVIS 2.0 \citep{bourque16} utilizing post-flashed CTE corrected darks. Then later in 2020 the dark reference files moved to contemporaneous darks. The pipeline still does not include improvements for hot pixels. For future reductions, we point to the hot pixel treatment in \citet{prichard22} and \citet{revalski23}, which also includes additional improvements to artifacts caused by CTE degradation such as matched amplifier backgrounds and read out cosmic rays (ROCRs).  

The UVIS image mosaics were created with \texttt{AstroDrizzle} and \texttt{TWEAKREG}. 
The images where corrected for cosmic rays and aligned to the NIR images. The final image drizzle parameters are set to have \texttt{final\_scale} of 0.04"/pixel,  \texttt{final\_pixfrac}=0.75, and The \texttt{final\_wht\_type}=IVM (inverse-variance weighting map). We also create NIR and UVIS image mosaics with matched plate scales (\texttt{final\_scale}) of 0.04"/pixel, 0.08"/pixel, and 0.13"/pixel. The use of \texttt{final\_scale} of 0.08"/pixel or 0.13"/pixel for UVIS data produces significantly undersampled images, and we therefore use the 0.04"/pixel NIR and UVIS mosaics for matched photometry. These images are then convolved with a kernal to match the point spread-function (PSF) of all the images to that of the F160W filter. We used the \texttt{IRAF} task \texttt{PSFMATCH} to calculate the matching kernals and convolve the images. Additionally, RMS images are created by taking $1/\sqrt{(\mathrm{IVM})}$ and cleaned images of the UVIS data are generated by masking edges, chip gaps, and bad pixels with randomly generated Gaussian noise. The pixel-matched imaging data in both WFC3/UVIS and WFC3/IR were made available in previous data releases from the WISP team via the survey website on MAST\footnote{\url{https://archive.stsci.edu/prepds/wisp/}}.

\subsection{Ground-based Imaging}
The ground-based data were taken between the 2010A-2019A semesters from various PIs across several independent programs. The data were taken with Palomar/LFC (31~nights), Palomar/WaSP (2~nights), WIYN/MiniMosaic (12~nights), WIYN/ODI (9.5~nights), and Magellan/Megacam (3.5~nights). A summary of the programs, including their IDs, PIs, dates, and the number of nights, are listed in Table~\ref{tab:obs_summary}.

\begin{table*}
\caption{Summary of ground-based imaging campaigns on WISP fields included in this release \label{tab:obs_summary}}
\begin{center}
\begin{tabular}{llllll}
 \hline \hline
ID & PI(s) & Telescope/Instr. & Semester & Bands & Nights \\ 
 \hline
P-23 & B. Siana & Palomar/LFC &  2010A & $gi$ &  4 \\ 
H-05 & B. Siana & Palomar/LFC &  2010B & $gi$ &  2 \\ 
J-06 & C. Scarlata & Palomar/LFC &  2010B & $gi$ &  3 \\ 
0438 & A. Henry & WIYN/MiniMo &  2010B & $gri$ &  4 \\ 
0160 & A. Henry & WIYN/MiniMo &  2011A & $gi$ &  5 \\ 
0222 & A. Henry & WIYN/MiniMo &  2011B & $g$ &  3 \\ 
J-18 & H. Telpitz & Palomar/LFC &  2012A & $gi$ &  3 \\ 
J-01 & M. Rafelski & Palomar/LFC &  2013A & $gi$ &  2 \\ 
J-10 & M. Rafelski & Palomar/LFC &  2013B & $gi$ &  3 \\ 
J-23 & M. Rafelski & Palomar/LFC &  2014A & $gi$ &  2 \\ 
J-05 & Y. Dai  & Palomar/LFC &  2014B & $gi$ &  2 \\ 
J-16 & J. Colbert & Palomar/LFC &  2015A & $gi$ &  3 \\ 
J-12 & Y. Dai & Palomar/LFC &  2016A & $g$ &  1 \\ 
J-14 & Y. Dai  & Palomar/LFC &  2016B & $gi$ &  3 \\ 
J-14 & I. Baronchelli & Palomar/LFC &  2017B & $gi$ &  3 \\ 
0298 & A. Battisti & WIYN/ODI &  2017B & $ugr$ &  2.5 \\ 
... & P. McCarthy & Magellan/Megacam &  2017B & $ugr$ &  1.5 \\ 
 & \& A. Battisti  & & & & \\
0275 & A. Battisti & WIYN/ODI &  2018A & $ugr$ &  2.5 \\ 
... & A. Battisti & Magellan/Megacam &  2018A & $ugr$ &  2 \\ 
0140 & A. Battisti & WIYN/ODI &  2018B/19A & $ugr$ &  4.5$^a$  \\ 
N-115 & Y. Dai & Palomar/WASP &  2019A & $g$ &  2 \\ 
 \hline
 \end{tabular}
\end{center}
 \textbf{Notes.} Columns list the (1) Program ID, (2) Principal Investigator, (3) Telescope and instrument used, (4) Semester observed, (5) Bands observed, (6) Number of nights where data was taken. $^a$Three of these nights were provided as additional technical time
\end{table*}

The observing campaign can be broadly divided into two categories: 
(1) programs with Palomar (LFC or WASP) and WIYN/MiniMosaic focused on observations in the $g$ and $i$ bands, where $i$-band was only obtained for fields without \hst /UVIS F814W or F600LP data (these filters have similar effective wavelengths), with the goal of improving stellar mass estimates for WISP galaxies; 
(2) programs with WIYN/ODI and Magellan/Megacam focused on observations in $u$, $g$, and $r$ bands, where $r$-band was only obtained for cases without \hst /UVIS F606W data (similar in wavelength), with the goal of providing rest-frame UV measurements to constrain reddening from dust attenuation \citep{battisti22}. The WIYN/ODI and Magellan/Megacam instruments have large fields-of-views (FOVs, 40\arcmin $\times$48\arcmin and 24\arcmin $\times$24\arcmin , respectively), which are significantly larger than the \hst\ WFC3/IR footprint (FOV = 2.3\arcmin $\times$2.2\arcmin). Whenever possible, observations were optimised to observe multiple WISP fields simultaneously.

The desired depths of the programs varied, but for WIYN/ODI and Magellan/Megacam they were AB mag=26, 26, 25.3 in $u$, $g$, and $r$ (5$\sigma$ point-source), respectively. These depths were typically achieved for the Magellan runs but not for most of the WIYN runs, due to poor weather and/or observing conditions (e.g., lunar phase). A detailed summary on the depths for each facility is given in Section~\ref{depth_summary}.

\subsubsection{Ground-based Data Reduction and Co-addition}\label{ground_reduction}
Below we describe the data reduction process for each telescope individually, but note that they follow a similar overall procedure that includes bias, flat, and dark-subtraction. All images were astrometrically aligned to the World Coordinate System (WCS) using the \texttt{astrometry.net} software package, unless otherwise specified. Flux calibrations were also performed in similar ways, unless otherwise specified, and is summarised in Section~\ref{ZP}.

\begin{itemize}
\item \textit{Palomar/LFC --}
The Palomar/LFC WISP observations spanned the largest number of different observers among our datasets. All raw images were 
re-reduced following the same procedure for self-consistency. The LFC consists of six chips (each 6.14\arcmin $\times$12.29\arcmin; referred to as \texttt{chip0} through \texttt{chip5}), covering a circular area with a $\sim$24\arcmin\ diameter, and the WISP field was always centered on \texttt{chip0}. Basic data reduction steps  were only performed on the \texttt{chip0} data. 
We note that dark frames were not always available for subtraction, however, this has a minimal impact on the final photometric quality because the LFC has very little dark current\footnote{\url{https://sites.astro.caltech.edu/palomar/observer/200inchResources/lfccookbook.html\#dark}}.

\item \textit{Palomar/WaSP --}
Basic reduction steps were identical to Palomar/LFC, except that WaSP has 4 large chips (square FOV: 18.4\arcmin $\times$18.5\arcmin) and all chips were reduced.

\item \textit{WIYN/Minimosaic --}
  the raw archival WIYN data were obtained via the NOAO Science Archive (now known as the NOIRLab Astro Data Archive\footnote{\url{https://astroarchive.noirlab.edu}}). Basic reduction steps were similar to Palomar, except that Minimosaic has 2 large chips (square FOV: 9.6\arcmin $\times$9.6\arcmin) and both chips were used. 

\item \textit{WIYN/ODI --}
  basic reduction, WCS-alignment, and photometric calibration was performed using the ODI Pipeline, Portal, and Archive (PPA) system\footnote{\url{https://portal.odi.iu.edu/index/front}}. PPA is a service provided by the WIYN Consortium, Inc., and hosted at Indiana University. The photometric calibration in the ODI PPA is based on measurements from the SDSS ($ugriz$) and Pan-STARRS ($grizy$). We note that for a few fields outside the SDSS footprint, the $u$-band ODI data required manual photometric calibration, which is discussed in Section~\ref{ZP}.

\item \textit{Magellan/Megacam --}
  images with basic reduction and WCS-alignment were provided by the OIR Telescope Data Center at Harvard's Center for Astrophysics (CfA), supported by the Smithsonian Astrophysical Observatory.
\end{itemize}

The process for co-adding images was the same for all ground-based data. First, images are normalised by exposure time. Second, normalised images are combined using the \texttt{SWarp} software \citep{bertin02, bertin10}. During \texttt{SWarp}, images are background subtracted using a 128 pixel background mesh unless saturated stars bleed onto a significant portion of field. In such cases, a 32 pixel background is adopted to minimise the area affected by saturation. During this step we also mask out satellite trails from individual exposures if they overlap with the WISP field (satellite trails outside the WISP area are ignored). For convenience, the publicly released images are cropped into 5\arcmin $\times$5\arcmin\ regions centered on the WISP fields. For reference, the \hst /UVIS FOV is 2.70\arcmin $\times$2.70\arcmin\ and the \hst /IR FOV is 2.05\arcmin $\times$2.27\arcmin .

\subsection{\spitzer\ IRAC Imaging}\label{spitzer}
\spitzer\ IRAC/channel~1, 3.6~\micron\ imaging (warm mission; PI: J Colbert) was obtained for $\sim$200 of the deepest WISP field over multiple cycles (GO 80134 (Cycle~8), 90230 (Cycle~9), 10041 (Cycle~10), 12093 (Cycle~12)), with the primary goal of providing accurate stellar masses for galaxies in the WISP survey. For a handful of WISP fields, IRAC/channel~2, 4.5\micron\ data were also obtained.

\subsubsection{\spitzer\ Data Reduction and Co-addition}\label{spitzer_reduction}
Roughly half of the WISP parallels are in low background fields ($\sim$0.08 MJy/sr at 3.6\micron ), while the remainder are in fields with medium backgrounds ($\sim$0.12 MJy/sr). To achieve similar sensitivity for all the targeted fields (5$\sigma$ depth of 0.9$\mu$Jy), we split the observations into two integration times, 25 versus 35 minutes on-source. The variation of the background with observing window does not significantly change the required exposure time. For the observations, we use a medium dither pattern with 100 second frames and either 15 or 21 exposures, depending on the total integration time. For the small subset of fields observed with the 4.5\micron\ filter, we always used the 21 exposure (35 minute) observation.

Individual calibrated IRAC exposures are referred to as Basic Calibrated Data, or BCDs. However, IRAC BCD data contain several artifacts, including mux-bleed, mux-stripe, column droop, and bright star ghosts\footnote{\url{https://irsa.ipac.caltech.edu/data/SPITZER/docs/irac/iracinstrumenthandbook/35/}}. The
IRAC pipeline therefore also produces corrected BCD images (cBCD), which attempt to mitigate these artifacts. We start with the cBCDs for all of our data reduction and analysis. 

All the IRAC mosaics are generated using the Spitzer MOsaicker and Point source EXtractor (\texttt{MOPEX}) package. Before generating mosaics, problematic cBCDs -- those with unusually high noise, extreme saturation, or other unexplained large artifacts -- are removed. Such problematic cBCDs only make up a small fraction (<5\%) of the input cBCDs and do not significantly affect total exposure times for any observations. The first frame of all IRAC Astronomical Observation Requests (AORs) has decreased sensitivity and was not included in the mosaics. The \texttt{MOPEX Overlap} pipeline subtracts the estimated background from Zodiacal light from each CBCD and then matches the background level in all frames with an additive correction. The \texttt{MOPEX Mosaic} pipeline then performs outlier rejection to remove cosmic rays, moving objects, and other artifacts, before re-sampling the images onto a common reference frame. Finally, the images are combined to produce a weighted mean and median mosaic along with associated coverage, uncertainty, and standard-deviation maps.

\section{WISP Photometric Pipeline}\label{photo_pipeline}
The methods to obtain zeropoints and photometry differ between the \hst\ data (high spatial resolution) and the ground-based and  \spitzer\ data (both low spatial resolution). In this section we describe the data processing workflow used to create the WISP photometric catalogue.

\subsection{Photometric Zeropoints}\label{ZP}
\subsubsection{\hst\ Zeropoints}
The \hst\ WFC3/UVIS and WFC3/IR zeropoints are taken from the STScI calibration website\footnote{\url{https://www.stsci.edu/hst/instrumentation/wfc3/data-analysis/photometric-calibration}}. We note that the WFC3 zeropoint solutions change slightly over time and different zeropoint values are applied to data from different \hst\ cycles over the course of the program.

\subsubsection{Ground-based Data Zeropoints}\label{ground_zeropoints}
For all ground-based imaging, the flux zeropoints were determined, in order of priority, from: (1) direct comparison to SDSS ($ugri$; $110\degree\lesssim RA\lesssim 265\degree $ and $Dec>0\degree $), (2) direct comparison to Pan-STARRS ($gri$; full sky for $Dec>-30\degree $), (3) using standard fields (i.e., fields with (1) or (2)) at different airmasses (atmospheric extinction) for individual nights of each observation. 
Option (3) was necessary when neither SDSS or Pan-STARRS are available for calibration. This was required for many of the southern WISP fields observed with Magellan, particularly for the $u$-band imaging. 

First, we extract \texttt{psfMag} values for stars (\texttt{PhotoType}=6) in the SDSS DR14 and Pan-STARRS DR2 public catalogues to use as our reference values.
Magnitudes in Pan-STARRS were converted into the SDSS system using the relationships in \citet{tonry12}. Second, we run \se\ \citep{bertin&arnouts96} with similar parameters across facilities, for consistency, on our reduced, co-added images (Section~\ref{ground_reduction}). A summary of the most relavant \se\ parameters are listed in Table~\ref{tab:wisp_sextractor}. Third, magnitude zeropoints are determined by cross-matching close sources and comparing \texttt{MAG\_AUTO} values from \se\ to the catalogue \texttt{psfMag} values (both represent total magnitudes). A distance threshold of $\sim$1\arcsec\ was typically adopted, but this varied with seeing conditions (maximum of 3\arcsec). 

The filters for all of the ground-based data (see Figure~\ref{fig:filters}b) are similar to SDSS. As a result, colour differences tend to be small, however, this was still accounted for during the ZP determination. In particular, the Magellan $g$ and $r$ and Palomar $i$ filters show the largest deviations from SDSS. 
These colour corrections are based on fitting the zeropoint magnitudes as a function of $u-g$, $g-r$, $g-r$, and, $r-i$ colours for the $u$, $g$, $r$, and $i$ filters, respectively.  

The typical precision of the ZPs are $\sigma_{ZP}\sim0.05$~mag for Magellan/Megacam ($ugr$), WIYN/ODI ($ugr$), WIYN/MiniMosaic ($gri$), and Palomar/WaSP ($g$). For Palomar/LFC ($gi$), the typical precision is slightly worse, $\sigma_{ZP}\sim0.1$~mag, due to the smaller FOV with fewer reference stars for calibration. 
We recommend adopting a minimum uncertainty of 0.1~mag for all ground-based optical data to account for potential systematic ZP offsets.

\begin{table}
\caption{Summary of \se\ parameters used on the ground-based data for determining zeropoints \label{tab:wisp_sextractor}}
\begin{center}
\begin{tabular}{ll}
 \hline \hline
Parameter & Value \\
 \hline
\texttt{DETECT\_THRESH} & 1.5 \\
\texttt{ANALYSIS\_THRESH} & 1.5 \\
\texttt{DEBLEND\_NTHRESH} & 32 \\
\texttt{DEBLEND\_MINCONT} & 0.005 \\
\texttt{DETECT\_MINAREA} & 6 \\
\texttt{BACK\_SIZE} & 64 \\
\texttt{BACK\_FILTERSIZE} & 3 \\
\texttt{FILTER} & Y \\
\texttt{FILTER\_NAME} & gauss\_2.0\_5x5.conv$^a$ \\[1ex] 
\hline
 \end{tabular}
\end{center}
\textbf{Notes.} The photometry from \se\ is used only for constraining magnitude zeropoints for the ground-based data. We use \texttt{TPHOT} to measure photometry for the WISP catalogue (see Section~\ref{ground_photo}).
$^a$5$\times$5 convolution mask of a Gaussian PSF with FWHM = 2.0 pixels. \\
\end{table}

For rare instances of overlapping data (e.g., both Magellan and Palomar data are available in the $g$ filter), we adopt the case with better seeing (typically also deeper), unless there is a dramatic difference in depth relative to the poorer seeing case ($\gtrsim1$~mag difference). A few comparisons between overlapping data from different telescopes with roughly similar imaging quality (depth, seeing) were possible and used as a check on zeropoints and colour-corrections. We find that cases of overlapping data agree to within 1$\sigma$ for $\sim$60-70\% of sources, as expected. 
We find that the agreement gets poorer (larger scatter) when comparing datasets with larger quality differences (e.g., 1" vs 1.7" seeing), however the median ZP offsets remain consistent with zero.

\begin{figure*}
\includegraphics[width=0.98\textwidth]{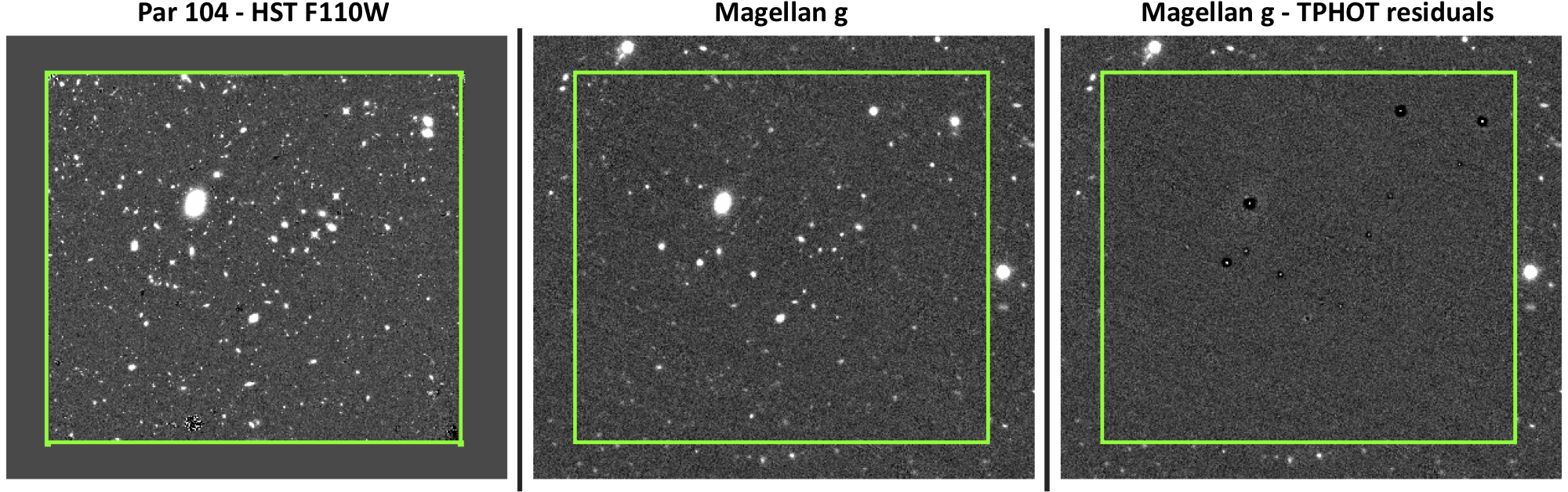}
\caption{\hst\ images and segmentation maps are used to perform `forced' photometry on the lower-resolution data (ground-based and \spitzer) by combining the PSFs of both \hst\ and the lower-resolution data (detailed in Section~\ref{ground_photo}). As an example, we show data from Par 104: (\textit{left}) the \hst\ F110W image, (\textit{middle}) the Magellan $g$ image, and (\textit{right}) the residual of the Magellan $g$ image after running through \texttt{TPHOT}. Instances of over-subtracted residuals (black circles) are expected to occur for sources that are saturated in the \hst\ image (mostly foreground stars); for these cases the \hst\ (deeper) light-profile models are less accurate relative to the same sources in the unsaturated (shallower) low-resolution data. The \hst\ WFC3/IR FOV (2.05\arcmin$\times$2.27\arcmin) is indicated by the green box in all panels.
 \label{fig:pipeline}}
\end{figure*}

\begin{table*}
\caption{Summary of the number of fields, $N$, with each filter in the WISP Photometric catalogue \label{tab:filter_field_summary}}
\begin{tabular}{|c|c|c|c|c|c|c|c|c|c|c|c|c|c|}
\hline
Filter & $u$ & $g$ & {\scriptsize F475X} & {\scriptsize F606W} & $r$ & {\scriptsize F600LP} & $i$ & {\scriptsize F814W} & {\scriptsize F110W} & {\scriptsize F140W} & {\scriptsize F160W} & I1 & I2  \\ $\lambda_\mathrm{eff}(\mu \mathrm{m})$ & 0.3551 & 0.4686 & 0.4939 & 0.5893 & 0.6165 & 0.7444 & 0.7481 & 0.8060 & 1.1534 & 1.3923 & 1.5396 & 3.550 & 4.493 \\
\hline
$N$ & 113 & 167 & 19 & 43 & 68 & 19 & 57 & 99 & 233 & 168 & 261 & 175 & 3 \\
\hline
\end{tabular} \\
\vspace{1mm}\textbf{Notes:} I1=IRAC1, I2=IRAC2. The catalogue contains data for 439 out of 483 WISP fields. For details on individual fields, refer to Table~\ref{tab:phot_depth_catalog}.
\end{table*}

\subsubsection{\spitzer\ Zeropoints}
Similar to \hst, the \spitzer\ zeropoints are taken from the IRAC instrument handbook website\footnote{\url{https://irsa.ipac.caltech.edu/data/SPITZER/docs/irac/iracinstrumenthandbook/14/}}. We note that the WISP data were obtained during the \spitzer\ warm-mission, which had a different zeropoint solution to the cold-mission.

\subsection{\hst\ PSF- and Aperture-matched Isophotal Photometry (high spatial resolution)}\label{hst_photo}
The methods used to obtain the \hst\ photometry are described in \citet{bagley17} and \citet{henry21}. In brief, a segmentation map is first generated from the F110W and F160W detections using \se\ on the 0.08"/pixel mosaics and then re-gridded to the 0.04"/pixel scale mosaics. The photometry is deteremined using \texttt{photutils} in Astropy \citep{astropy13, astropy18} to derive isophotal fluxes in all \hst\ bands using the re-gridded segmentation map and includes local sky subtraction. \se\ photometry is also performed on WFC3/IR (0.08"/pixel) images (prior to PSF-matching), in order obtain the total magnitudes (\texttt{MAG\_AUTO}). 

 For the WFC3/UVIS data, we determine aperture corrections, \texttt{APCOR}, using the difference between \texttt{MAG\_AUTO} and the isophotal magnitude, \texttt{MAG\_ISO}, in the F160W \hst\ filter and list these values in our catalogue (i.e., \texttt{APCOR\_UVIS}=\texttt{MAG\_AUTO\_F160W}-\texttt{MAG\_ISO\_F160W}). This is similar to the method used in \citet{henry21, battisti22}. We recommend total magnitudes to be estimated for UVIS filters using $m_\mathrm{tot}$=\texttt{MAG\_ISO\_[UVISFILTER]}+\texttt{APCOR\_UVIS}. 
 For the WISP subsample used in our results (Secton~\ref{results}), the median value of these aperture corrections (\texttt{APCOR\_UVIS}) is $\Delta m=-0.11$~mag.

\subsection{Ground-based and \spitzer\ Forced Photometry (low spatial resolution)}\label{ground_photo}

We performed template-fitting or `forced' photometric measurements on the lower-resolution data (ground-based and IRAC; both datasets follow the same procedure) 
by exploiting the coordinates of the \hst-detected sources as priors. Figure~\ref{fig:pipeline} demonstrates an example of the photometric fitting from the \texttt{TPHOT} WISP pipeline on the ground-based data.

The core of this computation is performed using the \texttt{TPHOT} software \citep[v2;][]{merlin15, merlin16}, which we summarise below, and provide in more detail in Appendix~\ref{appendix:tphot_config}.
In brief, \texttt{TPHOT} requires two images at different resolutions and a list of detected objects. The algorithm transforms the low resolution image into a (simulated) high-resolution image. To do this, \texttt{TPHOT} ``distributes'' the flux of the sources in the low-resolution image according to the flux distribution in the high-resolution image 
using a deconvolution kernel (provided by the user).  Compared to aperture photometry (and closest-counterpart associations), this technique allows us to obtain more accurate photometric estimates, especially when multiple \hst\ sources appear blended in the low-resolution images.
For all of the \hst\ sources undetected in the low-resolution 
bands, we set their fluxes as an upper limit to the 3$\sigma$ detection limit.

\section{WISP Photometric Catalogue}\label{photo_catalog}
\subsection{Photometric Catalogue Description}\label{photo_catalog_description}
The WISP photometric catalogue contains $\sim$230,000 sources in 439 WISP fields. These sources include stars, galaxies, and active galactic nuclei (AGN). Fields that are absent are cases where the WISP photometric pipeline failed, which most commonly occurred due to overcrowding, bright stars, or persistence. We note that the filter availability is quite inhomogeneous among WISP fields and entries for fields without data in a given filter have values of `-99' for their photometry.  We provide a summary of the number of fields with individual filters in this data release in Table~\ref{tab:filter_field_summary}. The photometry in this catalogue are \textit{not} corrected for foreground Milky Way extinction.

\begin{figure}
\includegraphics[width=0.48\textwidth]{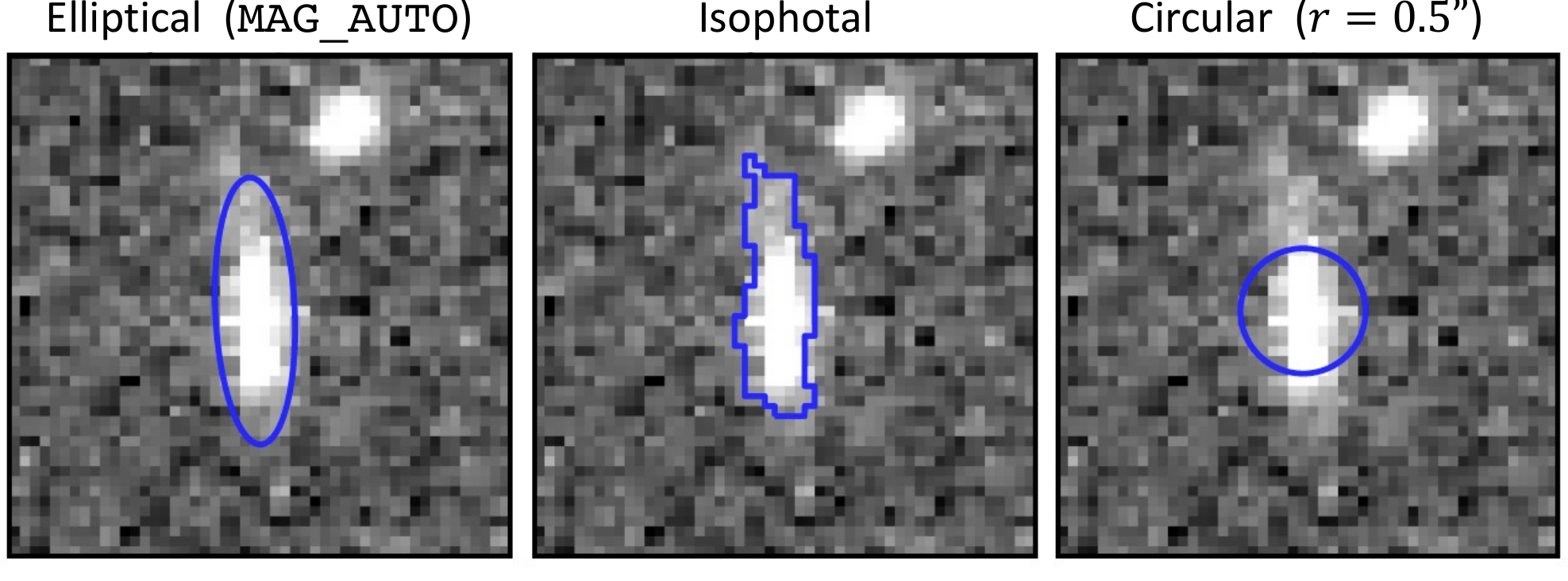}
\vspace{-5mm}
\caption{A demonstration of the three types of apertures available for \hst\ imaging data in the photometric catalogue.
 \label{fig:apertures}}
\end{figure}

We present an explanation of the WISP photometric catalogue entries in Table~\ref{tab:FITS_photo_catalog}. These include RA, Dec, sizes and position angles (based on \hst), and Kron-like elliptical aperture magnitude (\texttt{MAG\_AUTO}; only WFC3/IR filters) from \se. There are two additional aperture measurements available for \hst\ data: (1) isophotes (\texttt{ISO}) based on the F110W and F160W segmentation map and (2) fixed circular apertures (\texttt{APER}; radii $r$ = 0.2, 0.5, 1.0, 1.5\arcsec). A visual representation of these apertures is shown in Figure~\ref{fig:apertures}. For the ground-based and \spitzer\ data, photometry based on \texttt{TPHOT} is provided. \textit{For consistent `total' photometry, we recommend using} \texttt{MAG\_AUTO} \textit{for WFC3/IR,} \texttt{MAG\_ISO}+\texttt{APCOR\_UVIS} \textit{for WFC3/UVIS, and} \texttt{AB\_MAG} (\texttt{TPHOT} values) \textit{for ground-based and \spitzer\ data. } 

There are also two quality flags in the photometric catalogue. A description of these flags is provided in Appendix~\ref{appendix:phot_flags}. 
These flags should be carefully considered when selecting sources from the catalogue to avoid use of unreliable photometry.

\begin{table*}
\caption{Description of WISP Photometric catalogue \label{tab:FITS_photo_catalog}}
\begin{tabular}{ll}
\hline \hline
Title & Description \\ \hline

\texttt{PAR} & WISP field ID number \\
\texttt{OBJ} & Object ID number \\
\texttt{RA} & Decimal RA from HST images [deg] \\
\texttt{DEC} & Decimal Dec from HST images [deg] \\
\texttt{A\_WORLD} & \se\ profile RMS along major axis in world coordinates, measured on HST images [arcsec] \\
\texttt{B\_WORLD} & \se\ profile RMS along minor axis in world coordinates, measured on HST images [arcsec] \\
\texttt{THETA\_WORLD} & Position angle measured counter-clockwise from world x-axis, measured on HST images [deg] \\

\texttt{X\_IMAGE\_040} & Object x pixel coordinate in 0.04\arcsec/pixel scale image [pixels] \\
\texttt{Y\_IMAGE\_040} & Object y pixel coordinate in 0.04\arcsec/pixel scale image [pixels] \\
\texttt{A\_IMAGE\_040} & \se\ profile RMS along major axis measured on 0.04\arcsec/pixel HST images [pixels] \\
\texttt{B\_IMAGE\_040} & \se\ profile RMS along minor axis measured on 0.04\arcsec/pixel HST images [pixels] \\
\texttt{THETA\_IMAGE\_040} & Position angle counter-clockwise from x-axis, measured on 0.04\arcsec/pixel HST images [deg] \\

\texttt{X\_IMAGE\_080} & Object x pixel coordinate in 0.08\arcsec/pixel scale image [pixels] \\
\texttt{Y\_IMAGE\_080} & Object y pixel coordinate in 0.08\arcsec/pixel scale image [pixels] \\
\texttt{A\_IMAGE\_080} & \se\ profile RMS along major axis measured on 0.08\arcsec/pixel HST images [pixels] \\
\texttt{B\_IMAGE\_080} & \se\ profile RMS along minor axis measured on 0.08\arcsec/pixel HST images [pixels] \\
\texttt{THETA\_IMAGE\_080} & Position angle counter-clockwise from x-axis, measured on 0.08\arcsec/pixel HST images [deg] \\

\texttt{FLUX\_[APTYPE]\_[HSTFILTER]}$^a$ & Sky-subtracted signal within aperture [$\mu$Jy] \\
\texttt{FLUXERR\_[APTYPE]\_[HSTFILTER]}$^a$ & Aperture signal uncertainty [$\mu$Jy] \\
\texttt{MAG\_[APTYPE]\_[HSTFILTER]}$^a$ & Magnitude for \texttt{FLUX\_[APTYPE]\_[HSTFILTER]}; -99 for undetected sources [AB mag] \\
\texttt{MAGERR\_[APTYPE]\_[HSTFILTER]}$^a$ & Magnitude uncertainty; -99 for undetected sources [AB mag] \\

\texttt{FLUX\_SKY\_[HSTFILTER]}$^a$ & Background used for sky subtraction in all apertures; use 10\arcsec\ rectangle with sources masked [$\mu$Jy] \\
\texttt{FLUXERR\_SKY\_[HSTFILTER]}$^a$ & Sky aperture signal uncertainty [$\mu$Jy] \\

\texttt{MAG\_AUTO\_[NIRFILTER]}$^b$ & \se\ Kron-like elliptical aperture magnitude; limit reported for undetected sources [AB mag] \\
\texttt{MAGERR\_AUTO\_[NIRFILTER]}$^b$ & \se\ RMS error for AUTO magnitude; -99 for undetected sources [AB mag] \\
\texttt{FLAG\_[NIRFILTER]}$^b$ & \se\ extraction flags (8 bit flags; see Appendix~\ref{appendix:phot_flags}) \\

\texttt{AB\_MAG\_[FILTER]}$^c$ & Filter magnitude from \texttt{T-PHOT} output; limiting magnitude for undetected sources [AB mag] \\
\texttt{AB\_MAGERR\_[FILTER]}$^c$ & Filter magnitude uncertainty from \texttt{T-PHOT} output; -99 for undetected sources [AB mag] \\
\texttt{TPHOT\_FLAG\_[FILTER]}$^c$ & \texttt{T-PHOT} extraction flags (3 bit flags; see Appendix~\ref{appendix:phot_flags})
\\
\texttt{INSTR\_FLAG\_[GROUNDFILTER]}$^d$ & String indicating `Telescope-instrument' for ground-based data (e.g., `Magellan-Megacam'). \\
\texttt{APCOR\_UVIS} & Aperture correction to convert WFC3/UVIS isophotal magnitudes to total magnitude [AB mag] \\
\hline
\end{tabular} \\
\textbf{Notes:}  
$^a$\texttt{[ATYPE]} is one of: \texttt{ISO} (isophotes based on the F110W and F160W segmentation map) or \texttt{APER} (circular apertures of radii $r$ = 0.2, 0.5, 1.0, 1.5\arcsec; array has dimensions of ($n_\mathrm{obj}$, 4)). \texttt{[HSTFILTER]} is one of: F475X, F475Xc, F606W, F606Wc, F600LP, F600LPc, F814W, F814Wc, F110W, F110Wc, or F160W. `c' refers to photometry on images convolved to match the PSF of F160W. Only convolved fluxes are available for the \texttt{ISO} apertures. 
$^b$\texttt{[NIRFILTER]} is one of: F110W, F140W, or F160W.
$^c$\texttt{[FILTER]} is one of: $u$, $g$, $r$, $i$, I1, or I2. $^d$\texttt{[GROUNDFILTER]} is one of: $u$, $g$, $r$, $i$.
\end{table*}

\begin{table*}
\caption{Summary of WFC3/IR imaging depth and completeness for WISP fields in the photometric catalog \label{tab:wfc3ir_depth_completeness}} 
\begin{tabular}{|c|cccc|cccc|cccc|}
\hline
 & & \multicolumn{3}{c|}{{\scriptsize F110W} Depth (AB mag)} & & \multicolumn{3}{c|}{{\scriptsize F140W} Depth (AB mag)} & & \multicolumn{3}{c|}{{\scriptsize F160W} Depth (AB mag)} \\
Par & $N$ & Aperture & Point-source & Galaxy & $N$ & Aperture & Point-source & Galaxy & $N$ & Aperture & Point-source & Galaxy \\
\hline
  1 & 425 & 26.28 & 26.26 & 25.25 & 355 & 25.65 & 25.78 & 24.77 & $\bm{--}$ & $\bm{--}$ &   $\bm{--}$ &   $\bm{--}$ \\ \hline
  2 & $\bm{--}$ & $\bm{--}$ &   $\bm{--}$ &   $\bm{--}$ & 354 & 25.52 & 25.58 & 24.59 & $\bm{--}$ & $\bm{--}$ &   $\bm{--}$ &   $\bm{--}$ \\ \hline
  3 & 622 & 26.27 & 26.41 & 25.43 & 413 & 25.65 & 25.68 & 24.65 & $\bm{--}$ & $\bm{--}$ &   $\bm{--}$ &   $\bm{--}$ \\ \hline
  5 &  502 & 26.48 & 26.42 & 25.42 &  520 & 26.23 & 26.28 & 25.28 &  $\bm{--}$ & $\bm{--}$ & $\bm{--}$ & $\bm{--}$ \\ \hline 
  6 &  346 & 25.86 & 25.87 & 24.84 &  415 & 25.78 & 25.86 & 24.89 &  $\bm{--}$ & $\bm{--}$ & $\bm{--}$ & $\bm{--}$ \\ \hline 

\multicolumn{13}{|c|}{ ... } \\[1ex] 
\hline
\end{tabular} \\
\textbf{Notes:} $N$ is the number of detected objects in WISP photometric pipeline. `Aperture' is the average 5$\sigma$ depth within 0.5\arcsec\ sky regions. `Point-source' and `Galaxy' are the 50\% completeness depths for injected simulated point-sources and galaxies, respectively (see Section~\ref{depth_summary} and Figure~\ref{fig:phot_completeness_compare}). A full ASCII version of this table is available online. \\
\end{table*}

\setlength{\tabcolsep}{3pt} 
\begin{table*}
\caption{Summary of photometric availability and depth (AB mag) for WISP fields in this data release \label{tab:phot_depth_catalog}} 
\begin{tabular}{|c|c|cc|cc|ccc|c|cc|cc|}
\hline
Par  & $u$ & $g$ & {\scriptsize F475X} & {\scriptsize F606W} & $r$ & {\scriptsize F600LP} & $i$ & {\scriptsize F814W} & {\scriptsize F110W} & {\scriptsize F140W} & {\scriptsize F160W} & I1 & I2 \\
$\lambda_\mathrm{eff}(\mu \mathrm{m})$ & 0.3551 & 0.4686 & 0.4939 & 0.5893 & 0.6165 & 0.7444 & 0.7481 & 0.8060 & 1.1534 & 1.3923 & 1.5396 & 3.550 & 4.493 \\
\hline
\rowcolor{lightOrange} 1 &  {\color{DarkGreen} 25.06} &  {\color{DarkGreen} 26.29} &  {\color{red} $\bm{--}$} &  {\color{red} $\bm{--}$} &  {\color{DarkGreen} 25.82} &  {\color{red} $\bm{--}$} &  {\color{DarkGreen} 26.35} &  {\color{red} $\bm{--}$} &  {\color{DarkGreen} 26.02} &  {\color{DarkGreen} 25.59} &  {\color{red} $\bm{--}$} &  {\color{DarkGreen} 23.88} &  {\color{red} $\bm{--}$}  \\ \hline
2 &  {\color{red} $\bm{--}$} &  {\color{DarkGreen} 27.16} &  {\color{red} $\bm{--}$} &  {\color{red} $\bm{--}$} & {\color{red} $\bm{--}$} &  {\color{red} $\bm{--}$} &  {\color{DarkGreen} 26.28} &  {\color{red} $\bm{--}$} & {\color{red} $\bm{--}$} &  {\color{DarkGreen} 24.94} &  {\color{red} $\bm{--}$} &  {\color{red} $\bm{--}$} & {\color{red} $\bm{--}$}  \\ \hline
 3 &  {\color{red} $\bm{--}$} &  {\color{DarkGreen} 26.68} &  {\color{red} $\bm{--}$} &  {\color{red} $\bm{--}$} & {\color{red} $\bm{--}$} &  {\color{red} $\bm{--}$} &  {\color{DarkGreen} 26.14} &  {\color{red} $\bm{--}$} & {\color{DarkGreen} 25.91} &  {\color{DarkGreen} 25.42} &  {\color{red} $\bm{--}$} &  {\color{DarkGreen} 23.75} & {\color{red} $\bm{--}$} \\ \hline
\rowcolor{lightOrange} 5 &  {\color{DarkGreen} 24.48} &  {\color{DarkGreen} 26.90} &  {\color{red} $\bm{--}$} &  {\color{red} $\bm{--}$} &  {\color{DarkGreen} 24.64} &  {\color{red} $\bm{--}$} &  {\color{DarkGreen} 26.23} &  {\color{red} $\bm{--}$} &  {\color{DarkGreen} 26.36} &  {\color{DarkGreen} 25.97} &  {\color{red} $\bm{--}$} &  {\color{DarkGreen} 23.53} &  {\color{red} $\bm{--}$} 
 \\ \hline
6 &  {\color{red} $\bm{--}$} &  {\color{DarkGreen} 26.91} &  {\color{red} $\bm{--}$} &  {\color{red} $\bm{--}$} &  {\color{red} $\bm{--}$} &  {\color{red} $\bm{--}$} &  {\color{DarkGreen} 26.06} &  {\color{red} $\bm{--}$} &  {\color{DarkGreen} 25.68} &  {\color{DarkGreen} 25.38} &  {\color{red} $\bm{--}$} &  {\color{DarkGreen} 23.85} &  {\color{red} $\bm{--}$} 
 \\ \hline
 
\multicolumn{14}{|c|}{ ... } \\[1ex] 
\hline
\end{tabular} \\
\textbf{Notes:} The filters available for each field are indicated in green with the median 5$\sigma$ magnitude (AB) for detected sources (red dashes indicate no data). Rows highlighted in orange have `full' UV to near-IR coverage (6 or more filters). The $g$-band and F475X cover roughly similar wavelengths and typically only one of these were obtained. This similarly occurs for F606W/$r$-band, F600LP/$i$-band/F814W, and F140W/F160W. I1=IRAC1, I2=IRAC2. A full ASCII and PDF version of this table is available online. \\
\end{table*}

\subsection{Summary of Imaging Depth and Completeness}\label{depth_summary}

Since the segmentation maps of sources from WFC3/IR imaging data are used as priors for the shallower and/or lower-resolution imaging data, we only formally characterise the depth and completeness for the WFC3/IR images. This is because, by design, every source detected in the WFC3/IR catalogue has an entry for each additional filter with data (either a detection or upper limit; i.e., 100\% `complete'). Below we outline the method for estimating these values for the WFC3/IR images. We also provide a representative depth for each filter based on median 5$\sigma$ magnitude (AB) for detected sources, which we also describe below.

For the \hst\ imaging data, there are considerable differences in the depths and completeness achieved from field to field. This arises from the parallel nature of the program, where exposure times are set by the Primary observation. To estimate this, we use the code described in \cite{prichard22}\footnote{\url{https://github.com/lprichard/hst_sky_rms}} and briefly summarise here.
First, the code determines the rms background value in 1000 randomly selected empty sky regions, where the segmentation map from \se\ is used to exclude regions overlapping with sources. Next, it generates a histogram of the rms values and fits this distribution with a Gaussian profile to determine the sigma-clipped median. The median rms value is multiplied by the correlated pixel noise correction factor for that filter (for details, see Section 3.3 of the \texttt{DrizzlePac} Handbook\footnote{\url{https://hst-docs.stsci.edu/drizzpac}}). For the drizzle parameters of the WFC3/IR images, as described in Section~\ref{hst_reduction}, the correlated noise correction factor is 1.678. The corrected median rms is then converted into a limiting magnitude using the zero-point and a specified aperture size. 
We provide an estimate of depth for the WFC3/IR filters for a fixed-aperture size of 0.5\arcsec\ radius in Table~\ref{tab:wfc3ir_depth_completeness}. 

\begin{figure*}
$\begin{array}{ccc}
\includegraphics[width=0.31\textwidth]{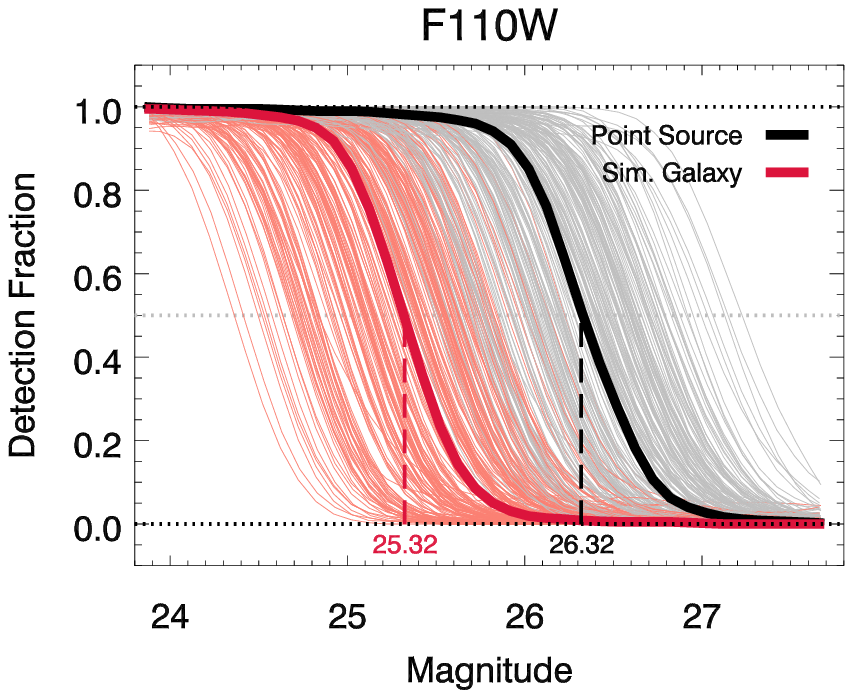} &
\includegraphics[width=0.31\textwidth]{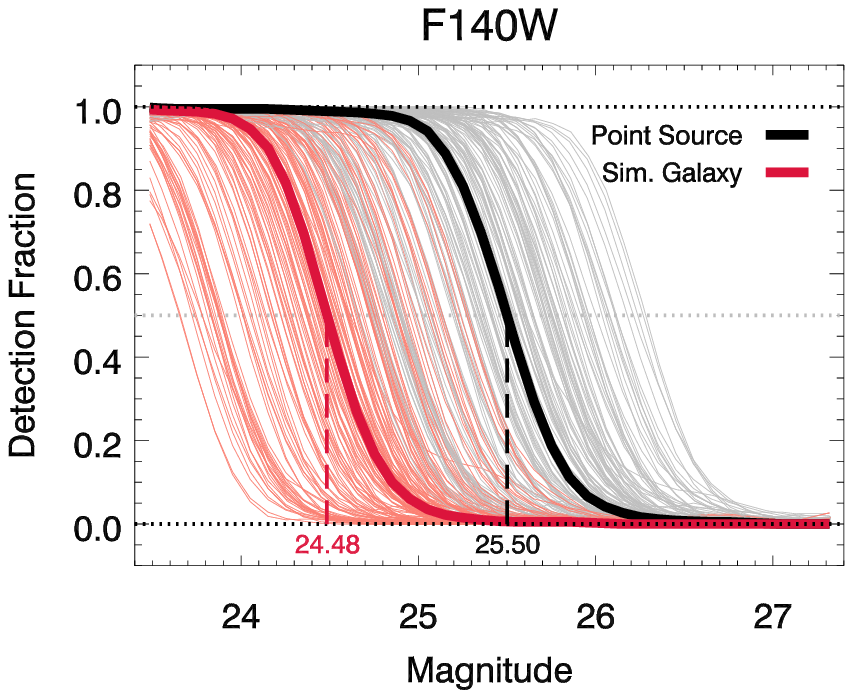} &
\includegraphics[width=0.31\textwidth]{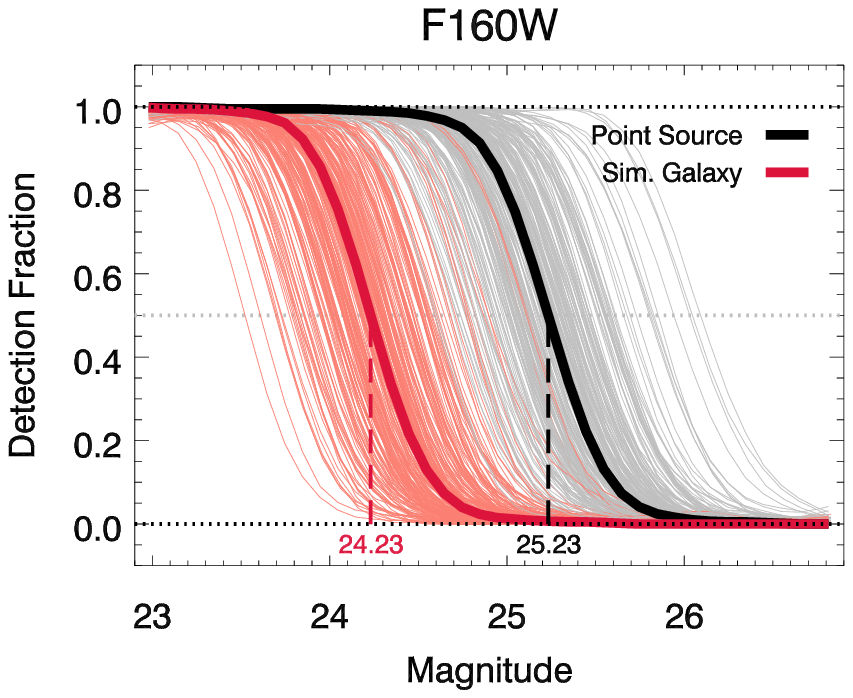} \vspace{2mm} \\ 
\includegraphics[width=0.32\textwidth]{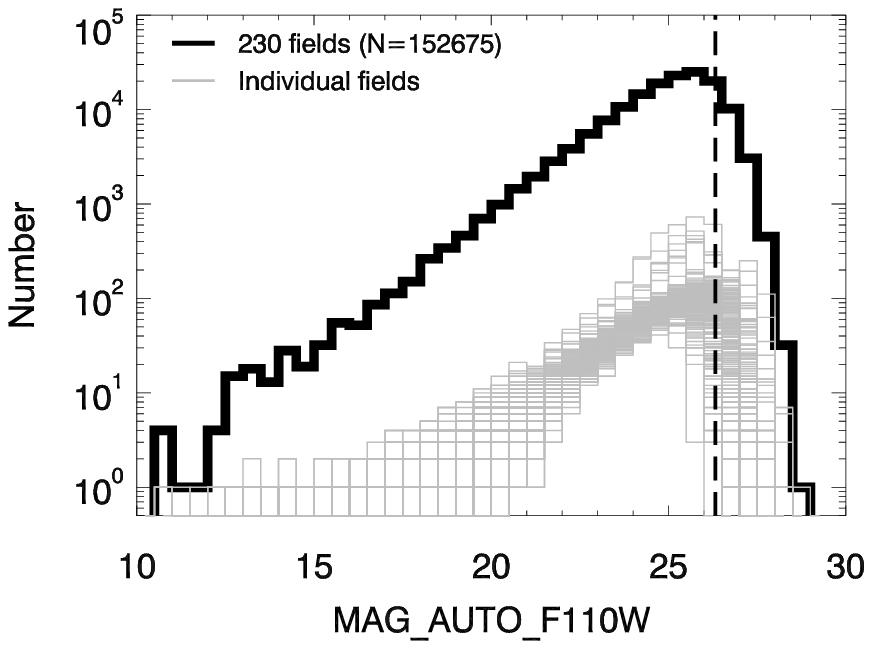} &
\includegraphics[width=0.32\textwidth]{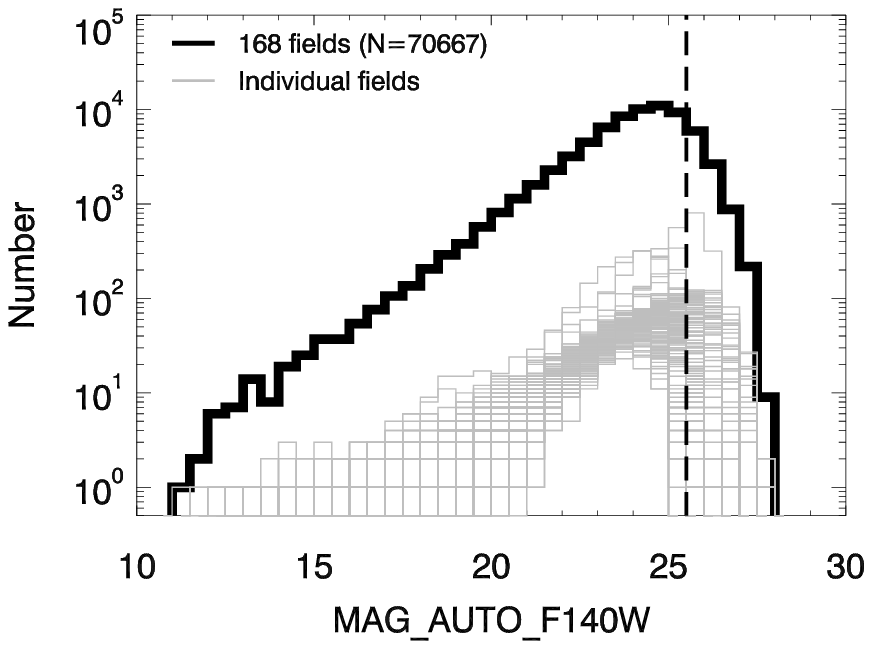} &
\includegraphics[width=0.32\textwidth]{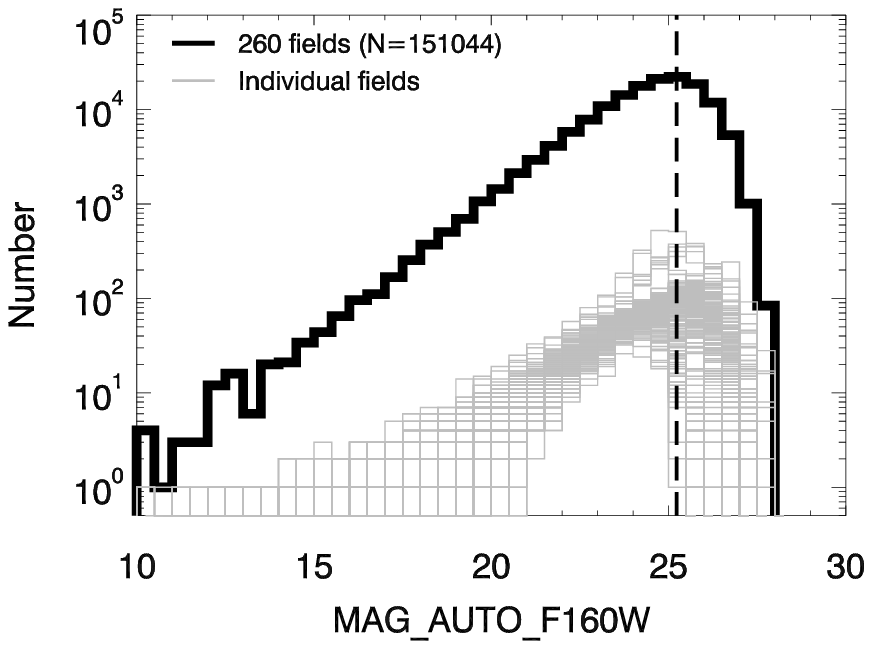} \\
\end{array}$
\vspace{-1mm}
\caption{\textit{Top row:} Comparison of the fraction of artificial sources that are recovered as a function of magnitude for the WISP WFC3/IR images. The thin gray and red lines show the simulated point-source and galaxy completeness, respectively, for individual fields. The thick black and red lines are the medians of the point-source and galaxy completeness, respectively, across all fields. The completeness (adopted as the 50\% detection fraction) varies by up to 1.8~mag across the sample due to the varying exposure length of the visits. The completeness values for each individual field are provided in Table~\ref{tab:wfc3ir_depth_completeness}. The median 50\% completeness for each source type is indicated at the bottom of each panel. 
\textit{Bottom row:} Histograms of \se\ \texttt{MAG\_AUTO} values for the WFC3/IR images in the WISP catalog. The black vertical dashed line denoted the median 50\% completeness of point sources for all fields (same as top panles).
 \label{fig:phot_completeness_compare}}
\end{figure*}

To measure the completeness, we inject simulated sources of varying brightness into the images and estimate their recoverability with \se, following methodology similar to \cite{revalski23}. First, we inject point-sources modeled as the PSF for each WFC3/IR filter. We adopt the \texttt{PSFSTD} models\footnote{\url{https://www.stsci.edu/hst/instrumentation/wfc3/data-analysis/psf}}, taking the average across the detector (3$\times$3 array), and resampling it to match the 0.08\arcsec\ resolution of the drizzled images. We inject 40 sources into each image and require that these do not overlap with themselves or with real sources, as determined from the \se\ segmentation map from the photometric pipeline. The magnitudes of the sources is scaled from [-3~mag,+1~mag], in steps of 0.1~mag (40 steps), relative to the filter zeropoint magnitude. This process is then repeated for a simulated galaxy, which is modeled as a S\'ersic profile with $n=2$, inclined at 45\degree, and an effective radii of 0.48\arcsec. This corresponds to the average size of a $\log[M_\star/M_\odot]=10$ star-forming galaxy at $z=1$ using the size-mass relation of \cite{vanderWel14}. The simulated galaxies are also convolved with each filter PSF. The resulting completeness measurements are shown in Figure~\ref{fig:phot_completeness_compare}. The point-source completeness values show close agreement ($\lesssim0.2$~mag) with the 5$\sigma$ depth based on 0.5\arcsec\ apertures. 

\begin{figure*}
$\begin{array}{cc}
\includegraphics[width=0.38\textwidth]{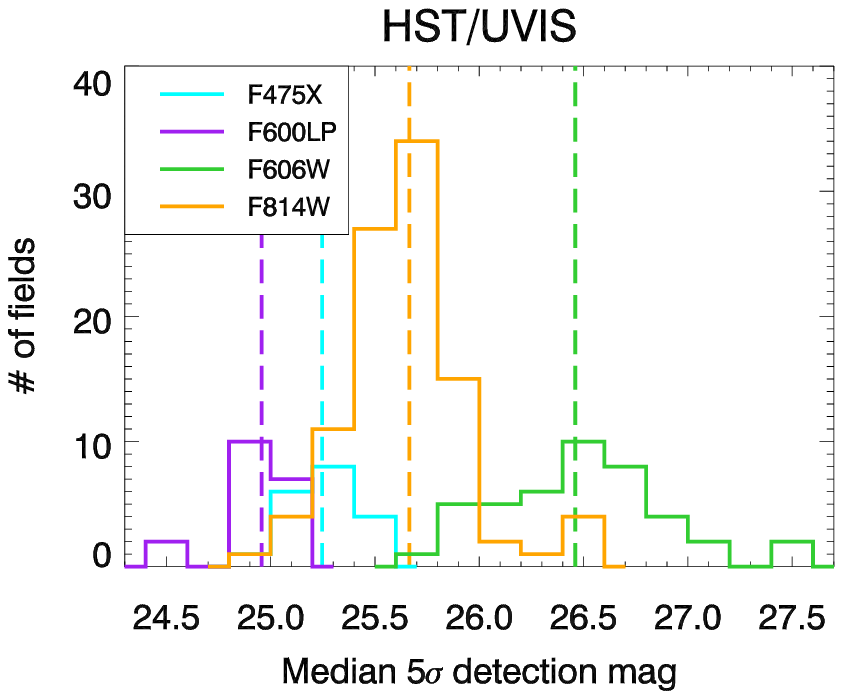} &
\includegraphics[width=0.38\textwidth]{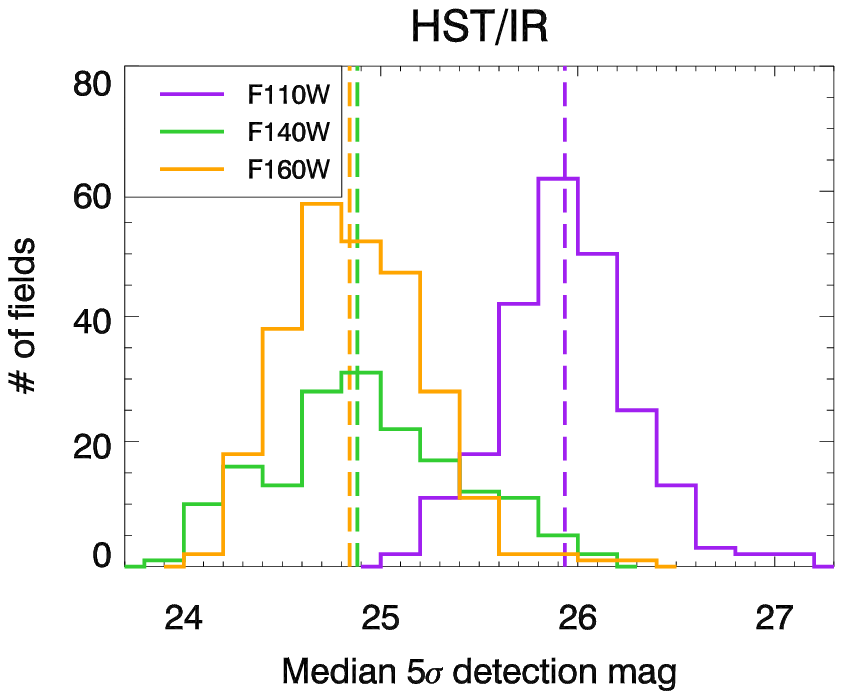}\vspace{2mm} \\
\includegraphics[width=0.38\textwidth]{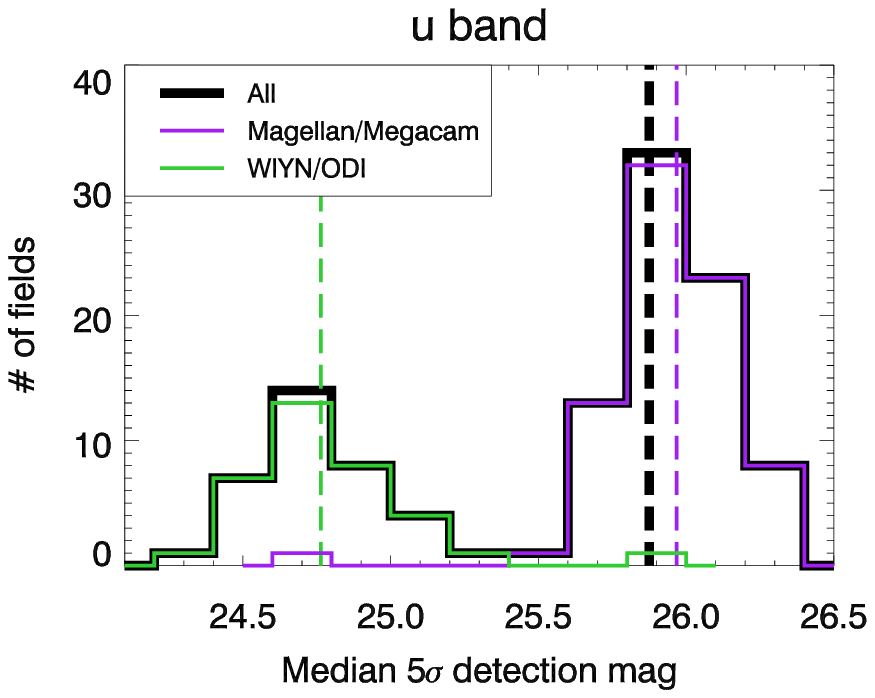} &
\includegraphics[width=0.38\textwidth]{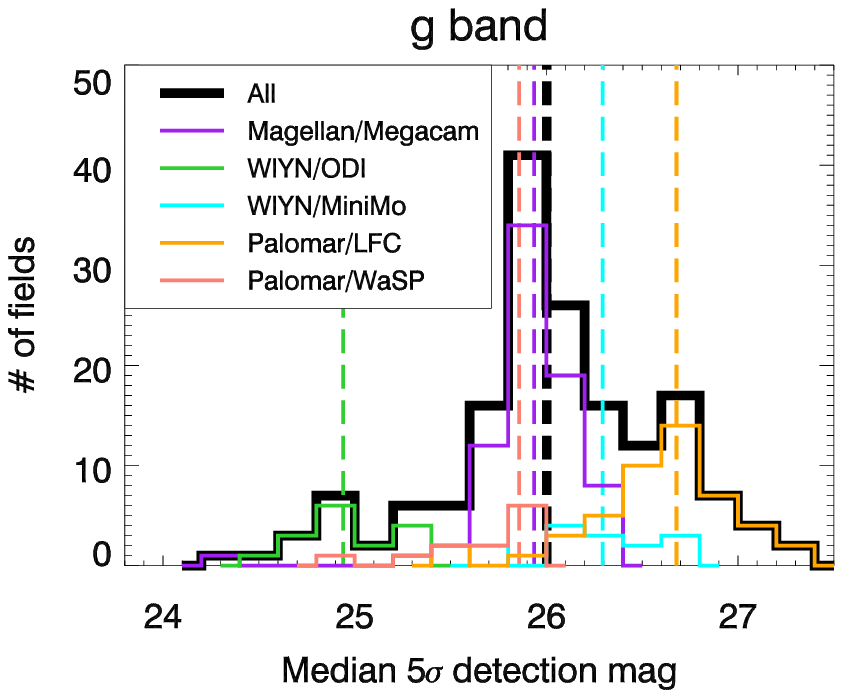}\vspace{2mm} \\
\includegraphics[width=0.38\textwidth]{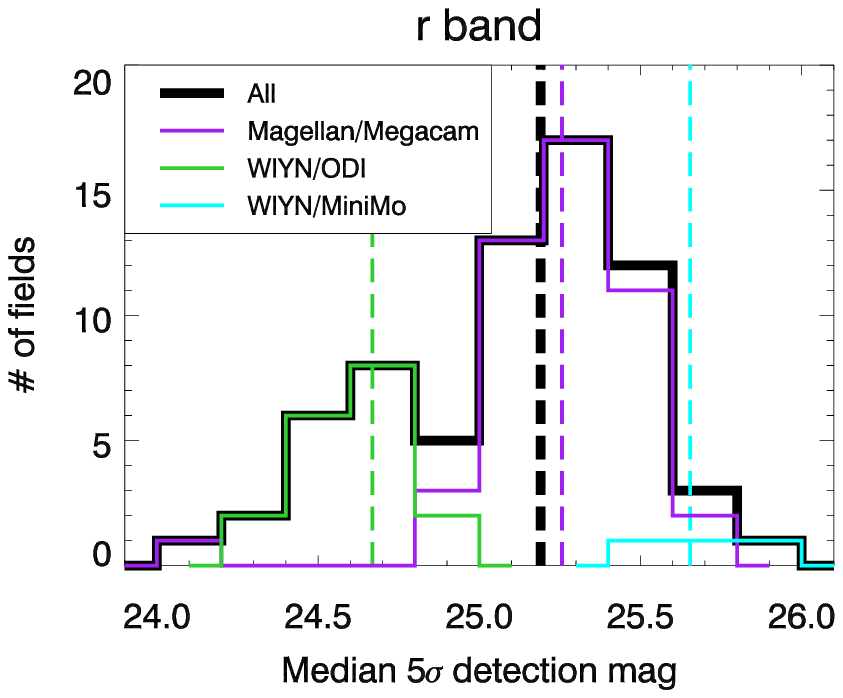} &
\includegraphics[width=0.38\textwidth]{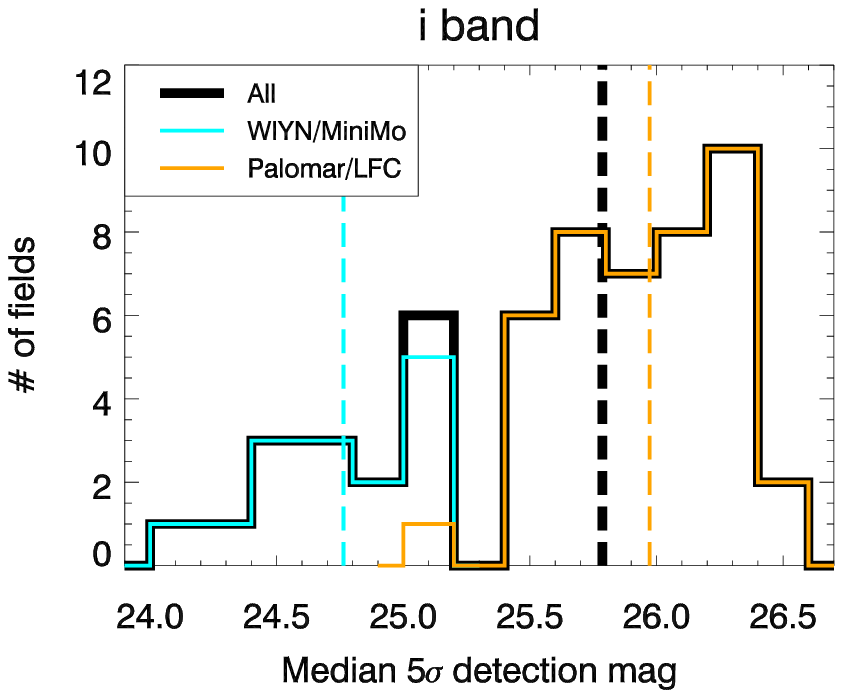}\vspace{2mm} \\
\multicolumn{2}{c}{\includegraphics[width=0.38\textwidth]{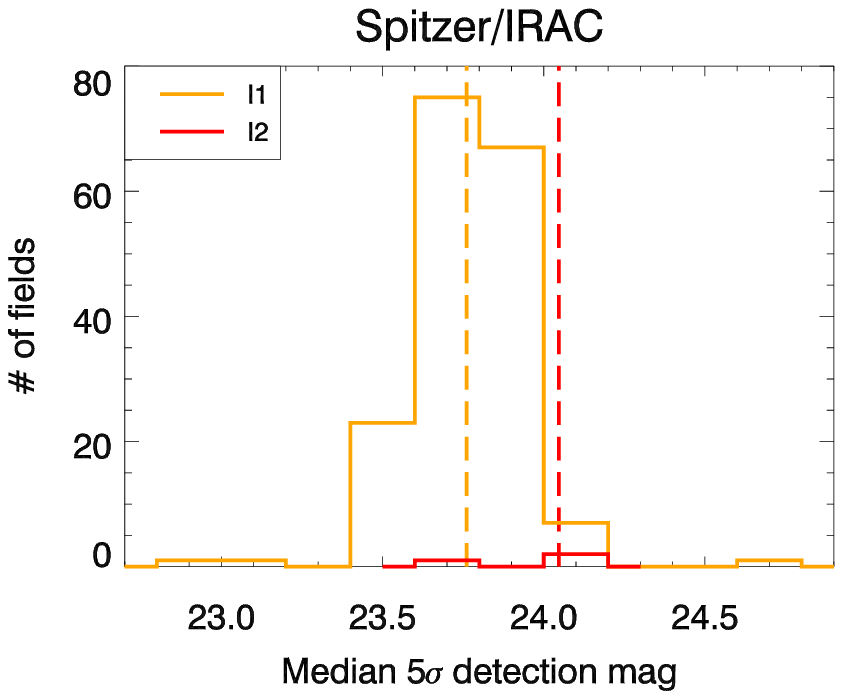}}\\
\end{array}$
\caption{Comparison of the median magnitude for sources detected at $\sim$5$\sigma$ significance (sources with $0.15<m_\mathrm{err}<0.25$~mag; $S/N\sim5$) for all WISP fields in this data release. Panels show (left to right, top to bottom) \hst\ WFC3/UVIS, WFC3/IR, ground-based $u$, $g$, $r$, $i$, and \spitzer/IRAC. The vertical dashed lines denote the median magnitude value of 5$\sigma$-detected sources for a given filter (i.e., image depth). The range in depths are mainly due to the varying length of exposure times, observing conditions, and/or field crowding (see Section~\ref{depth_summary}). 
 \label{fig:depth_compare} }
\end{figure*}

A full list of the median depth in each filter, based on median 5$\sigma$ magnitude (AB; $0.15<m_\mathrm{err}<0.25$~mag) for detected sources, are provided in Table~\ref{tab:phot_depth_catalog} and summarised in Figure~\ref{fig:depth_compare}. We note that this depth value is the median of the entire sample of 5$\sigma$ sources in each field, which span a range of sizes/surface brightness and hence is a not a uniform representation of depth. With regard to the WFC3/IR data, the depths reported here typically lie in-between the depth values for the point-source and simulated galaxies provided in Table~\ref{tab:wfc3ir_depth_completeness}. We provide some additional details on the variation in depth for the ground-based and \spitzer\ imaging data. 
For the ground-based data, there are considerable differences in the depths achieved from field to field. This difference is a result of the ground-based data being obtained through numerous programs, with different facilities/instruments and different science goals (see Table~\ref{tab:obs_summary}), and due to variations in the observing conditions. 
For \spitzer , the programs run over the different cycles (see Section~\ref{spitzer}) used slightly different exposures that were intended to achieve roughly uniform depth for each targeted field (see Section~\ref{spitzer_reduction}).

\subsection{Comparison of ground-based catalogue photometry to SDSS}\label{ancillary_comparison}
\begin{figure*}
$\begin{array}{cc}
\includegraphics[width=0.4\textwidth]{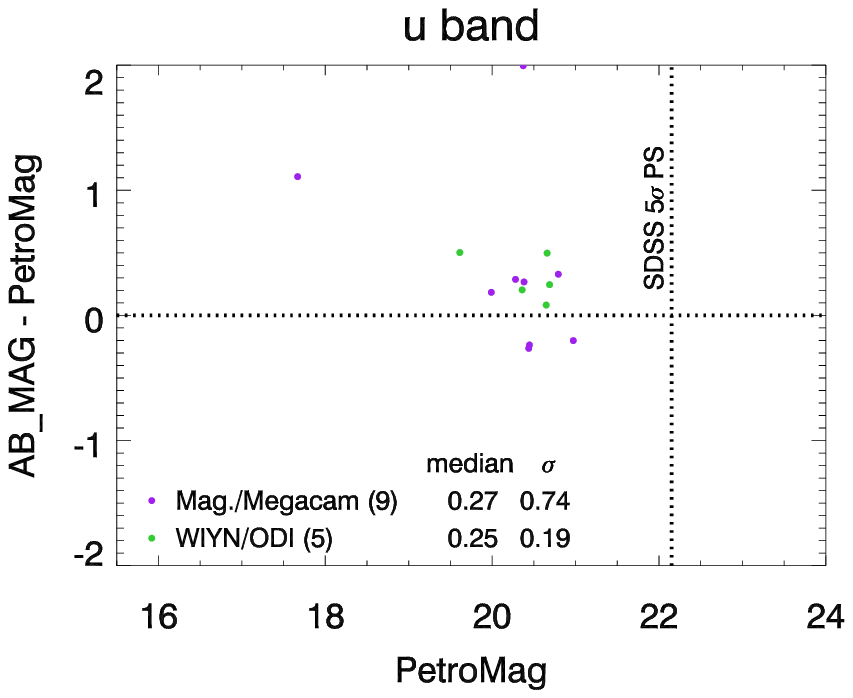} &
\includegraphics[width=0.4\textwidth]{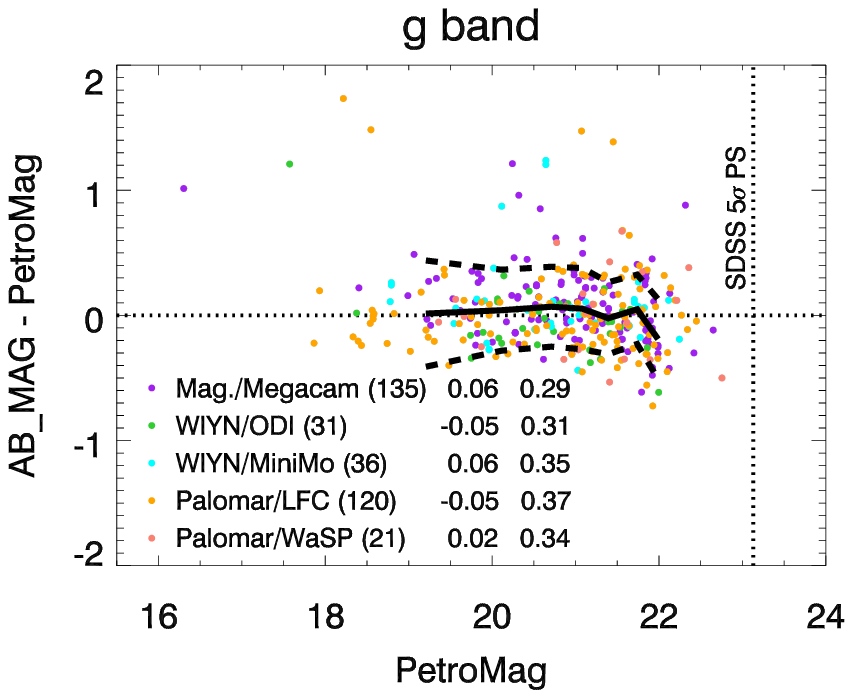} \vspace{2mm} \\
\includegraphics[width=0.4\textwidth]{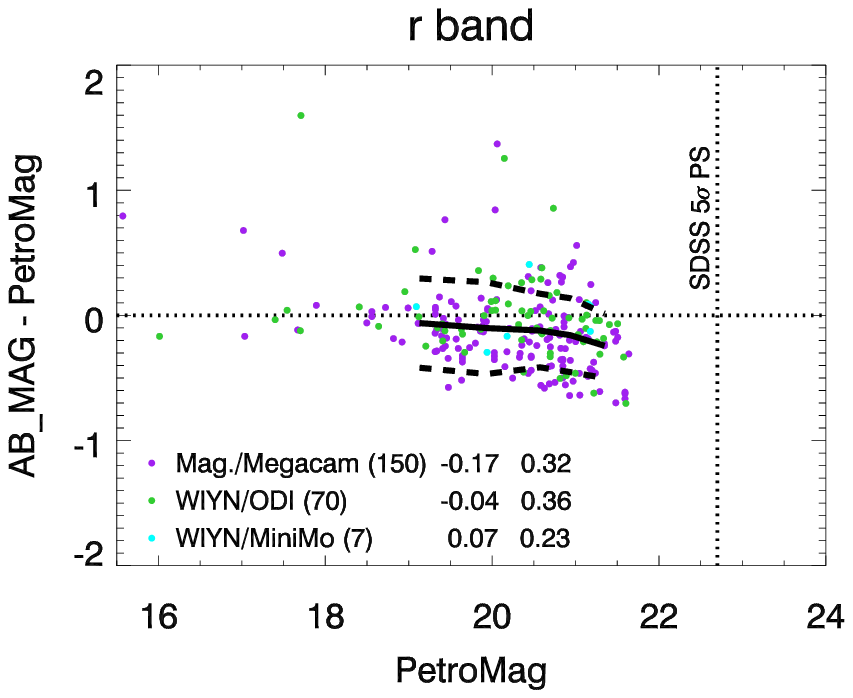} &
\includegraphics[width=0.4\textwidth]{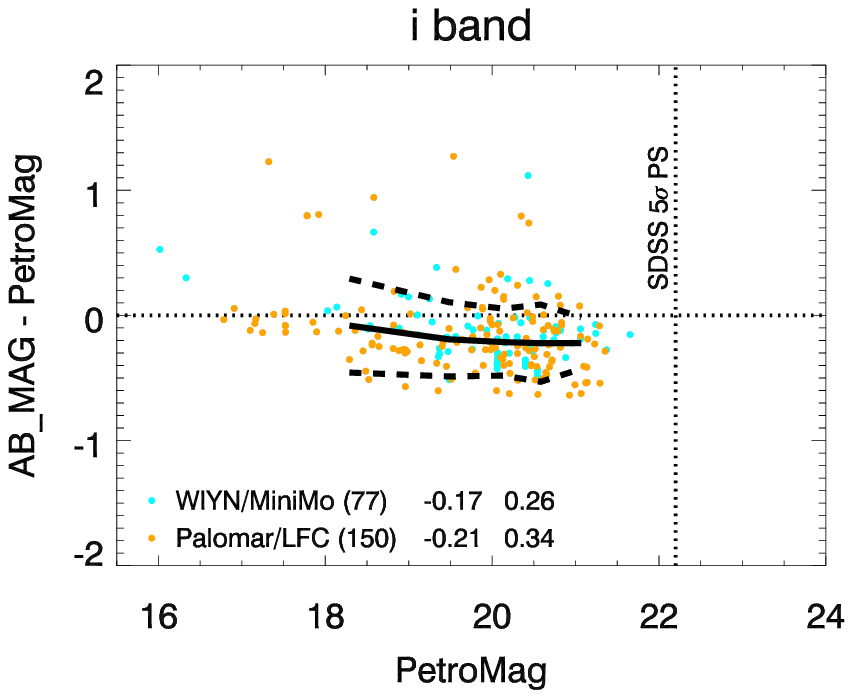}\\
\end{array}$
\vspace{-1mm}
\caption{Comparison of the magnitude offset for galaxies detected at $S/N \ge 5$ in optical bands from SDSS (\texttt{PetroMag}) and our WISP catalogue (\texttt{AB\_Mag}). The vertical dashed lines denote the median 5$\sigma$ point-source limit for SDSS. We note that the magnitude limit for galaxies is fainter than this (e.g., Figure~\ref{fig:phot_completeness_compare}). The median offsets are typically $\lesssim$0.2~mag and agree within the variance of the data. There are very few $u$-band galaxies bright enough for reliable comparison. The solid and dashed lines shows the median and $\pm1\sigma$ values combining all data in equal-number bins of 50 galaxies. For each instrument we provide the median magnitude difference (i.e., y-axis) and its variance in the legend, with the number of available sources shown in parentheses. 
 \label{fig:sdss_photo_compare}}
\end{figure*}

For the fraction of WISP fields that overlap with SDSS, we can make a comparison of the values in our photometric catalogue to values in the SDSS photometric catalogue. Due to SDSS being much shallower (5$\sigma$ point-source depth for $u,g,r,i$ is 22.15, 23.13, 22.70, 22.20~mag, respectively) than our data (see Figure~\ref{fig:depth_compare}), this comparison is quite limited (particulaly for $u$-band) and only possible for the brightest galaxies in each WISP field. We restrict our comparison to galaxies (SDSS \texttt{PhotoType}=4) and extract \texttt{PetroMag} values from SDSS DR14 public catalogues to use as our reference values. The \texttt{PetroMag} values are the most suitable for representing total galaxy flux that will be similar to our \texttt{TPHOT} photometry (\texttt{psfMag} was used to compare stars for the zeropoints in Section~\ref{ground_zeropoints}). However, the TPHOT photometry is based on priors from the deeper HST data such that the aperture sizes are likely to differ with respect to SDSS. A distance threshold of $\sim$0.5\arcsec\ is adopted, and we only compare sources with $S/N \ge 5$ in both datasets. A comparison for the $u,g,r,i$ filters is shown in Figure~\ref{fig:sdss_photo_compare}.

We find good general agreement in the photometry, with median offsets of $\lesssim$0.2~mag, although a slight bias (0.2~mag) may exist for the $r$ and $i$ bands. However, these offsets are within the variance for each band indicating they may not be significant. These results support or previous claim that the optical zeropoints have a typical accuracy of $\sim$0.1~mag.  
Individual sources with large differences ($\Delta>0.5$~mag) are likely due to source blending and/or incorrect cross-matching.
We note that the sample available for this comparison is much smaller than the sample used in the zeropoint determination for each image because the zeropoint estimate made use of the entire instrument FoV (>15$\times$ the area of WFC3/IR) whereas the catalogue only includes the small region overlapping with the WFC3/IR data.

\section{WISP Spectroscopic Pipeline}\label{spec_pipeline}
The WISP emission line catalogue is created in three steps. First, an  automatic detection algorithm examines every extracted spectrum to identify emission line candidates. Next, each candidate is independently inspected by two reviewers both for confirmation and to measure the source redshift and emission line properties. These first two steps are performed by the \texttt{wisp\_analysis}\footnote{\url{https://github.com/HSTWISP/wisp_analysis}} software package. Finally, the output from the two reviewers is combined to create one catalogue entry for each emission line source. We describe each step in this process in the following sections.

\subsection{Emission Line Candidate Identification}\label{line_identification}
We identify emission line candidates with a peak detection algorithm that uses a continuous wavelet transform (CWT) to select appropriately-shaped peaks in one-dimensional spectra. A wavelet transform breaks a signal into its base components, each of which is a modified version of the same ``mother'' wavelet function. This process is similar to Fourier analysis, but rather than sinusoidal  components of varying frequencies, the base components identified by a  wavelet analysis are scaled or shifted versions of the mother wavelet. The CWT is an improvement over the previous WISP line detection method,  which identified emission lines as contiguous pixels above a signal-to-noise (S/N) threshold \citep{colbert13}. The previous amplitude-based peak finding process was strongly dependent on the fit to the continuum and the amount of smoothing applied to the spectrum. It also resulted in many spurious detections as noise spikes can be misidentified as faint emission lines. The CWT technique identifies emission line features in the spectrum based on their shape as well as their amplitude, and therefore reduces the number of spurious emission line candidates that reviewers need to inspect. A detailed description of the steps in applying our CWT algorithm and using the results to identify emission line candidates is provided in Appendix~\ref{appendix:cwt}.

Emission line identification using the CWT algorithm is performed on the one-dimensional spectra extracted and calibrated by \texttt{aXe} \citep{kummel09}. We first remove from each spectrum the flux introduced by the overlapping continua from nearby objects. For this purpose we use the contamination model created by \texttt{aXe} during spectral extraction, where all objects detected in the direct images are approximated as two-dimensional Gaussians defined by the size and shape measured by \se. This process is performed on the one-dimensional spectra of each grism individually, allowing for 500\AA\ of overlap between the two grisms. We next require that at least two reviewers inspect each emission line candidate, both to confirm the candidates and to measure the line properties.

\subsection{Emission Line Candidate Inspection}\label{line_inspection}
The original, amplitude-based WISP line detection software resulted in a false detection rate of $\gtrsim70$\%, depending on the depth of the fields. Although the improvements presented here --- identifying emission lines using both amplitude and shape combined with the EW criterion --- reduce this rate significantly to approximately 55\%, the visual inspection remains necessary for constructing a clean catalogue. This inspection is especially crucial for undithered parallel data, which cannot be properly cleaned of artifacts.

\begin{figure}
\includegraphics[width=0.48\textwidth]{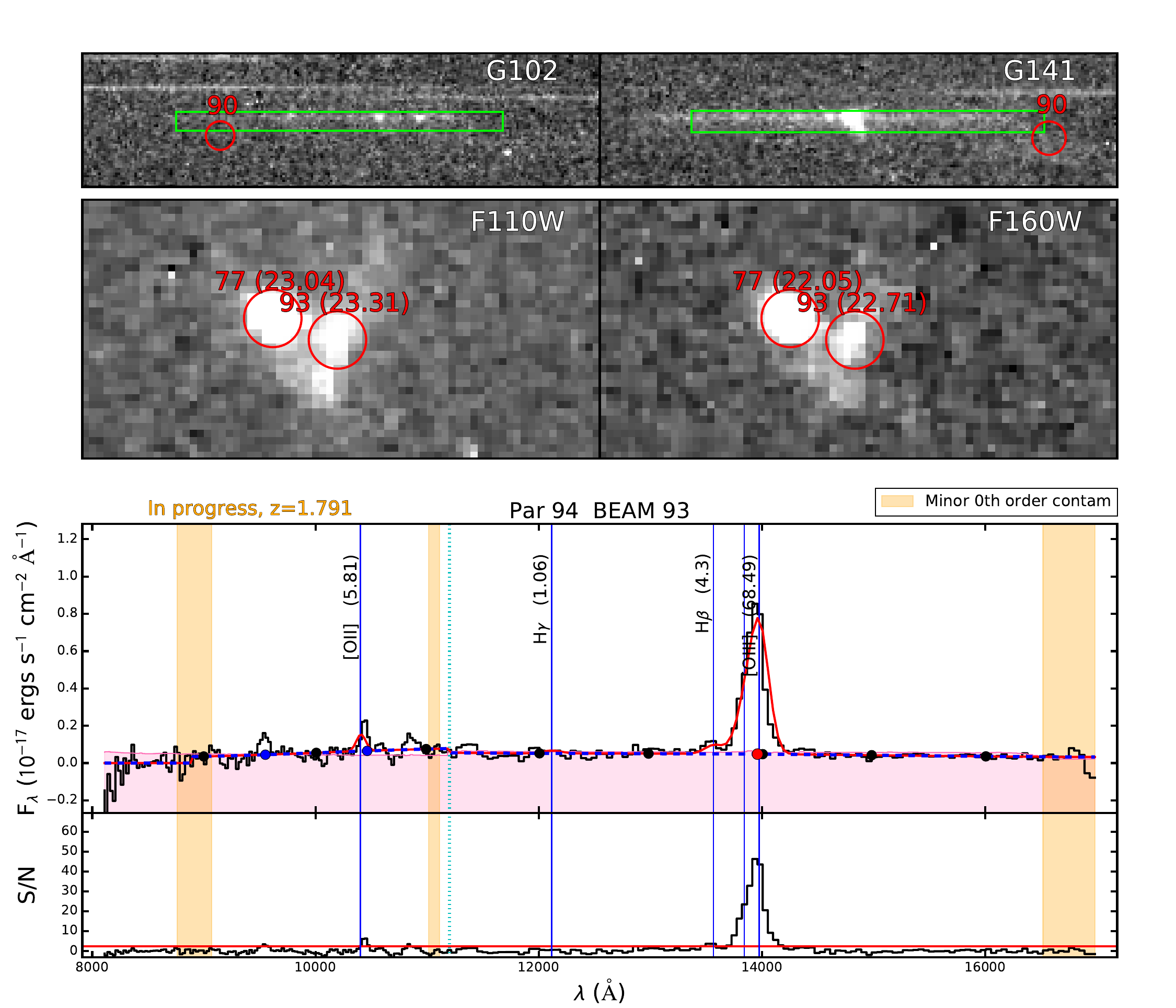}
\vspace{-6mm}
\caption{The display from the interactive portion of the emission line finding process. Reviewers inspect all available information for a given object.
\textit{Top row:} The two-dimensional spectral stamps for G102 (left) and G141 (right) are displayed with a green box indicating the trace of the spectra. Red circles identify the positions of zeroth orders of bright ($m<23$) objects that may contaminate the spectra. 
\textit{Middle row:} The direct images in all available filters, F110W (left) and F160W (right), with the positions of all objects and their corresponding magnitudes marked. The source of interest, here object 93, is centered in the direct images. 
\textit{Bottom panel:} The one-dimensional spectrum is displayed in both flux units ($f_{\lambda}$, top) and signal-to-noise (bottom). In the example shown here, the emission features identified in the spectrum of object 93 are not centered in the green box because they are coming from object 77. The emission line offset is clearest in the G102 spectrum. Object 93 was rejected as an emission line candidate and these emission features were instead measured for object 77. 
\label{fig:interactive_display}}
\end{figure}

For each emission line candidate, reviewers simultaneously inspect the direct images of the source, the two-dimensional spectral stamps, the one-dimensional extracted spectrum in units of $f_{\lambda}$, and a spectrum in units of S/N. An example of the full display from the interactive portion of the process is shown in Figure~\ref{fig:interactive_display}. 

We now briefly describe the inspection process and the series of checks reviewers perform for each candidate. The source displayed in Figure~\ref{fig:interactive_display}, object 93 from WISP field Par94 (hereafter 94-93), is used as an example to illustrate the process. The two main questions reviewers must answer are whether the emission line candidate identified by the detection algorithm is real, and whether it belongs to the object to which it is associated. In the first case, reviewers are validating the results of the detection software. In the second, they are considering and ruling out possible sources of contamination.

Candidate confirmation is required because spurious or false emission lines are occasionally identified by the detection program. False emission lines typically arise in two cases, when detector artifacts are identified as lines and when the continuum is improperly fit. The first case is a particular problem for a pure parallel survey such as WISP, since the telescope is not dithered between exposures. Combining multiple exposures therefore does not remove all cosmic rays, hot pixels, and other artifacts. Those that remain are sometimes selected as emission line candidates. Reviewers can often reject these by comparing the shape of the emission line candidate in the two-dimensional spectral stamp (top row of Figure~\ref{fig:interactive_display}) with the source shape in the direct image (middle row). Recall that an emission line in the spectral stamp is an image of the source at the given wavelength. The size, ellipticity, and position angle of the source are expected to be reflected to some degree in the emission line. This comparison is approximate, however, since the emission regions in a galaxy need not directly correlate with the broadband continuum emission detected in the imaging filters.

In the second case, false emission line candidates are identified in objects with continua that are poorly fit by the automatic software, which uses a cubic spline to fit the spectrum at eight wavelengths nodes. A steep rise in the continuum of an object, often caused by contamination from the spectrum of a nearby object, that is not reflected in the model fit can be incorrectly selected as a spectral peak. Given the large range of object sizes, fluxes, and levels of spectral confusion, one set of parameters will not work perfectly for all objects. The software's continuum fit is therefore treated as a first pass. The dashed blue line in  Figure~\ref{fig:interactive_display} is the continuum model for object 94-93, which in this case represents a good fit to the observed continuum. For objects with improperly fit continua, reviewers can tweak the model by adding, removing, or changing the wavelengths of the nodes used in the spline fit (black circles).

The reviewers must next determine whether the identified emission feature belongs to the source in question. With only a single roll angle, WISP spectra are often contaminated by overlapping spectra from other sources along the dispersion direction. This check generally involves four parts. First, the comparison between source and emission line shape described above can help identify emission lines coming from another object. We do not expect a galaxy to be much larger or much brighter in an emission line than it is in imaging.

Second, the emission should be vertically centered in the trace of the spectral stamps, indicated by the green box in the top panels of Figure~\ref{fig:interactive_display}. 
We can see that the line candidates in the spectrum of 94-93 are not centered, evidence that they likely belong to the nearby, brighter object 77. We note that it is of course possible that the emitting region of a source may not be centered on the continuum emission, and by rejecting emission lines that are off the centre of the trace we may also be rejecting real emission line galaxies.

Third, if there are multiple emission lines visible in the spectrum, their relative wavelengths should match. The wavelength solution of the grism is determined by the source position in the direct image, and will therefore only be correct for the spectrum of that source. For example, in Figure~\ref{fig:interactive_display}, given the assumed redshift for this object, the spectral peak around $\lambda \sim 10500$\AA\ should be [OII]$\lambda3727$. However, it does not line up exactly with the expected wavelength for [OII] at this redshift (indicated by the blue vertical line), further indication that these emission lines are contaminants from object 77.

Finally, the reviewers must consider the position of zeroth orders. We consider a portion of a spectrum to suffer from ``major'' zeroth order contamination if it directly overlaps with a zeroth order from a bright source ($m<23$ magnitude). The position of these bright zeroth orders are indicated by red circles in the grism stamps in Figure~\ref{fig:interactive_display}. The chance alignment of a bright zeroth order, especially from a compact source, can appear as a very convincing emission line. Spectral peaks that suffer from this major contamination are automatically rejected by the automatic software. Meanwhile, ``minor'' zeroth order contamination could be caused by (1) bright zeroth orders that are close to but not directly overlapping the spectral trace or (2) from direct overlap with zeroth orders of fainter sources ($m>23$). In some cases, again especially for the most compact objects, these fainter zeroth orders can be bright enough to masquerade as emission lines, and so the reviewers must remain vigilant for this possibility. The minor zeroth order contamination regions for object 94-93 are shown as orange bands in the spectrum of Figure~\ref{fig:interactive_display}.

\subsection{Emission Line Measurements}\label{linefits}
Once an emission line candidate has been confirmed, the reviewers fit a model to the spectrum. The fitting is performed via Levenberg-Marquardt least-squares minimisation. It is implemented with the software \texttt{mpfit}, based on the \texttt{MINPACK-1} FORTRAN package \citep{more78} and translated to Python by Mark Rivers\footnote{University of Chicago, \url{http://cars9.uchicago.edu/software/python/mpfit.html}}. With \texttt{mpfit}, each parameter can be held fixed or can be constrained with upper and lower bounds. The full model includes over 20 parameters, which are described below and listed in Table~\ref{tab:linefits}.

\begin{table*}
\caption{Model Spectrum Parameters \label{tab:linefits}}
\begin{center}
\begin{tabular}{llll}
 \hline \hline
Parameter & Description & Initial Value & Limits\\
$n_{\mathrm{nodes}}$ & Number of spline nodes used in continuum fit & 8 & Fixed \\
$\Delta \lambda_{\mathrm{fitting}}$ & Size of region used for line fitting & 1500 \AA\ & Fixed \\
$z_{\mathrm{init}}$ & Input redshift & $\left(\frac{\lambda_{\mathrm{obs}}}{\lambda_{\mathrm{H}\alpha}} - 1\right)$$^a$ & $\pm 0.02$ \\
$\Delta z_{\mathrm{[OIII]}}$ & Shift in redshift allowed for [OIII] profile fit & 0 & $\pm 0.02$ \\
$\Delta z_{\mathrm{[OII]}}$ & Shift in redshift allowed for [OII] profile fit & 0 & $\pm 0.02$ \\
$\Delta z_{\mathrm{[SIII,HeI]}}$ & Shift in redshift allowed for [SIII] and HeI fits & 0 & $\pm 0.02$ \\
FWHM$_{\mathrm{init}}$ & Input FWHM & $ 2 a \: \Delta \lambda_{\mathrm{G141}}$ \AA $^b$ & $^{-0.3 \mathrm{FWHM}_{\mathrm{init}}} _{+2.0 \mathrm{FWHM}_{\mathrm{init}}}$ \\
$A_{\mathrm{line}}$ & Input amplitude for each emission line & $\dotsb$$^c$ & 0,1$^c$\\
$r_{\mathrm{[SIII]}\lambda9532 / \lambda9069}$ & Ratio of [SIII]$\lambda9532$ to [SIII]$\lambda9069$ fluxes & 2.48 & Fixed \\
$\lambda_{\mathrm{T}}$ & Grism transition wavelength & 11200 \AA\ & Fixed \\
$\lambda_{\mathrm{blue}}$ & Blue wavelength cutoff for the G102 grism & 8100 \AA\ & Fixed \\
$\lambda_{\mathrm{red}}$ & Red wavelength cutoff for the G141 grism & 17000 \AA\ & Fixed \\
\hline
\end{tabular} \\
\vspace{1mm}\textbf{Notes:} $^a$The input redshift is that which will give \halpha\ for the current line, or is a redshift guess provided by the user.\\
$^b$FWHM$_{\mathrm{init}}$ is taken as twice the source
semi-major axis (\texttt{A\_IMAGE}) multiplied by the dispersion in the red grism.\\
$^c$Emission line amplitudes are estimated as the flux value
at line centre and are constrained to be positive.
\end{center}
\end{table*}

The model spectrum fit to the data consists of a continuum with the following emission lines added:
[OII]$\lambda\lambda3727+3729$, \hgamma, \hbeta, [OIII]$\lambda\lambda4959+5007$, \halpha, [SII]$\lambda\lambda6716+6730$, [SIII]$\lambda9069$, [SIII]$\lambda9531$, and HeI$\lambda10830$. The continuum is modeled using a cubic spline fit to a series of $n_{\mathrm{nodes}}$ spectral nodes, and emission lines are modeled as Gaussians, where the line centre is determined by the redshift guess for the source, the input amplitude is estimated separately for each line within $\pm \Delta \lambda_{\mathrm{fitting}}$ of line centre, and the standard deviation depends on the source size and the dispersion of the grism in which the line appears.  Emission lines are not fit individually. The entire spectrum --- continuum plus all lines --- is fit simultaneously, and all line profiles are constrained to have the same FWHM \textit{in pixels}, not in \AA. Hence, lines in the higher dispersion G102 will have smaller FWHM by a factor of two. This approach is reasonable under the assumption that all emission lines are images of the same host source. The source redshift is determined by the centre of the profile fit to the \halpha\ line at $z \lesssim 1.6$ and [OIII] at $z \gtrsim 1.6$ (when \halpha\ has redshifted out of the G141 grism). The centres of each additional line are allowed to vary up to a maximum wavelength equivalent to $\Delta z_{\mathrm{line}} = 0.02$. In the absence of multiple emission lines, single lines are assumed to be \halpha\ unless the clear asymmetry of the [OIII]$+$\hbeta\ line profile is visible.

As a consequence of the use of a full spectral model, all emission lines listed above are fit provided they fall within the grism wavelength coverage at the assigned redshift. Emission lines that were not identified by the detection algorithm will therefore be measured along with the identified lines. We refer to the lines strong enough to have been identified by the detection algorithm as ``primary lines'', while the remaining lines are called ``secondary lines''. This distinction is an important one for the emission line catalogue completeness, which is discussed in the next section (Section~\ref{completeness}).

Lines with a flux S/N$<$1 are set at 1$\sigma$ and reported as upper limits in flux and therefore EW. These limits are calculated by summing in quadrature the error array within $2\times$FWHM of line center. We find, however, that the error arrays calculated by \texttt{aXe} are underestimates of the spectral noise properties. The 1$\sigma$ limits are all systematically lower than the sensitivity limits measured for the fields. We therefore apply a correction factor to the flux limits, correcting the amplitude of the limits while preserving the scatter in the measurements.

We note several emission lines --- \halpha\ and [NII], the [SII]$\lambda\lambda6716+6731$ doublet, the [OII]$\lambda\lambda3727+3730$ doublet, and the [OIII]$\lambda4959+5007$ doublet --- are blended at the resolution of the WFC3 grisms. The fluxes measured for \halpha\ therefore include the contribution from [NII]$\lambda6583$ and [NII]$\lambda6550$, and those for [OII], [SII], and [OIII] each include both doublet lines.
The flux ratio [SIII]$\lambda$9532/[SIII]$\lambda$9069 is fixed to 2.48:1.

While inspecting each spectrum, reviewers can change multiple parameters in order to improve the fit to the spectrum. In addition to moving, adding, or subtracting nodes for the continuum fit, they can provide a new redshift guess for the source; modify the wavelength ranges of each grism to fit emission lines at the grism edges (changing the transition between grisms, $\lambda_{\mathrm{T}}$, and their wavelength cutoffs $\lambda_{\mathrm{blue}}$ and $\lambda_{\mathrm{red}}$); or provide a new guess for the FWHM$_{\mathrm{init}}$, usually decreasing the default guess for sources where the dispersion axis is along the minor-axis. Finally, reviewers can mask regions of the spectrum that suffer from severe contamination from either zeroth orders or nearby continua, thereby making sure they do not affect the full spectral model. The emission line candidate vetting, spectral fitting and cataloguing are all performed as part of one streamlined process. The results from the two reviewers are then combined for each object, and a series of quality flags are assigned to the emission line object determined by the level of redshift, flux and EW agreement between the reviewers' classifications (see Section~\ref{spec_catalog} and Appendix~\ref{appendix:line_flags}).

\subsection{Completeness Corrections} \label{completeness}
The completeness of a survey or catalogue is a description of how accurately the detected sample represents the true population in the universe. Understanding a survey's completeness is necessary before the results can be used to conclude anything about the true underlying distribution of sources and source properties. A survey such as WISP can suffer from incompleteness for a variety of reasons. Sources may be lost amidst the noise in images if their fluxes are close to the detection limit. Some sources may not be detected, or their emission lines missed in their spectra, because they overlap or are blended with nearby bright objects. The completeness of a survey depends on the specific selection function used to detect sources. In the case of the WISP emission line catalogue, the selection function includes the detection of sources in the direct images, the identification of emission line candidates via the detection algorithm, and the acceptance during the visual inspection. We also only look for emission lines of continuum-detected sources in the images. We must understand the fraction of sources and emission lines that are not detected through this full process as a function of their properties such as size, shape, and the strength of their emission.

In quantifying the completeness corrections that must be applied to a catalogue, we are determining the types and numbers of sources that are missed. To do so, we create a simulated catalogue of 10,000 sources and their spectra, 5,000 each for the shallow and deep portions of the WISP Survey. The simulated sources have $H$-band magnitudes in the range $16.8 \leq m_H \leq 26.2 (27.6)$ and observed \halpha\ fluxes in the range of $5\times10^{-17} (1\times10^{-17}) \leq f \leq 1\times10^{-15}$ erg s$^{-1}$ cm$^{-1}$ for the shallow (deep) fields. See Table~\ref{tab:sim_inputs} for the full list of input parameters and values. We insert these simulated sources into real WISP images, 25 sources at a time, and run them through the full WISP pipeline and emission line detection software. The creation of the simulated data is described in Appendix~\ref{appendix:sims}. In order to save time and the effort required during the visual inspection stage, the reviewers only inspect the spectra of simulated sources that were identified by the line finding algorithm. Yet not all of the emission line candidates were real. Some were noise spikes, contamination, or the result of poorly fit continua. We note, however, that because of this choice we cannot use the simulations to measure the rates of contamination or redshift mis-identification in the catalogue.

\begin{figure}
\includegraphics[width=0.48\textwidth]{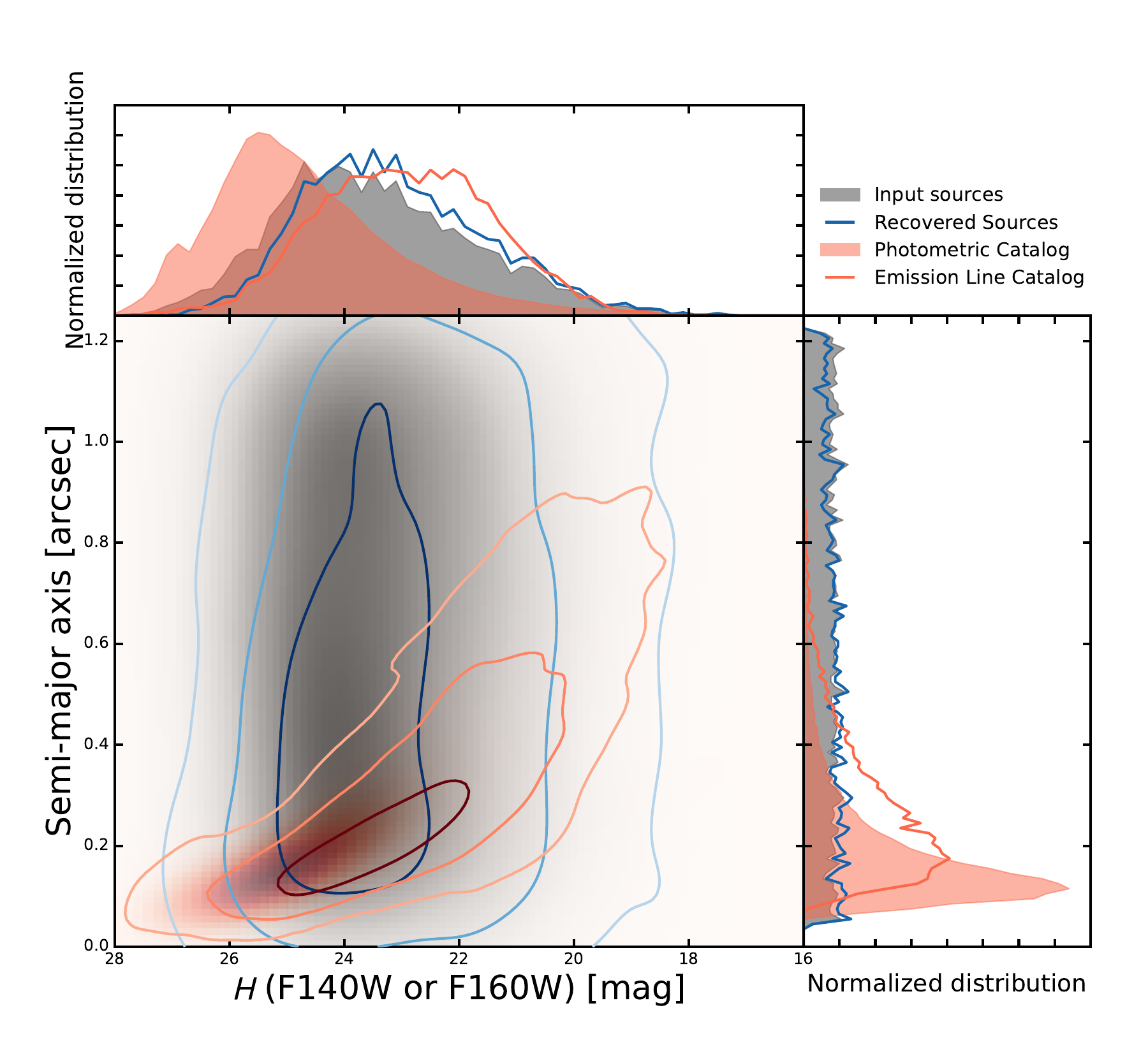}
\vspace{-6mm}
\caption{The semi-major axis as a function of the $H$-band magnitude for all input sources (grey) and those recovered in imaging (blue contours). The real sources from the photometric and emission line catalogues are shown in red. The imaging completeness is a function of the magnitude and size of the sources. The recovery fraction drops for large ($a \geq 0.7\arcsec$) and faint ($H>24.5$) sources.
\label{fig:recoveredmags}}
\end{figure}

Of the 10,000 input sources, 7,721 were recovered by the WISP reduction pipeline, with an equal number recovered in the shallow and deep fields.
This 77\% recovery fraction represents the overall imaging completeness given the imaging depths and our set of source detection parameters.
The majority of the sources that are not recovered in the imaging catalogue, and which therefore have no extracted spectra, are faint and/or extended. In Figure~\ref{fig:recoveredmags} the input semi-major axis sizes (before convolution with the PSF) are shown as a function of magnitude for the simulated sources that were input and recovered. The distributions of real sources are shown for reference. The sources that are not recovered in imaging mainly have a semi-major axis of $a \geq 0.7\arcsec$ and are fainter than 24.5 magnitudes in the $H$ band.

For the simulated sources recovered in imaging, we calculate the fraction of these that are recovered by the full line finding process. We find that the completeness depends on source size and shape and emission line EW and S/N. The S/N dependence is essentially a dependence on line flux, but includes the effects of the variable depths of the WISP fields (see Figure~\ref{fig:line_depths}). While we can measure the S/N of observed emission lines, there is no analogous definition of the input S/N for the simulated lines. The input template spectra do not include noise, and while we have added shot noise to the simulated grism data based on the integration times of the exposures into which they are added, this is not the only source of noise that will affect the flux measurements. The depths reached in each field depend also on the level of the zodiacal background for each pointing. We therefore instead characterise the completeness as a function of emission line ``scaled flux'', or the emission line flux (input or recovered) divided by the sensitivity limit of the field at the wavelength of the line.

\begin{figure}
\includegraphics[width=0.48\textwidth]{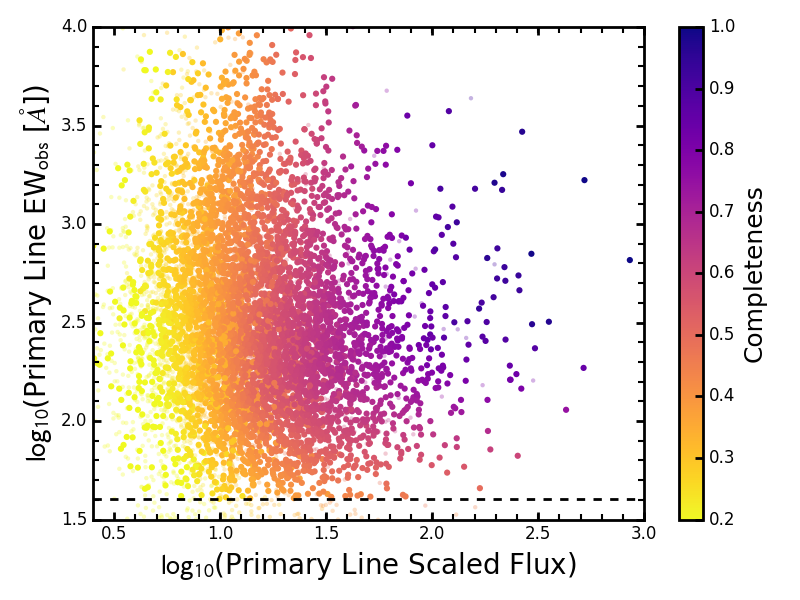}
\vspace{-6mm}
\caption{The completeness of the emission line catalogue as a function of the scaled flux and observed EW of the strongest line for each source. The dashed horizontal line indicates the EW completeness limit 40 \AA. The small, transparent points indicate the lines with EW$<$40 \AA\ and/or S/N$<$5.
\label{fig:line_completeness}}
\end{figure}

As discussed in Section~\ref{linefits}, a source will enter the catalogue because of the detection of the primary lines. We consider only one line per spectrum --- both for the input simulated and the output measured spectra --- taking the line with the brightest scaled flux as the source's primary line. As we are not attempting to quantify the rates of redshift mis-identification, we consider a detected emission line recovered regardless of whether the reviewers have properly identified it (i.e., regardless of what redshift is assigned to the object). We recover 868 of the 5000 simulated sources added to deep WISP fields covered by both grisms and 1541 of the 5000 added to shallow, G141-only fields. 
This recovery reflects the completeness due to both the imaging and the spectroscopic selection functions, and is heavily influenced by source size and shape as discussed below.

The object size and shape will strongly affect the completeness, as large, low surface brightness emission lines may be missed by the peak finder. However, the large sources that suffer from the highest levels of incompleteness, those with $a \geq 0.7\arcsec$, constitute less than 1\% of the total catalogue. We simulate sources with a uniform distribution of sizes, but then weight the input sources by the distribution of observed sizes in the emission line catalogue. This step both reflects the observed distribution and allows us to consider a two-parameter completeness correction, maintaining sufficient number counts for the completeness analysis without requiring reviewers to visually inspect tens of thousands of sources.

The completeness is calculated in four bins of scaled flux and five bins of EW. The bin edges are determined by the distribution of sources in the real WISP emission line catalogue such that there are an approximately equal number of real sources in each bin. The one exception is the bin of lowest EW, which we add in order to probe an area of the parameter space with low completeness \citep[EW$_{\mathrm{obs}} < 40$\AA, see ][]{colbert13}. We use a radial basis function to approximate the three-dimensional surface formed by the bin centers and recovery fractions calculated in each bin. The resulting completeness corrections calculated for each source in the WISP emission line catalogue are shown in Figure~\ref{fig:line_completeness}. They are applied according to the scaled flux and observed EW of the strongest line in the spectrum, which is most often \halpha\ or [OIII] but is occasionally [OII] when \halpha\ or [OIII] are masked out due to overlap with a bright zeroth order or other major contaminant. These completeness corrections are applicable at the source level or for the primary lines for each source. They are not appropriate for secondary lines. As expected, very few low-EW emission lines were recovered in the simulations, making the completeness corrections calculated in bins with EW$_{\mathrm{obs}} < 40$\AA\ very uncertain. We therefore find that the EW criterion presented in \cite{colbert13} applies to the new version of the line finding process as well (see \citealt{bagley20} for a comparison of these completeness corrections with those calculated by \citealt{colbert13}).
We also recommend applying an emission line flux S/N cut at S/N$>$5 when using this catalogue.

\section{WISP Emission Line Catalogue}\label{spec_catalog}
\subsection{Emission Line Catalogue Description}\label{line_catalog_description}
\begin{figure}
\includegraphics[width=0.48\textwidth]{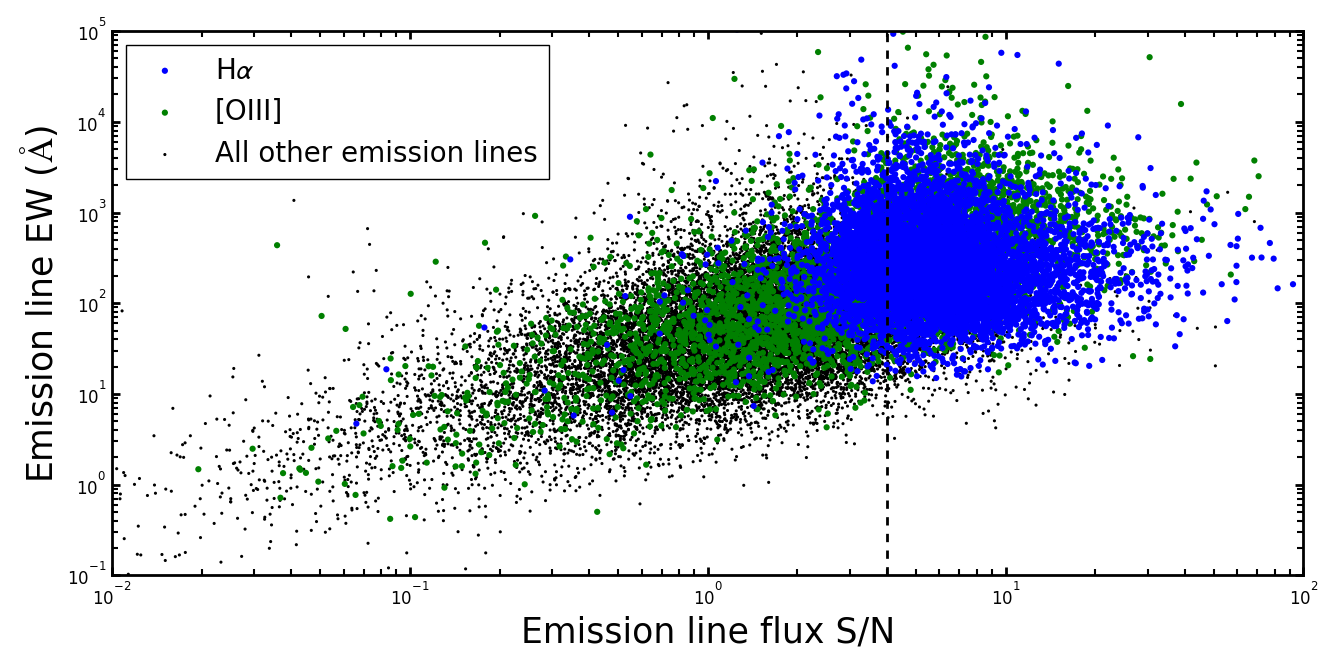}
\vspace{-6mm}
\caption{The emission line S/N and EW of all emission lines in the catalogue. \halpha\ is the two most commonly identified ``primary line'' and therefore the majority of \halpha\ in the catalogue has a S/N$>$4. The majority of the lower-S/N lines were fit as a consequence of the redshift assigned by identification of a primary line.
\label{fig:snr}}
\end{figure}

The WISP emission line catalogue contains 8,192 sources in 419 WISP fields. These sources include only galaxies and AGN (spanning $0.137<z_\mathrm{grism}<2.785$; no stars included). Fields that are absent, relative to the photometric catalogue, fall into one of a few categories: they were too crowded (a few WISP fields captured portions of dwarf galaxies), heavily contaminated by bright stars, or suffered from poor and uneven background subtraction in the grism data. We note that line availability is dependent on the grism coverage and galaxy redshift and that sources without data in a given emission line have values of `-1' for their entries.  The S/N and EW of all emission lines in the catalogue are shown in Figure~\ref{fig:snr}. The vast majority of the \halpha\ emission lines (blue circles) have a S/N greater than 4 (vertical dashed line), indicating that \halpha\ is most often the primary line in a spectrum. Emission line fluxes to the left of the dashed line were likely fit as secondary lines and are therefore at a lower S/N. The emission lines in this catalogue are \textit{not} extinction-corrected (neither foreground Milky Way nor internal extinction applied).

We present an explanation of the WISP emission line catalogue entries in Table~\ref{tab:FITS_line_catalog}. These include RA, Dec, WFC3/IR total magnitudes, sizes (based on \hst), spectroscopic redshift, and the FWHM used for all emission lines. There are also emission line flux, flux error, EW, and observed wavelength, for nine lines ([OII], \hgamma, \hbeta, [OIII], \halpha+[NII], [SII], [SIII]$\lambda$9069, [SIII]$\lambda$9532, and HeI$\lambda$10830).

There are also nine quality flags in the emission line catalogue. A description of these flags is provided in Appendix~\ref{appendix:line_flags}. These flags should be carefully considered when selecting sources from the catalogue. For the most robust selection of emission line sources, we recommend only using cases with \texttt{REDSHIFT\_FLAG}=0 (5054 sources) to avoid cases with ambigious redshift determinations. Cases with \texttt{REDSHIFT\_FLAG}$>$0 indicate some type of redshift disagreement between the reviewers or redshift uncertainty for single line emitters. The majority of sources with \texttt{REDSHIFT\_FLAG}$>$0 are those that were identified by only a single reviewer (\texttt{REDSHIFT\_FLAG}$\geq$16; 2061 sources). These cases are likely marginal detections where one reviewer did not consider the identified emission feature to be real above the noise or rejected the emission feature as some type of artifact or contamination. Sources with \texttt{REDSHIFT\_FLAG}$\geq$16 should therefore be considered with caution. 

\begin{table*}
\caption{Description of WISP Emission Line catalogue \label{tab:FITS_line_catalog}}
\begin{tabular}{ll}
\hline \hline
Title & Description \\ \hline

\texttt{PAR} & WISP field ID number \\
\texttt{OBJ} & Object ID number \\
\texttt{RA} & Decimal RA from HST images [deg] \\
\texttt{DEC} & Decimal Dec from HST images [deg] \\
\texttt{JMAG} & J-band magnitude; \se\ \texttt{MAG\_AUTO\_F110W} from WISP Photometric Catalogue [mag] \\
\texttt{HMAG} & H-band magnitude; \se\ \texttt{MAG\_AUTO\_[F140W} or \texttt{F160W]} from WISP Photometric Catalogue [mag] \\

\texttt{A\_IMAGE} & \se\ profile RMS along major axis measured on 0.08\arcsec/pixel HST images [pixels] \\
\texttt{B\_IMAGE} & \se\ profile RMS along minor axis measured on 0.08\arcsec/pixel HST images [pixels] \\

\texttt{FILTER\_FLAG} & Flag identifying filter coverage for field (10 bit flags; see Appendix~\ref{appendix:line_flags})\\
\texttt{GRISM\_FLAG} & Integer flag identifying grism coverage for field; (1=G102, 2=G141, 3=G102+G141)\\
\texttt{[GRISM\_FILTER]\_FLAG}$^a$ & Flags identifying any problems with the grism data for this field (5 bit flags; see Appendix~\ref{appendix:line_flags})\\
\texttt{EDGE\_FLAG} & Flag identifying objects close to edges of direct image (5 bit flags; see Appendix~\ref{appendix:line_flags})\\

\texttt{NNEIGHBORS} & Number of sources in WISP Photometric Catalogue within 1\arcsec\ of object \\
\texttt{NLINES} & Number of lines detected at $> 2\sigma$ for object \\
\texttt{REDSHIFT} & Redshift from simultaneous fit to all lines in spectrum \\
\texttt{REDSHIFT\_ERR} & Redshift 1$\sigma$ uncertainty \\
\texttt{dz\_OIII} & Shift in redshift of the [OIII] profile fit compared with reported \texttt{REDSHIFT} for object \\
\texttt{dz\_OII} & Shift in redshift of the [OII] profile fit compared with reported \texttt{REDSHIFT} for object \\
\texttt{dz\_SIII\_HE1} & Shift in redshift of the [SIII] and HeI profile fit compared with reported \texttt{REDSHIFT} for object \\
\texttt{DELTA\_REDSHIFT} & Difference in redshift fits of the two reviewers \\
\texttt{REDSHIFT\_FLAG} & Flag identifying quality of redshift fit (6 bit flags; see Appendix~\ref{appendix:line_flags}) \\
\texttt{FWHM\_OBS} & FWHM used for all emission line profile fits; observed frame using the 46.5 \AA/pix dispersion in G141 grism [\AA] \\
\texttt{FWHM\_OBS\_ERR} & FWHM 1$\sigma$ uncertainty [\AA] \\
\texttt{FWHM\_FLAG} & Flag identifying quality of reported FWHM (5 bit flags; see Appendix~\ref{appendix:line_flags})\\
\texttt{COMPLETENESS} & Source completeness reflecting the selection function of the full line-finding procedure (See Section~\ref{completeness}) \\
\texttt{[LINE]\_NREVS}$^b$ & Number of reviewers who measured emission line \\
\texttt{[LINE]\_FLUX}$^b$ & Emission line flux from profile fit [\esc] \\
\texttt{[LINE]\_FLUX\_ERR}$^b$ & Emission line flux 1$\sigma$ uncertainty [\esc] \\
\texttt{[LINE]\_DELTA\_FLUX}$^b$ & Difference in emission line fluxes of the two reviewers [\esc] \\
\texttt{[LINE]\_EW\_OBS}$^b$ & Emission line equivalent width reported in observed frame; -1 for undetected lines [\AA] \\
\texttt{[LINE]\_DELTA\_EW}$^b$ & Difference in emission line EWs of the two reviewers [\AA] \\
\texttt{[LINE]\_FLAG}$^b$ & Flag identifying quality of emission line measurements (6 bit flags; see Appendix~\ref{appendix:line_flags}) \\
\texttt{[LINE]\_CONTAM}$^b$ & Flag identifying the contamination noted by each reviewer (`a' and `b'); String has form `a.b'
(4 bit flags; see Appendix~\ref{appendix:line_flags}) \\
\texttt{[LINE]\_WAVEOBS}$^b$ & Observed wavelength of emission line [\AA] \\
\texttt{[LINE]\_EDGE\_FLAG}$^b$ & Flag identifying emission lines close to edges of grism where sensitivity decreases (4 bit flags; see Appendix~\ref{appendix:line_flags}) \\


\hline
\end{tabular} \\
\vspace{1mm}\textbf{Notes:}  
$^a$\texttt{[GRISM\_FILTER]} is one of: G102 or G141. 
$^b$\texttt{[LINE]} is one of: OII, Hg, Hb, OIII, HaNII, SII, SIII\_9069, SIII\_9532, or He1\_10830.
\end{table*}

\subsection{Grism Redshift Accuracy and Precision}\label{redshift_accuracy}

The redshift accuracy of the grism data are primarily driven by the number of available lines, with a greater number of lines generally providing more reliable redshifts. For example, \citet{baronchelli20} show that the default choice of assuming WISP single-line emitters are \halpha\ is incorrect for $\sim$30\% of cases, where most of these are likely to be the [OIII]$\lambda$5007 emission line. Therefore, in the absence of other independent information that can inform how to break single-emitter degeneracies, such as photometric redshifts (photo-$z$) and/or machine learning \citep[e.g.,][]{baronchelli20, baronchelli21}, we urge caution in using sources with only a single emission line (\texttt{NLINES}$\leq$1; 3350 sources in catalogue). It is also important to consider the redshift quality flags (\texttt{REDSHIFT\_FLAG}; see Appendix~\ref{appendix:line_flags}) that indicate potential disagreements between emission line reviewers.

For sources with multiple detected emission lines (i.e., \texttt{NLINES}$>$1; 4842 sources in catalogue), the reliability of the redshifts is expected to higher, however this is difficult to quantify without independent metrics or follow-up higher resolution spectroscopy. We refer readers to \cite{baronchelli20, baronchelli21} for a detailed discussion on the issues pertaining to contamination and purity in a grism spectroscopic sample. \cite{masters14} presented a subset of 26 emission-line galaxies (from 23 WISP fields) where follow-up observations with Magellan/FIRE ($R\sim6000$; 0.8-2.5\micron) were made on sources with $S/N \gtrsim 10$ (from grism) in [OIII] and/or \halpha. They found very good agreement in line identification (close to 100\%) for these sources, although these cases have higher $S/N$ than most of the WISP sources. A larger sample with follow-up observations from VLT/FORS2 ($R\sim1200$; 0.51-0.85\micron) will be presented in Boyett et al. (in prep), which we briefly highlight here. This program observed 85 emission line galaxies out of 138 potential emission sources in the 4 targeted WISP fields. Emission lines were detected in the FORS2 data for 38/85 cases, with agreement in the line identification between the grism and FORS2 data seen for 36 of those galaxies (suggesting 95\% accuracy of line identification). Of the 38 galaxies with detected lines, 17 are cases where the grism redshift is based on a single line. All cases where lines were not detected with FORS2 are either cases where no lines were predicted to lie in the FORS2 window (15/85) or the grism redshift is based on a single line (32/85). In summary, these findings add further support that selecting sources with multiple emission lines is necessary to ensure robust redshift estimates.

The precision of the redshifts (relating to \texttt{REDSHIFT\_ERR}) are a separate metric from accuracy, with the former mainly driven by the spectral resolution of the grism data. For reference, the WFC3/IR detector using the G141 grism has a sampling of 46.5~\AA/pix and a FWHM$\sim$110\AA. We characterise the precision as $\sigma_z/(1+z)$ (i.e., \texttt{REDSHIFT\_ERR} /(1+\texttt{REDSHIFT}), which has a median value of 0.00088 ($\sim$0.09\%) and the 16th and 84th percentiles are 0.00045 and 0.00154, respectively. \cite{bagley20} performed an independent test of the precision using the 36 WISP fields that overlap to some degree with each other. These fields result in $\sim$140 sources that were observed multiple times, often with very different exposure times (field depths) and roll angles. This comparison showed an empirical precision of $\sigma_z/(1+z)=0.00136$, which is in close agreement with the median value in the catalogue when accounting for the fact that the test combines the uncertainty of two line measurements.

\section{Results}\label{results}

\begin{figure*}
\includegraphics[width=0.7\textwidth]{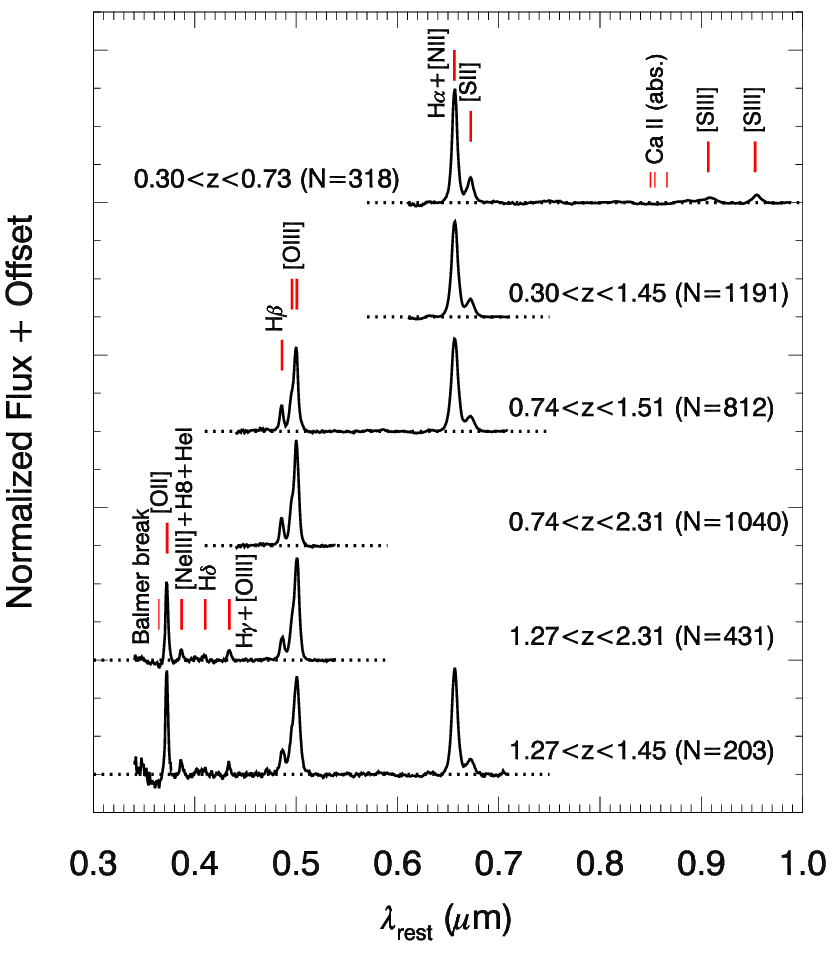}
\vspace{-2mm}
\caption{Demonstration of the wavelength coverage of WISP stacked spectra for different redshift windows. Multiple emission lines are simultaneously available for several windows in the redshift range from $z=0.3$ to 2.3. The sample size indicated corresponds to sources that satisfy the criteria for our science results (criteria 1-8; see Section~\ref{selection}).
\label{fig:stack_window_compare}}
\end{figure*}

We demonstrate the utility of combining the WISP photometric and emission line catalogues to study galaxy evolution by dividing the sample into four grism windows (redshift regimes) where different sets of emission lines are available in G102 and/or G141 (i.e., do not always require both) to constrain properties of the interstellar medium:
\begin{itemize}
\item grism coverage of \halpha+[NII], [SII], and [SIII] doublet ($0.30<z<0.73$)
\item grism coverage of \halpha+[NII] and [SII] ($0.30<z<1.45$)
\item grism coverage of \hbeta\ and [OIII] ($0.74<z<2.31$)
\item grism coverage from [OII] to [SII] ($1.27<z<1.45$; maximum $\lambda$-coverage)
\end{itemize}

We do not examine the window where \hbeta, [OIII], and \halpha+[NII] are simultaneously covered to constrain the Balmer decrement (\halpha/\hbeta; $0.74<z<1.51$), which is a subset of sources in the third window above,  because this was the focus of \citet{battisti22}. Similarly, we do not examine the window where [OII], \hbeta, and [OIII] are covered ($1.27<z<2.31$; wider redshift than fourth window above) to constrain the metallicity via the $R_{23}\equiv (\mathrm{[OII]}\lambda\lambda$3726, 3729 + [OIII]$\lambda\lambda$4959, 5007)/\hbeta\ diagnostic because this was the focus of \citet{henry21}. Both of those studies also supplement their samples with other \hst\ grism surveys such that they have larger sample sizes than are available from WISP alone. A visual demonstration of the spectral coverage from WISP for these windows is shown in Figure~\ref{fig:stack_window_compare}, together with the sample size satisfying the criteria for our science results (Section~\ref{selection}).

\subsection{Sample Selection Criteria}\label{selection}
\begin{figure}
\begin{center}
\includegraphics[width=0.5\textwidth]{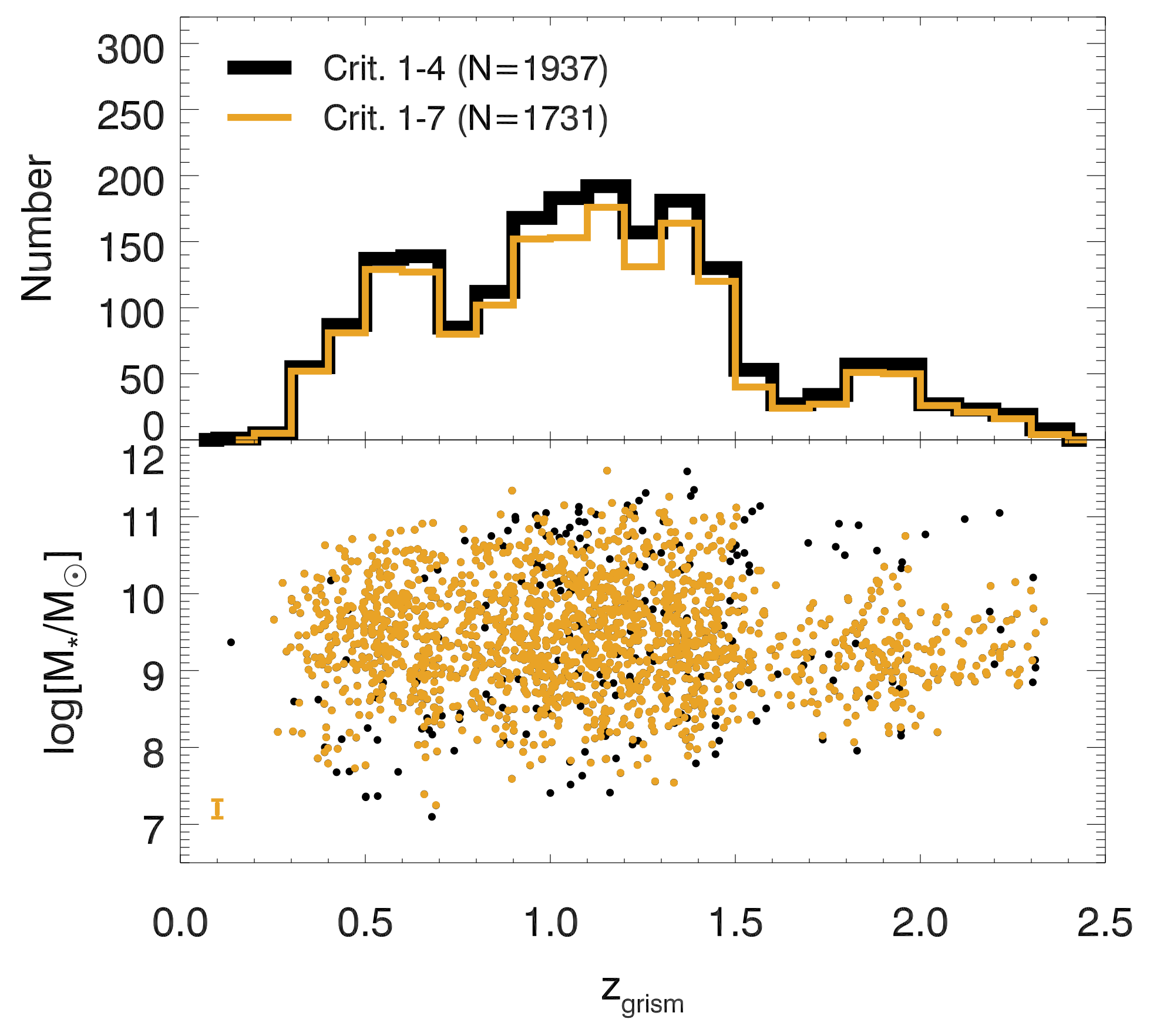}
\end{center}
\vspace*{-0.5cm}
\caption{\textit{Top panel:} grism redshift, \zgrism , distribution for galaxies in the WISP sample that have robust photometric and spectrocopic data (criteria 1-4; black) and the subset that also have good SED fits, accurate stellar masses, and are not AGN candidates (criteria 1-7; orange), as described in Section~\ref{selection}. \textit{Bottom panel:} stellar mass vs. redshift. A representative median 1$\sigma$ errorbar for the criteria (1)-(7) sample (orange) is indicated in the lower-left, with values of $\langle \sigma( z_\mathrm{grism}) \rangle = 0.0017$ (i.e., negligible) and $\langle \sigma(\log[M_\star/M_\odot])= 0.12$~dex. No significant selection effects on stellar mass are apparent with redshift. 
 \label{fig:z_hist}}
\end{figure}

Combining the overlapping sources in both the WISP photometric and emission line catalogues provides us with a parent sample of 1937 galaxies that satisfy the following spectroscopic and photometric selection criteria:
\begin{enumerate}
\item[(1)] One emission line with $S/N \ge 3$ and one additional line with $S/N \ge 2$
\item[(2)] At least three bands of photometry with $S/N>3$
\item[(3)] Independent redshift agreement between the two reviewers (\texttt{REDSHIFT\_FLAG=0})
\item[(4)] Emission line $\mathrm{FWHM}<600$\AA\ (\texttt{FWHM\_OBS}$<$600)
\end{enumerate}
Criteria (1) ensures an accurate \zgrism, which is important for optimally aligning the spectra, as well as reducing false identifications. 
Criteria (2) ensures that we have adequate SED coverage for characterising stellar masses, $M_\star$. Criteria (3) removes ambiguous sources where the WISP pipeline reviewers have either assigned different redshift solutions or the spectrum was rejected by one of the reviewers (no confidence in redshift solution). We note that criteria (3) removes $\sim$1/3 of the potential sample, which indicates the difficulty in determining reliable redshifts from low signal-to-noise spectra. Criteria (4) removes very broad emission line sources ($\sim$1\% of sample). These broad profiles may be due to an AGN, which we want to exclude for our analysis. The redshift and stellar mass distribution of sources satisfying criteria (1)-(4) are shown in Figure~\ref{fig:z_hist}.

For creating stacked spectra according to stellar mass, we also impose additional cuts on SED goodness of fit and uncertainty on stellar mass:
\begin{enumerate}
\item[(5)] $\chi_\mathrm{red}^2<3$
\item[(6)] $\sigma(\log M_\star)\,< 0.3$~dex
\end{enumerate}
which is detailed in Section~\ref{SED_fit}.  Together, criteria (5) and (6) remove 147 sources (6 fail both criteria) or 7.6\% of the sample. For reference, width of the stellar mass bins in our analysis are roughly 0.3~dex or larger to improve the reliability of bin assignment for each galaxy.

For the subset sources with grism coverage and detection of [OIII] and \hbeta, we also exclude AGN candidates based on the Mass-Excitation diagram (log[[OIII]/\hbeta] vs. log[$M_\star/M_\odot$]; \citealt{juneau14}): 
\begin{enumerate}
\item[(7)] non-AGN in MEx diagram
\end{enumerate}
which is detailed in Section~\ref{oiii_hb_window}. Criteria (7) removes an additional 59 sources. 
This leave 1731 sources that satisfy criteria (1)-(7), and this sample is also shown in Figure~\ref{fig:z_hist}. 

Finally, a criteria limiting the contamination of the main emission lines if that line is the focus of the stacking analysis:
\begin{enumerate}
\item[(8)] neither reviewer marked line contamination (\texttt{[LINE]\_CONTAM<4} from both reviewers)
\end{enumerate}
For example, we require \hbeta\ and [OIII] are uncontaminated for the stacks in the \hbeta\ and [OIII] window (Section~\ref{oiii_hb_window}). Depending on the window considered, this criteria typically removes an additional 15-20\% of sources in that window. This highlights the high rate of contamination that can occur in single-orientation grism data.

\subsection{SED modelling for stellar masses}\label{SED_fit}
To estimate stellar masses, we perform SED-fitting on our galaxy sample using the \magphys\ (high-$z$) spectral modelling code \citep{daCunha15, battisti20}, adopting the grism redshift as the input redshift (i.e., fixed-$z$). Prior to fitting, the photometry is corrected for foreground Milky Way extinction using the Galactic dust extinction maps from \citet{schlafly&finkbeiner11} via the NASA/IPAC Infrared Science Archive\footnote{\url{https://irsa.ipac.caltech.edu/applications/DUST/}}. 
\magphys\ uses spectral population synthesis models of \cite{bruzual&charlot03}, and we refer readers to that paper for details on the stellar tracks, templates, and isochones adopted in that model, also noting that it does include a prescription for thermally pulsing AGB stars. \magphys\ adopts a \citet{chabrier03} initial mass function (IMF) and has 14 free model parameters (high-$z$ version), and for full details we refer readers to the documentation on the \magphys\ website\footnote{\url{http://www.iap.fr/magphys/index.html}}. In brief, they include: a uniform prior in metallicity from 0.2 to 2 times solar (1 parameter); a parametric star-formation history (SFH; 3 parameters), which rises linearly at early ages and then declines exponentially (delayed-tau model) with additional instantaneous bursts of star formation; the dust model of \citet{charlot&fall00} (4 parameters) for which the interstellar dust is distributed into two components, one associated with star-forming regions (migration time of 10~Myr) and the other with the diffuse interstellar medium (ISM), with the addition of the 2175\AA\ absorption feature \citep{battisti20}; the dust emission models of \cite{daCunha08}, which uses templates based on four components (5 parameters); and a normalisation that sets the stellar mass and star formation rate (SFR) from the SFH (1 parameter).
The adopted SFH parametrisation may introduce systematic biases to the stellar mass estimates \citep[e.g.,][]{leja19}. For reference, when comparing overlapping WISP galaxies in \citet{henry21}, who use similar photometry but adopt a non-parametric SFH for their SED modelling, we find their stellar masses are systematically larger than the \magphys-derived values by 0.2~dex. 

\begin{figure*}
\includegraphics[width=0.95\textwidth]{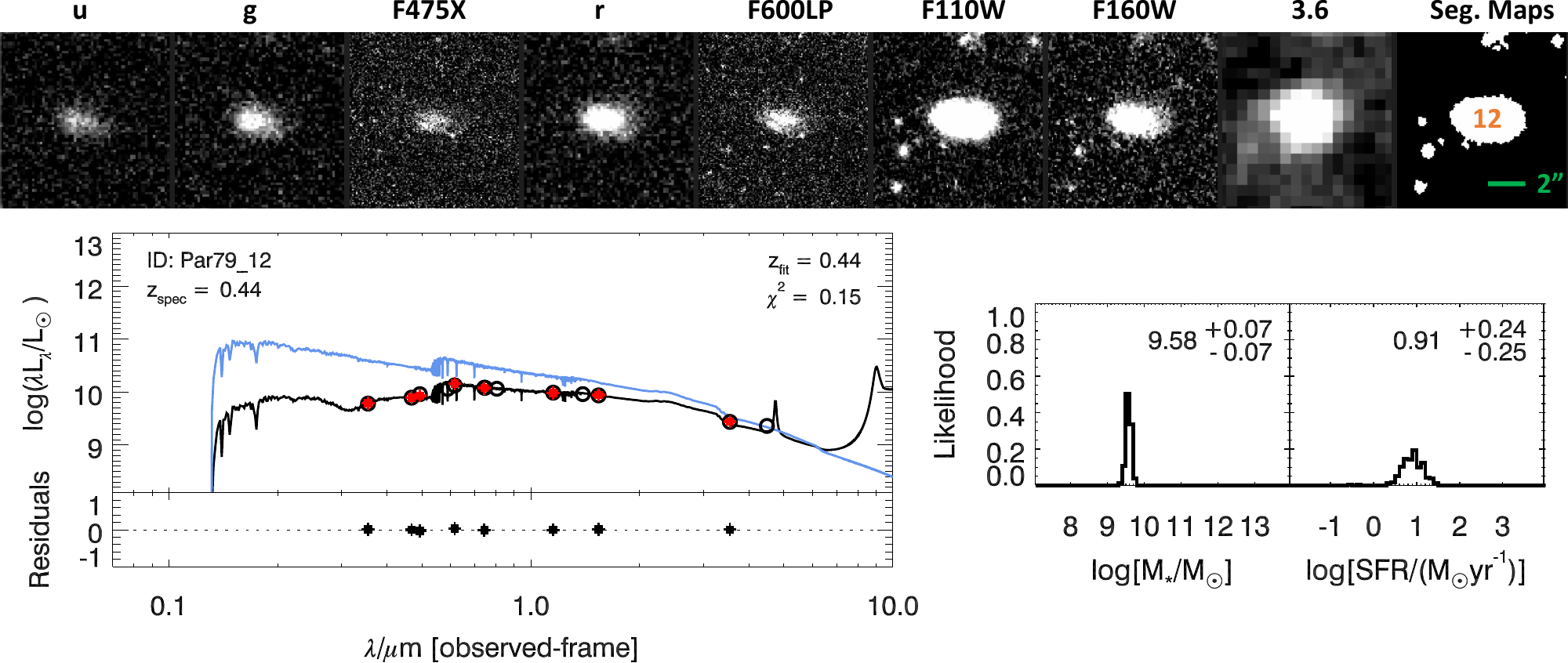}
\caption{\textit{Top row:} Native resolution $10\arcsec\times10\arcsec$ image thumbnails (linear-scale) of the available photometry for Par79\_12 (i.e., parallel field 79, object 12). The last panel shows the \se\ segmentation map, based on F110W and F160W, which is an input for \texttt{TPHOT}, with object 12 indicated. \textit{Bottom-left:} The \magphys\ best-fit SED (black line) for Par79\_12. The red squares are the observed photometry and the black circles are the corresponding model values. The blue line shows the predicted intrinsic stellar population SED (without attenuation). \textit{Bottom-right:} The posterior probability distribution functions (PDFs) for stellar mass and SFR are also shown. The available photometric coverage is sufficient to accurately constrain stellar masses ($\sim$0.1~dex), however SFRs have poorer constraints due to the lack of rest-frame IR data (i.e., age/dust reddening degeneracy; $\sim$0.3~dex). 
 \label{fig:wisp_example}}
\end{figure*}

\magphys\ does not include templates for emission line fluxes and therefore we perform emission line subtraction prior to SED fitting, when available. This is especially important for this study because we are using an emission-line-selected sample. If one assumes a roughly flat continuum (in $F_\lambda$), the average flux density measured in the photometry can be approximated as \citep[e.g.,][]{whitaker14}
\begin{equation}    
F_\lambda \simeq F_{\lambda,\mathrm{cont}} + F_\mathrm{line}/\Delta \lambda \,,
\end{equation}
where $F_{\lambda,\mathrm{cont}}$ is the continuum-only flux density and $\Delta \lambda$ is the width of the filter, which we take to be its FWHM. We subtract emission line fluxes for all lines with $S/N>2$ from the photometric data. The impact of emission lines on photometry are the largest for galaxies with fainter continuum emission (typically lower $M_\star$) and higher equivalent widths ($EW=F_\mathrm{line}/F_{\lambda,\mathrm{cont}}$). An example of a \magphys\ fit for a WISP galaxy is shown in Figure~\ref{fig:wisp_example}. 

\begin{table*}
\caption{Stellar mass and SFR percentiles from \magphys\ for the 1937 WISP sources with `robust' spectroscopy and photometry (criteria 1-4 in Section~\ref{selection}) \label{tab:magphys_output}} 
\begin{center}
\begin{tabular}{ccccccccccccccccc}
 \hline \\
\multirow{2}{*}{Par}  &  \multirow{2}{*}{Obj} &  \multirow{2}{*}{\zgrism} &  \multirow{2}{*}{$z_\mathrm{grism,err}$} & \multirow{2}{*}{AGN\_flag$^a$} & \multicolumn{5}{c}{$\log[M_\star/ M_\odot]$} & \multicolumn{5}{c}{$\log[\mathrm{SFR}(\mathrm{SED})/ (M_\odot \mathrm{yr}^{-1})]$} & \multirow{2}{*}{$\chi^2_\mathrm{red}$} & \multirow{2}{*}{$N_\mathrm{bands}$}\\
 & & & & & 2.5 & 16 & 50 & 84 & 97.5 & 2.5 & 16 & 50 & 84 & 97.5 & & \\
 \hline
  1 & 10  & 0.5084  & 0.0013 & 0 &  9.92   &  10.03   &  10.15   &  10.26  &   10.40 &   0.01 & 0.73 & 1.16 & 1.58 & 1.90 &   0.011 & 5 \\
  1 & 13  & 0.5309 & 0.0011 & 0 &  9.80 & 9.94  &   10.02   &  10.13   &  10.21 &  -0.60 & 0.26 & 0.92 & 1.28 & 1.74 &   0.055 & 7 \\
  1 & 15 & 0.6699 & 0.0011 & 0 &  9.86 & 9.96  &   10.04  &   10.13    & 10.22 &   0.68 & 0.95 & 1.32 & 1.73 & 2.08 &   0.464 & 7 \\
  1 & 28 & 1.3443 & 0.0028 & 0 & 9.41 & 9.50 & 9.61 & 9.68 & 9.80 &  1.07 & 1.16 & 1.32 & 1.48 & 1.58 &   0.297 & 7 \\
  1 & 41 & 1.3065 & 0.0018 & 0 & 9.67 & 9.81 & 9.89 & 10.00 & 10.13 &  0.91 & 1.21 & 1.53 & 1.77 & 1.94 &   0.326 & 7 \\
  
\multicolumn{17}{c}{...} \\[1ex] 
\hline
 \end{tabular}
\end{center}
\textbf{Notes.} A full ASCII version of this table is available online. Percentiles provided are 2.5\%, 16\%, 50\%, 84\%, and 97.5\%. SFR(SED) corresponds to the average SFR over the last 100~Myr of the SFH. $\chi_\mathrm{red}^2=\chi^2/N_\mathrm{bands}$ is the reduced $\chi^2$ of the best-fit model and $N_\mathrm{bands}$ is the number of bands observed with non-zero flux. \\
$^a$AGN\_flag=1 indicates sources that lie in the AGN region of the Mass-Excitation diagram (Section~\ref{oiii_hb_window}).
\end{table*}

For each \magphys\  fit, a goodness-of-fit is determined based on the best-fit model using a reduced $\chi^2$ metric, $\chi_\mathrm{red}^2=\chi^2/N_\mathrm{bands}$, where $N_\mathrm{bands}$ is the number of bands observed with non-zero flux. We exclude cases of poor-quality fits by removing galaxies with $\chi_\mathrm{red}^2>3$ from our analysis (criteria 5), which removes 140 of the criteria (1)-(4) sources. Cases of poor fits may be associated with poor/inconsistent photometric data and/or AGN contamination (AGN models are not included in the SED fitting). 
We also require accurate stellar masses because we will bin our data according to stellar masses, and exclude sources with $\sigma(\log M_\star)> 0.3$~dex (based on the 16th and 84th percentiles of the posterior PDF; $\sigma(\log M_\star)$=($\log M_{\star\mathrm{,p84}}-\log M_{\star\mathrm{,p16}})/2$; criteria 6). This occurs for 13 of the criteria (1)-(4) sources. The median values of the remaining WISP sample is $\bar{\chi}_\mathrm{red}^2=0.32$. The low $\chi^2$ values are due to the fact that the models tend to `over-fit' the data (more free parameters than datapoints). 

Table~\ref{tab:magphys_output} provides the $\log M_\star$ and $\log \mathrm{SFR(SED)}$ percentiles from \magphys\ for our sample of 1937 galaxies that satisfy criteria (1)-(4) in Section~\ref{selection}. SFR(SED) represents the average SFR over the last 100~Myr of the SFH. For reference, the median 1$\sigma$ uncertainty on $\log M_\star$ and $\log \mathrm{SFR(SED)}$ (taken as (84th-16th percentile)/2) is 0.12 and 0.25~dex, respectively. We note that our uncertainties may be underestimated due to the SFH parametrisation used in \magphys. For overlapping WISP galaxies in \citet{henry21}, which uses a non-parametric SFH for SED modelling, their median 1$\sigma$ uncertainties on $\log M_\star$ and $\log \mathrm{SFR(SED)}$ are larger by 0.03 and 0.04~dex, respectively. 
SED-derived SFRs have very large uncertainty when relying only on UV through near-IR data. For this reason, we avoid using SED-derived SFRs for our analysis and instead use \halpha-based estimates whenever possible. We note that the median number of bands available for the subsample is 5, with 85\% of the sample having \textit{coverage} in IRAC (i.e., detection or an upper limit). The median uncertainty on $\log M_\star$ for sources with 3, 4, 5, and 6+ bands available is 0.19, 0.13, 0.11, and 0.10~dex, respectively. We find that fields without IRAC coverage have only marginally higher stellar mass uncertainty (0.01~dex) and no significant bias. The difference in median value for the two samples is $\log M_\star(\mathrm{w/~IRAC})-\log M_\star(\mathrm{no~IRAC})=0.045$~dex and below the typical uncertainty and our binning size. We stress that these are formal uncertainties and that the true mass uncertainty is higher when accounting for systematic uncertainties arising from model assumptions (e.g., SFH). We include the number bands available in the SED modelling for each source as a column in Table~\ref{tab:magphys_output}.

\subsection{Spectral Stacking and Emission Line Fitting of Stacked Spectra}\label{method_stacking_fitting}
All galaxy spectra are stacked and fit following similar methods to those described in \citet{henry21, dai21, battisti22}. In brief, we use the continuum-subtracted spectra (using a cubic spline; see Section~\ref{linefits}) and normalise them by the `typical' brightest line in the spectral window considered, which for the windows we consider is either \halpha+[NII] (low-$z$ windows) or [OIII] (high-$z$ windows). The spectra are de-redshifted using a linear interpolation to shift them onto a common grid of rest wavelengths and we take the median of the normalised fluxes at each wavelength.

The procedure to fit the stacked spectra differs slightly from the method used for individual sources for the emission line catalogue (Section~\ref{linefits}) and this is due to the fact that the stacks reach greater depth such that more parameters are generally required for good fits (e.g., narrow+broad components). To fit the stacked spectra, we fit a set of Gaussian profiles to the emission lines in the region of interest. We adopt two Gaussian components for each line, one narrow and one broad component. Multiple components can arise due to kinematic differences among ionising sources (e.g., HII vs AGN), but can also occur in grism spectra due to line profiles having a dependence on the spatial distribution of the emitting sources. The FWHM of the broad component is fixed to be the same for all of the lines and also required to be between 1-4x the FWHM of the narrow components. The amplitudes of the broad components for each line are allowed to vary independently (among positive values).

The emission lines are simultaneously fit with the following assumptions/restrictions: (1) the ratio of [OIII]$\lambda$5007/[OIII]$\lambda$4959 is fixed to 2.98:1 \citep{storey&zeippen00} and [SIII]$\lambda$9532/[SIII]$\lambda$9069 is fixed to 2.47:1 \citep{berg21}, (2) single profiles are used for the closely spaced blends of [OII]$\lambda\lambda$3727, 3729, \halpha +[NII]$\lambda\lambda$6548, 6583, and [SII]$\lambda\lambda$6716, 6731, (3) we require the narrow component FWHM of close pairs to match (e.g., \hbeta\ and [OIII], \halpha +[NII], and [SII]). We do not match all components in order to account for the effect that the spectral resolution difference between the G102 and G141 grisms can have on the profiles, (4) we require the FWHM of the narrow components to be within a factor of 2 with each other, (5) we allow a $\pm10\text{\AA}$ shift (rest frame) of the emission line centroids to accommodate systematic uncertainties in the grism wavelength solution, and (6) we account for any (small) residual continuum offsets due to imperfect continuum subtraction by including free parameters for the spectra amplitudes (i.e., constant offsets) in the regions near emission line groups (e.g., $4400\text{\AA}<\lambda_\mathrm{rest}<5500\text{\AA}$ for \hbeta\ and [OIII]; $6000\text{\AA}<\lambda_\mathrm{rest}<7100\text{\AA}$ for \halpha +[NII] and [SII]). 

Line flux measurements for the stacks are based on scaling the average flux of the normalised line in each bin (i.e., reversing the normalisation). Line flux measurement uncertainties on the stacked spectra are obtained by bootstrapping with replacement. In brief, for each sample of $N$ galaxies that are stacked, we draw $N$ random galaxies from that sample, allowing individual objects to be selected more than once. We create a new stack from these objects and measure the lines and repeat this procedure 1000 times and calculate the standard deviation on the line fluxes from these realisations and adopt this as the uncertainty.
  
\subsection{Stellar Mass Stack Results}\label{stack_results}
For all stellar mass bins we require that they contain $N\gtrsim100$ galaxies to ensure reliable recovery of faint emission lines and ensure stacks are not sensitive to any individual outlier galaxies in the stack. This also ensures that corrections, which are based on averages, are reasonable for the sample (e.g. Section \ref{ha_sii_window}).
Below we show our results using stellar mass stacks for four redshift windows.

\subsubsection{\hbeta\ and [OIII] Window ($0.74<z<2.31$)}\label{oiii_hb_window}

\begin{figure*}
$\begin{array}{c}
\includegraphics[width=0.95\textwidth]{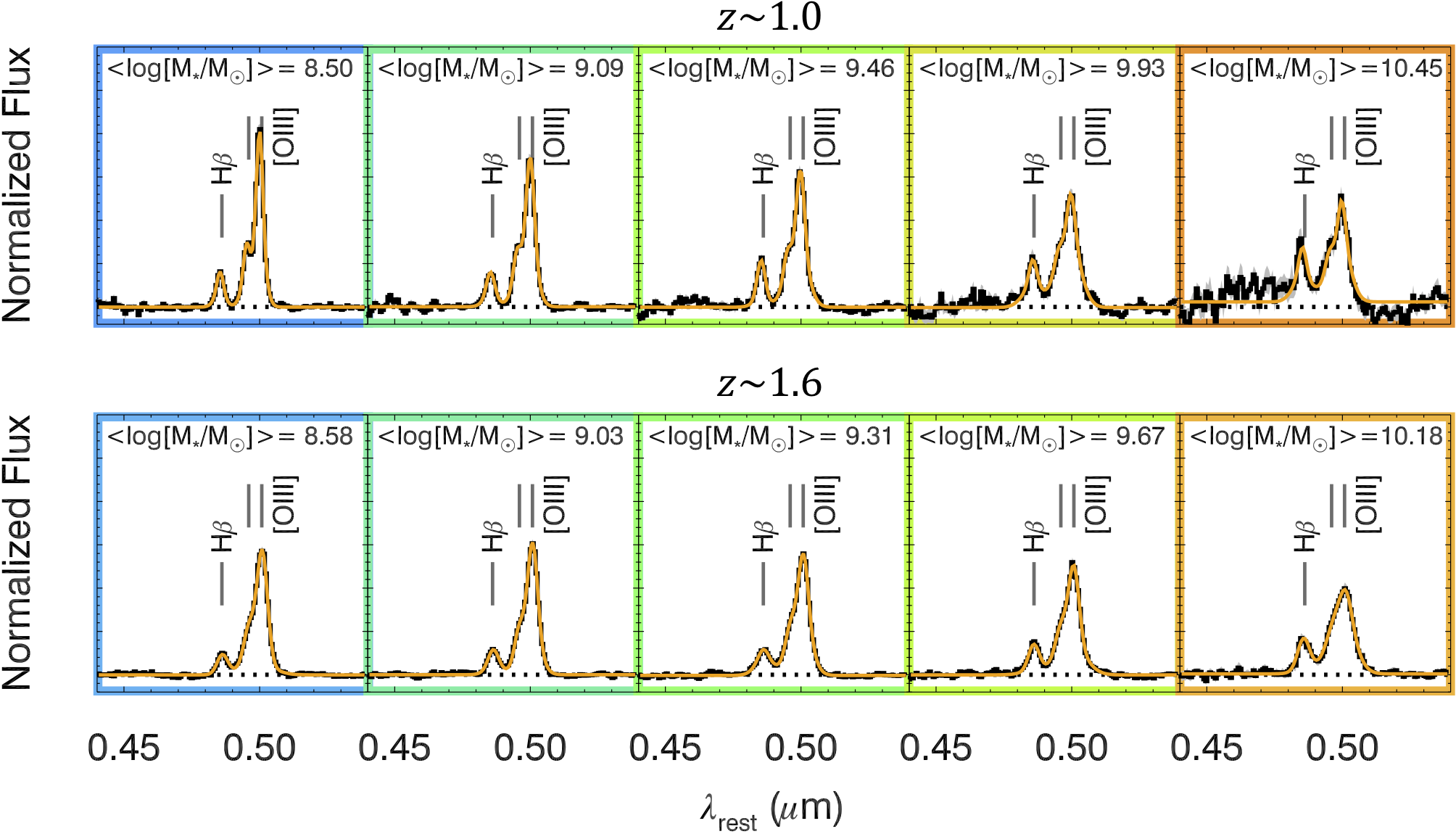}\vspace{1mm} \\
\includegraphics[width=0.7\textwidth]{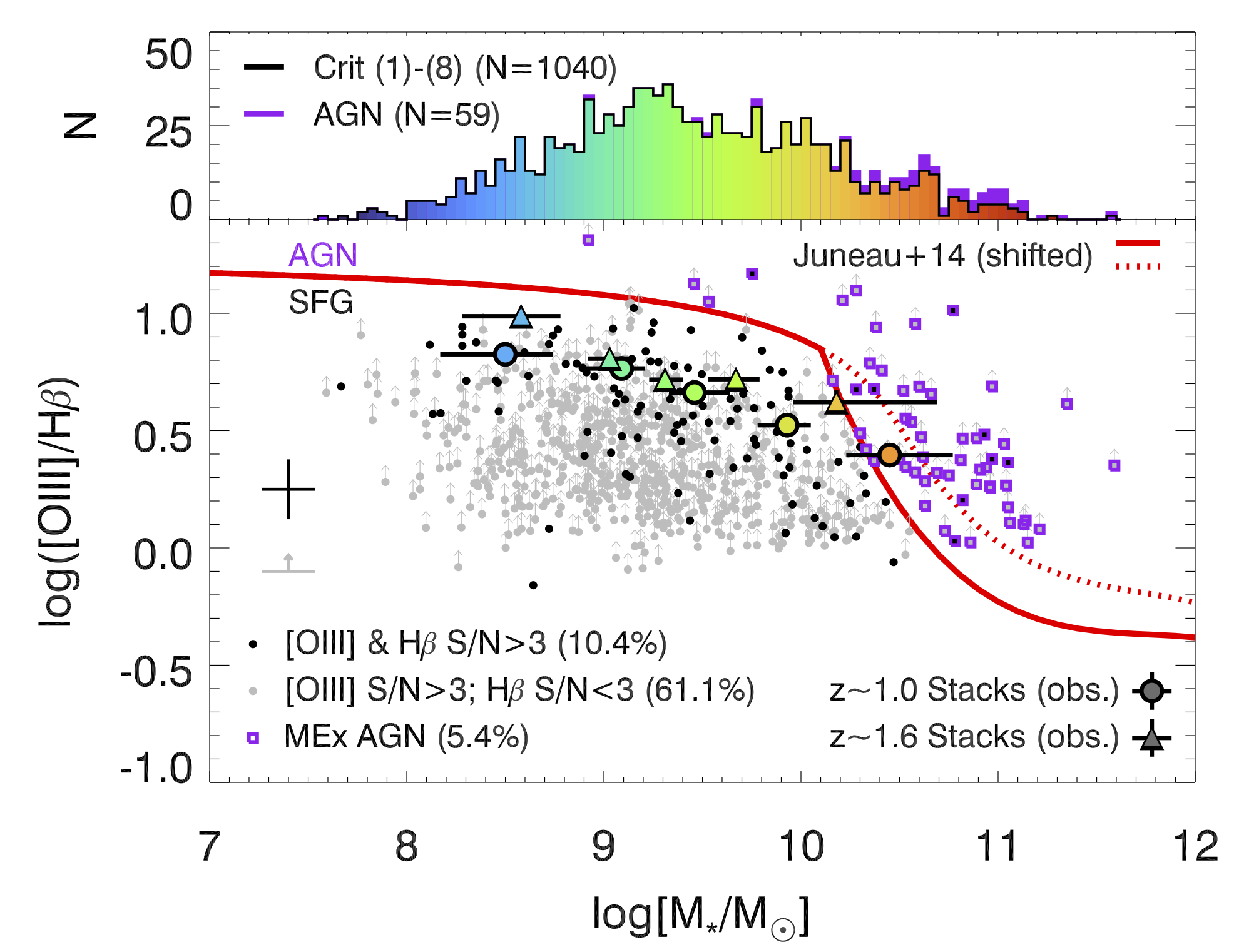} \\
\end{array}$
\vspace{-4mm}
\caption{\textit{Top:} Stacked spectra in the \hbeta\ and [OIII] window normalised to [OIII] total flux in bins of stellar mass for our subsample at $0.74<z<2.31$ ($N$=1040 with criteria 1-8). The median $\log M_\star$ is indicated in each panel. The spectral fits are shown by the orange lines. 
\textit{Bottom:} Mass-Excitation (MEx) diagram for our sample. The solid (dashed) red lines corresponds to the lower (upper) boundaries of \citet{juneau14} shifted to higher masses by 0.5~dex (i.e., to the right), corresponding to the relation found at $z\sim1.6$ by \citet{kashino19}, and up by 0.13~dex to account for using total [OIII] instead of only [OIII]$\lambda$5007. Small black and gray points denote individual galaxies where MEx positions are constrained (71.5\% of sample), with median errorbars given on the middle-left, and AGN candidates are indicated by purple squares. \hbeta\ undetected cases are treated as lower-limits (see Section~\ref{oiii_hb_window}). 
The large circles/triangles (colour based on median $\log M_\star$), correspond to the \textit{observed} values from the stacked spectra shown at the top (i.e, \hbeta\ is not corrected for stellar absorption). Stack x-axis errorbars denote the 1$\sigma$ mass range spanned by each bin (Table~\ref{tab:stack_hb_oiii} lists full mass range). The y-axis errorbars denote the line ratio error on the stacked spectra, which are smaller than the symbol sizes in this case (<0.05~dex; see Table~\ref{tab:stack_hb_oiii}). Despite excluding AGN candidates in the stacking, the position of our highest-mass stacks remain on/above the lower AGN boundary. We attribute this to the 28.5\% of our individual sources being unconstrained ($S/N<3$ in both [OIII] and \hbeta), with larger $\log M_\star$ galaxies preferentially lacking detection in these lines (see Section~\ref{oiii_hb_window}).
 \label{fig:oiii_hb_compare}}
\end{figure*}

We start with galaxies with both \hbeta\ and [OIII] grism coverage ($0.74< z_\mathrm{grism} < 2.31$) because it is one of the largest groups we consider ($N$=1040 with criteria 1-8) and the ratio [OIII]/\hbeta\ as a function of stellar mass, known as a the Mass-Excitation (MEx) diagram, has been established as a reliable tool for distinguishing star forming galaxies (SFGs) and AGN at both low- \citep{juneau14} and high-redshifts \citep{coil15, kashino19}. The results of \citet{kashino19} and \citet{coil15} suggest that the demarcation line to distinguish SFGs and AGN should shift to higher stellar masses with increasing redshift, requring a 0.5~dex and 0.75~dex shift at $z\sim1.6$ and $z\sim2.3$, respectively. We use the demarcation from \citet{kashino19} as our reference because we are examining similar redshifts ($z\sim1.0$ and $z\sim1.6$). The position on the MEx diagram of individual galaxies in our subsample is shown in Figure~\ref{fig:oiii_hb_compare}, bottom. We find that 10.4\% of the WISP sources in this subsample are detected in both \hbeta\ and [OIII] ($S/N>3$), 61.1\% detected only in [OIII], and the rest are undetected in both lines ($S/N<3$; 28.5\%). For \hbeta\ undetected cases, we use the 3$\sigma$ \hbeta\ error and treat the ratio as a lower-limit. We find 59 individual sources above the AGN lower boundary line, which are excluded from the stacks.

We sub-divide galaxies in this window into two redshift ranges, $0.74<z<1.22$ and $1.22\leq z<2.31$ (z=1.22 is the median for this window), and each of those into 5 equal-number bins in stellar mass (10 bins total). The spectra of these stellar mass stacks and their position on the MEx diagram are shown in Figure~\ref{fig:oiii_hb_compare}. The average galaxy properties and emission line values for these stacks are listed in Table~\ref{tab:stack_hb_oiii}. As expected, most of the stacks lie in the SFG region of the diagram. However, despite excluding all individual galaxies with emission line ratios or upper limits that lie in the AGN region of the MEx diagram prior to stacking (purple squares), the position of the highest-mass stacks for each redshift remains on/above the lower AGN boundary. We attribute this result to the 28.5\% of individual sources that are unconstrained in the MEx diagram (i.e., those with $S/N<3$ for both [OIII] and \hbeta ), together with the fact that higher stellar mass galaxies have preferentially weaker [OIII] and \hbeta\ relative to \halpha\ than lower stellar mass galaxies \citep[e.g., Figure~5 of ][]{battisti22}. As a reference, the lowest stellar mass bins for each redshift range have $\sim$90\% of individual sources detected in [OIII] or \hbeta, whereas the highest stellar mass bins have 30-40\% of individual sources detected in [OIII] or \hbeta\ (for these cases, usually \halpha\ and [SII] are detected). Thus, a majority of sources in the higher mass bins might be AGN that we are unable to identify individually using the MEx diagram. This indicates that there may be significant AGN contamination for WISP emission line sources at high stellar masses. 

\renewcommand{\arraystretch}{1.2} 
\begin{table}
\caption{Average properties, emission line luminosities, and line ratios for $N$=1040 stacked spectra in the \hbeta\ and [OIII] window ($0.74<z<2.31$) that satisfy criteria (1)-(8) \label{tab:stack_hb_oiii}} 
\begin{center}
\begin{tabular}{cccccc}
 \hline \\[-1em]
\multirow{2}{*}{$N$} & 
$\log M_\star$ & 
\multirow{2}{*}{$\left\langle \log M_\star \right\rangle$} & 
\hbeta &
[OIII] &
\multirow{2}{*}{$\dfrac{\mathrm{[OIII]}}{\mathrm{H}\beta}$} \\
\vspace{0.04in}  & range & & \multicolumn{2}{c}{($10^{41}$~erg s$^{-1}$)} \\
 \hline
\multicolumn{6}{c}{$0.74 < z_\mathrm{grism} < 1.22$ \, ($z\sim 1.0$)} \\
     104 & [ 7.59, 8.82] &  8.50 &  1.09$\pm$0.06 &  7.27$\pm$0.11 &  6.69$\pm$0.37 \\ 
     104 & [ 8.83, 9.27] &  9.09 &  1.24$\pm$0.06 &  7.24$\pm$0.11 &  5.82$\pm$0.31 \\ 
     104 & [ 9.29, 9.71] &  9.46 &  1.09$\pm$0.06 &  5.01$\pm$0.09 &  4.60$\pm$0.28 \\ 
     104 & [ 9.72,10.12] &  9.93 &  1.42$\pm$0.09 &  4.74$\pm$0.11 &  3.34$\pm$0.23 \\ 
     104 & [10.12,11.60] & 10.45 &  1.17$\pm$0.11 &  2.92$\pm$0.15 &  2.49$\pm$0.27 \\ 
\hline
\multicolumn{6}{c}{$1.22 \le z_\mathrm{grism} < 2.31$ \, ($z\sim 1.6$)} \\
     104 & [ 7.77, 8.85] &  8.58 &  2.11$\pm$0.10 & 20.56$\pm$0.25 &  9.72$\pm$0.49 \\ 
     104 & [ 8.86, 9.19] &  9.03 &  3.60$\pm$0.15 & 23.11$\pm$0.18 &  6.42$\pm$0.27 \\ 
     104 & [ 9.19, 9.47] &  9.31 &  4.70$\pm$0.16 & 24.50$\pm$0.18 &  5.22$\pm$0.19 \\ 
     104 & [ 9.47, 9.89] &  9.67 &  3.82$\pm$0.25 & 20.00$\pm$0.31 &  5.24$\pm$0.35 \\ 
     104 & [ 9.89,11.26] & 10.18 &  2.60$\pm$0.29 & 10.89$\pm$0.30 &  4.18$\pm$0.48 \\ 
\hline
 \end{tabular}
\end{center}
\textbf{Notes.} \hbeta\ is not corrected for stellar absorption. [OIII] is the sum of [OIII]$\lambda\lambda$4959, 5007.
\end{table}

These WISP results are consistent with \citet{forsterSchreiber19}, who examined AGN occurrence rates, $f_{AGN}$, in a sample of 600 galaxies at $0.6<z<2.7$ in KMOS$^\mathrm{3D}$. They found that galaxies with $\log (M_\star /M_\odot)<10.2$ have an AGN occurrence rate of $f_{AGN}\lesssim10$\%, with $f_{AGN}$ increasing dramatically with increasing mass (e.g., $\sim$60\% at $\log (M_\star /M_\odot)=11$; see their Figure~6). These results are also similar to findings in \citet{henry21}, which included both the WISP and CLEAR+3D-HST surveys. 
For the subsequent analysis, we do not exclude galaxies based on a stellar mass threshold \citep[e.g.,][]{battisti22}, but caution that AGN may contaminate bins above $\log (M_\star /M_\odot)\gtrsim10.2$.

\subsubsection{\halpha+[NII] and [SII] Window ($0.30<z<1.45$)}\label{ha_sii_window}
Next, we consider galaxies with grism coverage of \halpha+[NII] and [SII] ($0.30< z_\mathrm{grism} < 1.45$; $N$=1191 with criteria 1-8), which is our largest group size. This group can be used to examine how representative our sample is of typical SFGs at these redshifts. To do this, we characterise WISP galaxies relative to the star-forming galaxy Main Sequence \citep[MS; logSFR vs. log$M_\star$; e.g.,][]{brinchmann04, speagle14, leslie20}, using SFRs based on \halpha\ (described below). 

\begin{figure*}
$\begin{array}{c}
\includegraphics[width=0.84\textwidth]{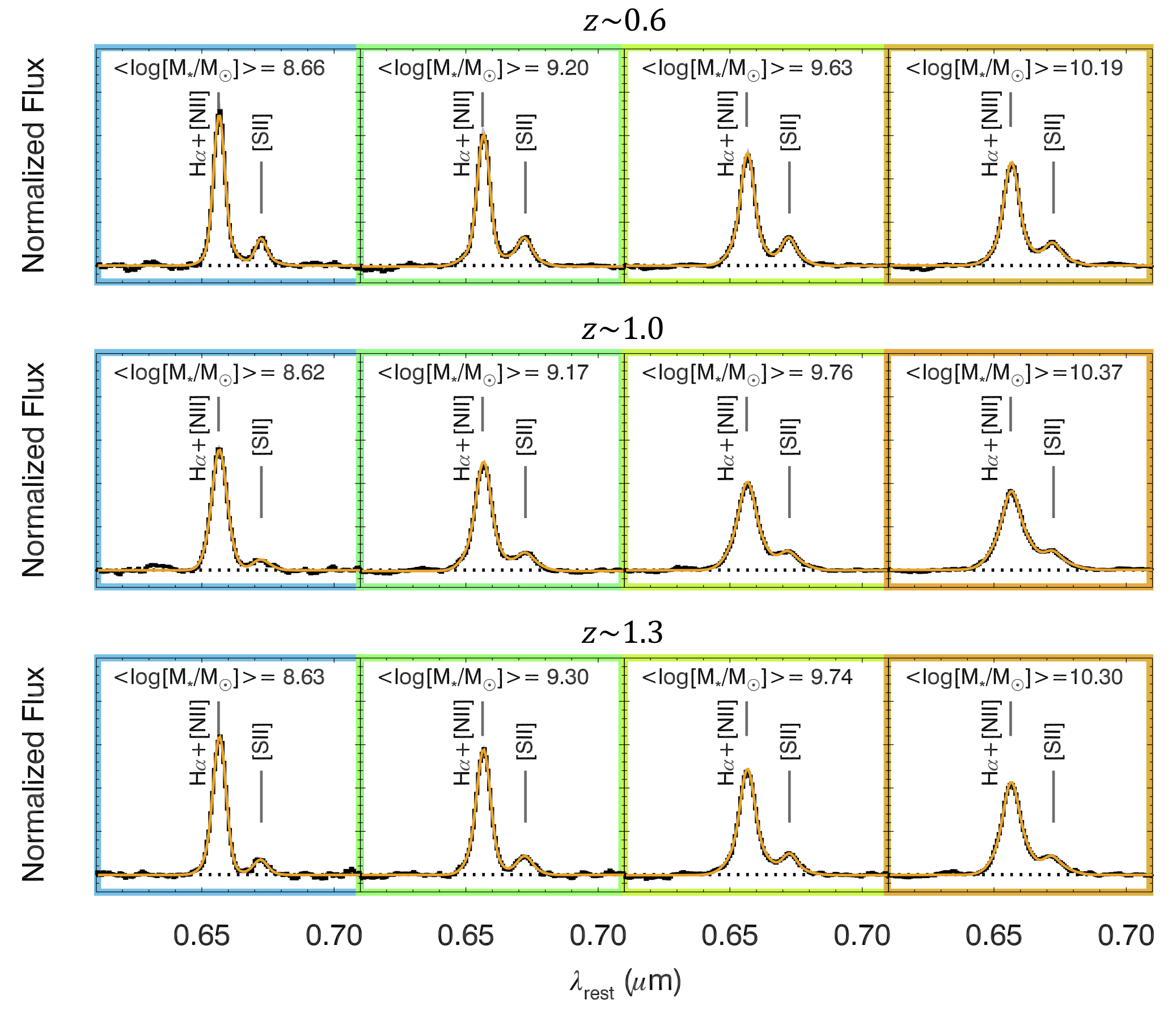}\vspace{1mm} \\
\includegraphics[width=0.6\textwidth]{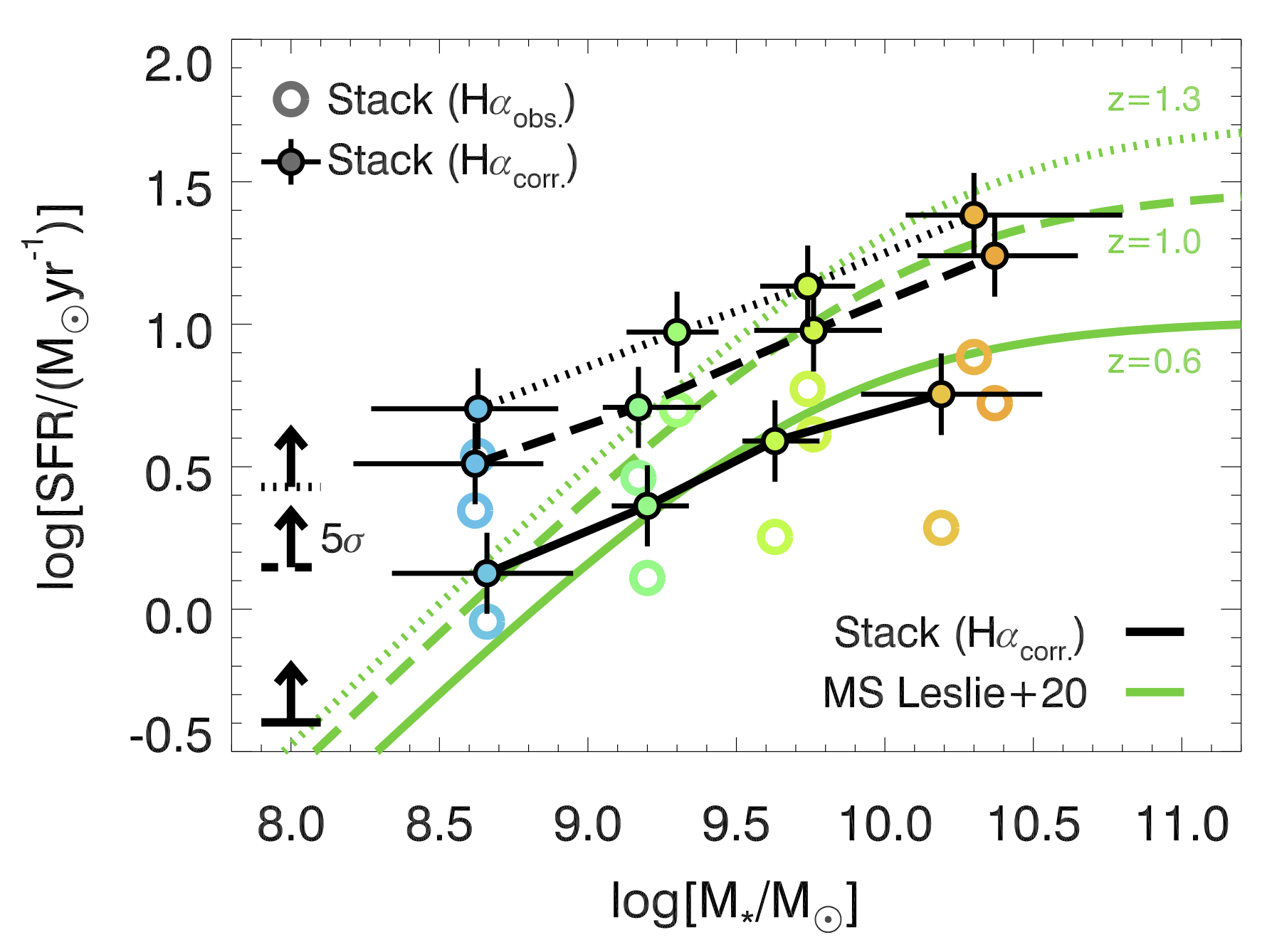} \\
\end{array}$
\vspace{-2mm}
\caption{\textit{Top:} Similar to Figure~\ref{fig:oiii_hb_compare}, but now for the \halpha+[NII] and [SII] window and the stacked spectra are normalised to \halpha+[NII] total flux.
\textit{Bottom:} The galaxy main-sequence (MS; logSFR vs. log$M_\star$), for our sample binned by stellar mass (large coloured symbols) across three redshift ranges, which are distinguished by the line style connecting the filled datapoints (not a fit; $z\sim0.6$, 1.0, and 1.3 are solid, dashed and dotted, respectively). Bins are based on stacked \halpha\ without dust corrections ($\mathrm{SFR(H}\alpha_\mathrm{obs})$) and with dust corrections ($\mathrm{SFR(H}\alpha_\mathrm{corr})$), where the latter is based on the log$M_\star$-$\tau_B^l$ relation \citep{battisti22}.
The green lines are the galaxy MS from \citet{leslie20} at the median redshift of the groups (line styles match as above). Our sample mostly coincides with the MS except for a bias at low masses due to sensitivity limits. The median \halpha\ line detection threshold (5$\sigma$) for each redshift range is indicated (bottom-left), but we note this has large variation due to differing opportunity lengths.
\label{fig:ha_sii_compare}}
\end{figure*}

We divide galaxies in this window into three redshift ranges, $0.30<z\leq 0.83$, $0.83<z\leq 1.15$ and $1.15\leq z<1.45$ (equal-number in each), and each of those into 4 equal-number bins in stellar mass (12 bins total). The spectra of these stellar mass stacks are shown in Figure~\ref{fig:ha_sii_compare}. The average galaxy properties and emission line values for these stacks are listed in Table~\ref{tab:stack_ha_sii}. 
  
\renewcommand{\arraystretch}{1.2} 
\begin{table*}
\caption{Average properties, emission line luminosities, and line ratios for $N$=1191 stacked spectra in the \halpha+[NII] and [SII] window ($0.30<z<1.45$) that satisfy criteria (1)-(8) \label{tab:stack_ha_sii}} 
\begin{center}
\begin{tabular}{ccccccccc}
 \hline \\[-1em]
\multirow{2}{*}{$N$} & 
$\log M_\star$ & 
\multirow{2}{*}{$\left\langle \log M_\star \right\rangle$} & 
\multicolumn{2}{c}{$\left\langle \log \mathrm{SFR} \right\rangle$} & 
\halpha+[NII] &
[SII] & \halpha$_\mathrm{obs}$ & \halpha$_\mathrm{corr}$ \\
\vspace{0.04in}  & range & & $\mathrm{H}\alpha_\mathrm{obs}$ & $\mathrm{H}\alpha_\mathrm{corr}$ & \multicolumn{4}{c}{($10^{41}$~erg s$^{-1}$)} \\
 \hline
\multicolumn{9}{c}{$0.30<z_\mathrm{grism} \leq 0.83$ \, ($z\sim 0.6$)} \\
     100 & [ 7.25, 9.01] &  8.66 & -0.04$\pm$0.01 &  0.13$\pm$0.14 &  1.81$\pm$0.02 &  0.35$\pm$0.02 &  1.68$\pm$0.04 &  2.49$\pm$ 0.82 \\ 
      99 & [ 9.01, 9.44] &  9.20 &  0.11$\pm$0.01 &  0.36$\pm$0.14 &  2.73$\pm$0.03 &  0.79$\pm$0.02 &  2.40$\pm$0.06 &  4.29$\pm$ 1.41 \\ 
      99 & [ 9.45, 9.84] &  9.63 &  0.25$\pm$0.02 &  0.59$\pm$0.14 &  4.23$\pm$0.05 &  1.13$\pm$0.03 &  3.34$\pm$0.13 &  7.25$\pm$ 2.39 \\ 
      99 & [ 9.84,10.92] & 10.19 &  0.29$\pm$0.02 &  0.75$\pm$0.14 &  5.28$\pm$0.06 &  1.28$\pm$0.04 &  3.59$\pm$0.16 & 10.58$\pm$ 3.49 \\ 
 \hline
\multicolumn{9}{c}{$0.83<z_\mathrm{grism} \leq 1.15$ \, ($z\sim 1.0$)} \\ 
     100 & [ 7.59, 8.93] &  8.62 &  0.35$\pm$0.01 &  0.51$\pm$0.14 &  4.41$\pm$0.04 &  0.43$\pm$0.03 &  4.13$\pm$0.08 &  6.03$\pm$ 1.98 \\ 
      99 & [ 8.94, 9.44] &  9.17 &  0.46$\pm$0.01 &  0.71$\pm$0.14 &  6.01$\pm$0.06 &  1.16$\pm$0.05 &  5.38$\pm$0.11 &  9.51$\pm$ 3.12 \\ 
      99 & [ 9.45,10.04] &  9.76 &  0.61$\pm$0.03 &  0.98$\pm$0.14 &  9.60$\pm$0.08 &  2.08$\pm$0.06 &  7.65$\pm$0.48 & 17.73$\pm$ 5.90 \\ 
      99 & [10.04,11.34] & 10.37 &  0.72$\pm$0.02 &  1.24$\pm$0.14 & 15.04$\pm$0.14 &  3.31$\pm$0.09 &  9.84$\pm$0.47 & 32.40$\pm$10.71 \\ 
\hline
\multicolumn{9}{c}{$1.15\leq z_\mathrm{grism} <1.45$ \, ($z\sim 1.3$)} \\ 
     100 & [ 7.54, 9.02] &  8.63 &  0.54$\pm$0.01 &  0.70$\pm$0.14 &  6.83$\pm$0.07 &  0.84$\pm$0.07 &  6.42$\pm$0.12 &  9.41$\pm$ 3.08 \\ 
      99 & [ 9.04, 9.50] &  9.30 &  0.70$\pm$0.01 &  0.97$\pm$0.14 & 10.49$\pm$0.10 &  1.80$\pm$0.09 &  9.36$\pm$0.17 & 17.47$\pm$ 5.72 \\ 
      99 & [ 9.50, 9.96] &  9.74 &  0.77$\pm$0.02 &  1.13$\pm$0.14 & 13.35$\pm$0.14 &  2.65$\pm$0.11 & 11.04$\pm$0.39 & 25.33$\pm$ 8.33 \\ 
      99 & [ 9.97,11.60] & 10.30 &  0.89$\pm$0.04 &  1.38$\pm$0.15 & 21.22$\pm$0.17 &  4.29$\pm$0.12 & 14.29$\pm$1.32 & 45.02$\pm$15.30 \\ 
\hline
 \end{tabular}
\end{center}
\textbf{Notes.} [SII] is the sum of [SII]$\lambda\lambda$6716, 6731. \halpha$_\mathrm{obs}$ are values after correcting for [NII] blending, where the uncertainty is the $1\sigma$ dispersion in correction values for individual galaxies and the line measurement uncertainties added in quadrature. \halpha$_\mathrm{corr}$ are values after also correcting for stellar absorption and dust extinction, with the latter introducing significant uncertainty (see Section~\ref{ha_sii_window}).
\end{table*}
  
In order to estimate SFRs from \halpha, we need to apply several corrections. We note that \textit{we only apply correction factors on stacked spectra} based on groups of $N>100$ galaxies to minimise the impact of intrinsic variation of the correction factors. 
First, we deblend the \halpha+[NII] line using the stellar mass- and redshift-dependent functional relation from \citet{faisst18},  which used $\sim$190,000 SDSS galaxies combined with the observed BPT locus evolution of SFGs from $0<z<2.5$. These corrections were derived over the range of $0<z<2.7$ and $8.5 < \log(M_\star/ M_\odot) < 11.0$ and have an intrinsic scatter of $\sim$0.2~dex. A small fraction of our stellar masses are $\log(M_\star/ M_\odot)<8.5$ but extrapolating the \citet{faisst18} relation to lower masses should have minimal impact as the expected [NII] contribution is $\lesssim5\%$ at all redshifts in this mass regime. Based on this relation, the contribution of [NII] range from $\sim$6\% for the lowest mass bins, up to $\sim$35\% for the most massive bins.

Second, we correct \halpha\ for stellar absorption, which cannot be directly fit in the low resolution grism spectra. To correct for stellar absorption, we use the stellar mass- and SFR-dependent functional relation from \citet{kashino&inoue19}, which is based on trends observed for $\sim$190,000 SDSS galaxies. The fractional corrections were determined over the range of $7.2 < \log(M_\star/ M_\odot) < 11.4$, with an intrinsic scatter of $\sim$10-20\%. For each stack, we use the median value of stellar masses and SFRs from \magphys. However, if the inferred SFR from \magphys\ is lower than SFR(\halpha$_\mathrm{obs}$) (corrected for [NII] blending), then we adopt the latter as the input to determine the fraction of stellar absorption because SFR(\halpha$_\mathrm{obs}$) can be considered a lower limit. Due to the relatively high specific-SFR (SFR/$M_\star$) of WISP emission-line galaxies, the \halpha\ absorption correction factor is negligible for most of the sample ($\sim1\%$).

Lastly, we correct for dust extinction. Due to the fact that \hbeta\ is not available for most galaxies in this sample (due to coverage and/or depth), we cannot use the Balmer decrement for dust corrections. Instead, we use the $\tau_B^l$-$\log M_\star$ relation of SDSS galaxies \citep[$z\sim0$;][]{battisti22}, where $\tau_B^l$ is the Balmer optical depth, and assume a 0.15~dex uncertainty on $\tau_B^l$ for a given $\log M_\star$. 
This relation appears to show minimal evolution with redshift \citep[e.g.,][]{shapley22, battisti22}. SFRs based on these corrections were found to be roughly consistent with SED-based SFRs in \cite{battisti22} for a similar sample.
We adopt the MW extinction curve from \citet{fitzpatrick19}, but note that most extinction curves have shapes that are very similar in the optical/near-IR regime such that this choice has a smaller impact relative to the uncertainty on $\tau_B^l$.

A comparison of our sample, both before and after dust corrections, to the galaxy MS at these redshifts from \citet{leslie20} is shown in Figure~\ref{fig:ha_sii_compare}. The \citet{leslie20} relations are based on stacked radio data from $\sim$200,000 galaxies in the COSMOS field (we use their `All' sample). For our SFRs, we adopt the conversion from \cite{kennicutt&evans12}:
\begin{equation}
\log \left[\frac{\mathrm{SFR(H}\alpha)}{M_\odot~\mathrm{yr}^{-1}}\right]=\log \left[\frac{L(\mathrm{H}\alpha)}{\mathrm{erg~s}^{-1}}\right]-41.27 \,
\end{equation}
which assumes the IMF of \citet[][this is comparable to IMF used in \magphys]{kroupa&weidner03}. All of the dust-corrected SFR values with $\log (M_\star/M_\odot) \gtrsim 9$ appear roughly consistent with the main sequence, indicating they are fully representative of `normal' star-forming galaxies at these redshifts. Even though our WISP galaxies are selected by emission line strength, this provides a sample similar to those selected by traditional broad-band continuum photometry. This reflects the fact that
all star-forming galaxies at $z \gtrsim 0.5$ have strong emission lines which WISP can detect.

At the lowest stellar masses ($\log (M_\star/M_\odot) \lesssim 9$), however, WISP galaxies tend to reside above the MS, which we attribute to the line detection threshold of WISP \citep[$\sim5\times 10^{-17}$~erg~s$^{-1}$~cm$^{-2}$;][]{atek10}. In other words, WISP can only detect emission lines with high equivalent widths at low stellar masses. We show the lower limit on SFR(\halpha$_\mathrm{obs}$) for this line sensitivity at $z=0.6$, 1.0, and 1.3 (median redshifts of our bins) in the lower-left of Figure~\ref{fig:ha_sii_compare}.

\subsubsection{\halpha+[NII], [SII], and [SIII] doublet Window ($0.30<z<0.73$)}\label{S32_window}
Next, we consider galaxies with grism coverage of \halpha+[NII], [SII], and [SIII] doublet ($0.30< z_\mathrm{grism} < 0.73$; $N$=318 with criteria 1-8). The $S_{32}\equiv \mathrm{[SIII]}\lambda\lambda$9069,9531/[SII]$\lambda\lambda$6716,6731 ratio is a proxy for the ionisation parameter of a galaxy and is relatively insensitive to the gas-phase metallicity and ISM pressure \citep{kewley19}. However, there is considerable uncertainty regarding the exact conversion of $S_{32}$ to ionisation parameter because [SII] lines are typically underestimated by photoionisation models \citep{kewley19}. For this reason, we simply report the $S_{32}$ line ratios instead of ionisation parameter.
For reference, 16.0\% of the WISP sources in our subsample are detected in both [SII] and [SIII] ($S/N>3$), 26.5\% detected only in [SII], and the rest are undetected in both lines ($S/N<3$; 57.5\%). 

\begin{figure*}
$\begin{array}{c}
\includegraphics[width=0.98\textwidth]{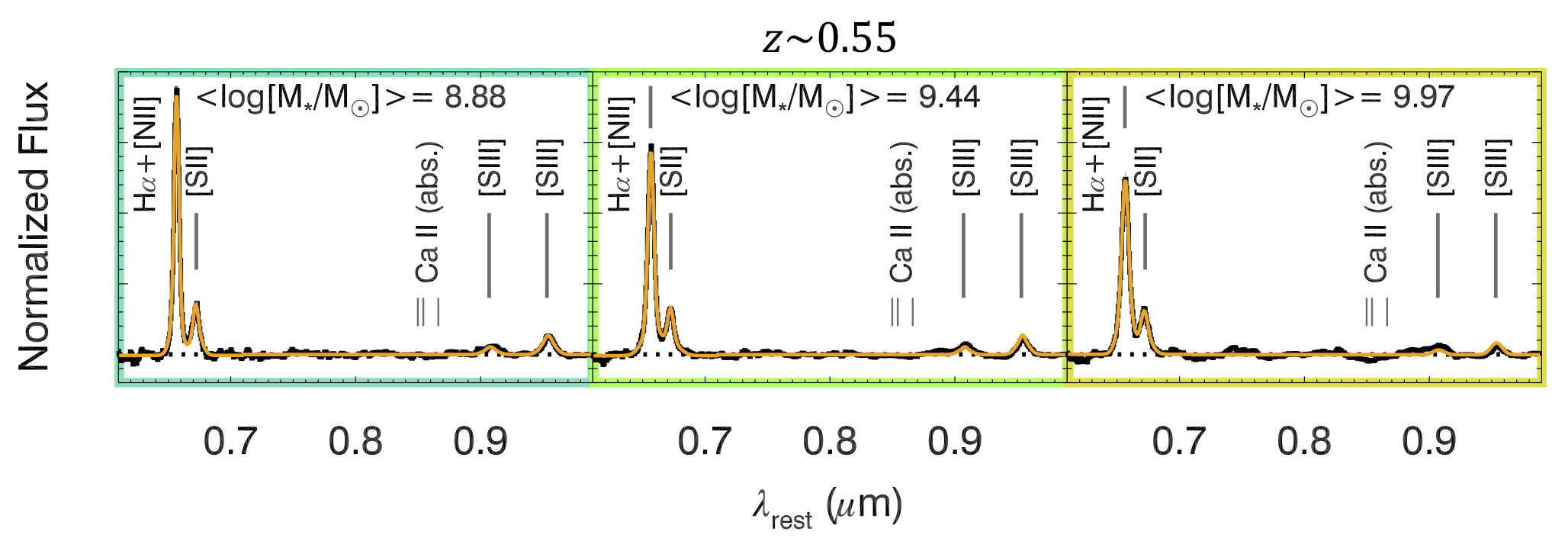}\vspace{1mm} \\
\includegraphics[width=0.6\textwidth]{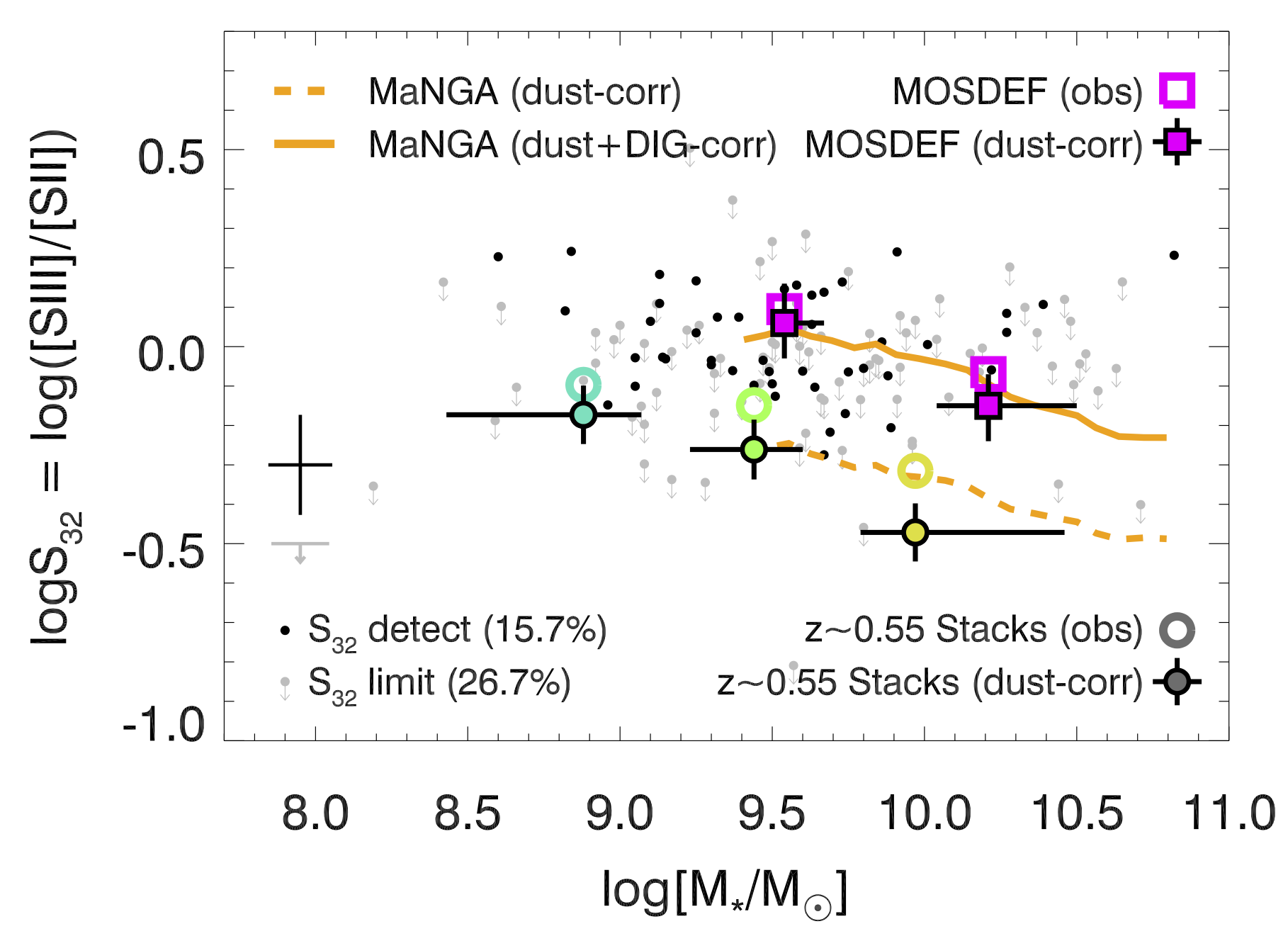} \\
\end{array}$
\vspace{-2mm}
\caption{\textit{Top:} Similar to Figure~\ref{fig:oiii_hb_compare}, but now for the \halpha+[NII], [SII], and [SIII] window ($0.30<z<0.73$) and the stacked spectra are normalised to \halpha+[NII] total flux. The CaII triplet absorption features are also indicated and appear to affect the stacked spectrum in the region near the [SIII]$\lambda$9069 line, particularly for the largest stellar mass bin. \textit{Bottom:} We compare the $S_{32}$ ratio vs stellar mass from WISP ($z\sim0.55$) to the results in \citet{sanders20} for the MaNGA ($z\sim0$) and MOSDEF ($z\sim1.5$) samples. Small black and gray points denote individual galaxies where $S_{32}$ positions are constrained (42.4\% of sample), with median errorbars given on the middle-left. Coloured open and closed symbols correspond to the stellar mass stacks before and after dust extinction correction, labeled `obs' and `dust-corr', respectively (see Section~\ref{S32_window}). The WISP stacks appear most similar to the MaNGA relation before correcting for DIG, perhaps indicating a non-negligible DIG contribution in the WISP sample. All samples show a mildly decreasing $S_{32}$ with increasing $\log M_\star$.
 \label{fig:S32_compare}}
\end{figure*}

We divide galaxies in this window into 3 equal-number bins in stellar mass over the full redshift range and the spectra of these stacks are shown in Figure~\ref{fig:S32_compare},~Top. We label the region where the CaII triplet absorption features occur, which are not accounted for in the continuum fitting (simple polynomial spline), and it appears that this may affect the stacked spectrum in the region near the [SIII]$\lambda$9069 line, particularly for the largest stellar mass bin. We enforce a fixed [SIII]$\lambda$9532/[SIII]$\lambda$9069 ratio such that the impact should be mitigated, however the WISP values should be treated with caution. Attempting to account for this is beyond the scope of our results. The average galaxy properties and emission line values for these stacks are listed in Table~\ref{tab:stack_s23}. 

We show a comparison of the $S_{32}$ ratio vs $\log M_\star$ for WISP relative to the results for the MaNGA ($z\sim0$) and MOSDEF ($z\sim1.5$) samples \citep{sanders20} in Figure~\ref{fig:S32_compare},~Bottom. We show both the observed and extinction corrected line ratios. For MaNGA and MOSDEF, these extinction corrections are based on direct measurements of Balmer decrements (\halpha/\hbeta) and assuming a MW extinction curve \citep{cardelli89}. For WISP, we do not have \hbeta\ in our redshift window and, similar to the previous section, we adopt the $\tau_B$-$\log M_\star$ relation from \citet{battisti22} (with 0.15~dex uncertainty) and the MW extinction curve from \citet{fitzpatrick19}. We again note that different extinction curves give a similar outcome because of the similarity in their shape in the near-IR (due to using a line ratio). The WISP sample shows similar values to the MaNGA sample relation prior to making corrections for diffuse-ionised gas (DIG). If the WISP galaxies ($z\sim0.55$) have a similar component of DIG as MaNGA ($z\sim0$), then this would imply they have comparable ionisation parameter values as the MaNGA and MOSDEF samples. 
We also find a decreasing value $S_{32}$ with increasing $\log M_\star$, similar to findings of \citet{sanders20} at both lower and higher redshifts, indicating a lower ionisation parameter with increasing stellar mass. We highlight that $S_{32}$ is relatively insensitive to the shape of the ionizing spectrum \citep[e.g.,][]{sanders20}, and that studies favour a redshift evolution in the hardness of the ionizing spectrum \citep[e.g.,][]{steidel16}.

\renewcommand{\arraystretch}{1.2} 
\begin{table*}
\caption{Average properties, line luminosities, and line ratios for $N$=318 stacked spectra in the \halpha+[NII], [SII], and [SIII] doublet window ($0.30<z<0.73$) that satisfy criteria (1)-(8) \label{tab:stack_s23}} 
\begin{center}
\begin{tabular}{cccccccccccc}
 \hline \\[-1em]
\multirow{2}{*}{$N$} & 
$\log M_\star$ & 
\multirow{2}{*}{$\left\langle \log M_\star \right\rangle$} &  
\halpha+[NII] &
[SII] & [SIII] &
\halpha$_\mathrm{obs}$ & \halpha$_\mathrm{corr}$ & 
[SII]$_\mathrm{corr}$ & [SIII]$_\mathrm{corr}$ & 
\multirow{2}{*}{$\dfrac{\mathrm{[SIII]}}{\mathrm{[SII]}}$} & 
\multirow{2}{*}{$\dfrac{\mathrm{[SIII]}_\mathrm{corr}}{\mathrm{[SII]}_\mathrm{corr}}$} \\
\vspace{0.04in} & range & & \multicolumn{7}{c}{($10^{41}$~erg s$^{-1}$)} & & \\
 \hline
     106 & [7.25, 9.13] &  8.88 &  1.72$\pm$0.02 &  0.44$\pm$0.01 &  0.35$\pm$0.01 &  1.57$\pm$0.04 &  2.49$\pm$ 0.82 &  0.69$\pm$0.22 &  0.47$\pm$0.09 &  0.80$\pm$0.04 &  0.67$\pm$0.11 \\
     106 & [9.13, 9.66] &  9.44 &  3.19$\pm$0.05 &  0.84$\pm$0.04 &  0.60$\pm$0.02 &  2.67$\pm$0.11 &  5.29
$\pm$ 1.75 &  1.63$\pm$0.52 &  0.89$\pm$0.17 &  0.71$\pm$0.04 &  0.55$\pm$0.10 \\
     106 & [9.66,10.83] &  9.97 &  4.31$\pm$0.05 &  1.15$\pm$0.03 &  0.55$\pm$0.02 &  3.06$\pm$0.17 &  7.93$\pm$ 2.63 &  2.87$\pm$0.91 &  0.97$\pm$0.19 &  0.48$\pm$0.02 &  0.34$\pm$0.06 \\
\hline

 \end{tabular}
\end{center}
\textbf{Notes.} [SII] is the sum of [SII]$\lambda\lambda$6716, 6731. [SIII] is the sum of [SIII]$\lambda\lambda$9069, 9532. \halpha$_\mathrm{obs}$ and \halpha$_\mathrm{corr}$ have same meaning as in Table~\ref{tab:stack_ha_sii}. [SII]$_\mathrm{corr}$ and [SIII]$_\mathrm{corr}$ are values after correcting for dust extinction (see Section~\ref{ha_sii_window}).
\end{table*}
  
\subsubsection{[OII] to [SII] Window ($1.27<z<1.45$)}\label{full_window}
Finally, we consider galaxies with grism coverage from [OII] all the way to [SII], which occurs for only a narrow redshift window ($1.27< z_\mathrm{grism} < 1.45$; $N$=203 with criteria 1-8). This window is unique in that it provides coverage across most strong optical emission lines simultaneously, which is beneficial for characterising various properties of the ISM, including: $O_{32}\equiv \mathrm{[OIII]}\lambda\lambda$4959,5007/[OII]$\lambda\lambda$3726,3729, a proxy for the ionisation parameter \citep{kewley19}; $R_{23} \equiv (\mathrm{[OII]}\lambda\lambda$3726, 3729 + [OIII]$\lambda\lambda$4959, 5007)/\hbeta, a proxy for gas-phase metallicity \citep[e.g.,][]{curti17}; and \halpha/\hbeta, a proxy for dust attenuation \citep{calzetti01}. We note that unlike $S_{32}$, the relation between $O_{32}$ and ionisation parameter is more sensitive to gas-phase metallicity and ISM pressure \citep{kewley19}. 

\begin{figure*}
$\begin{array}{cc}
\multicolumn{2}{c}{\includegraphics[width=0.98\textwidth]{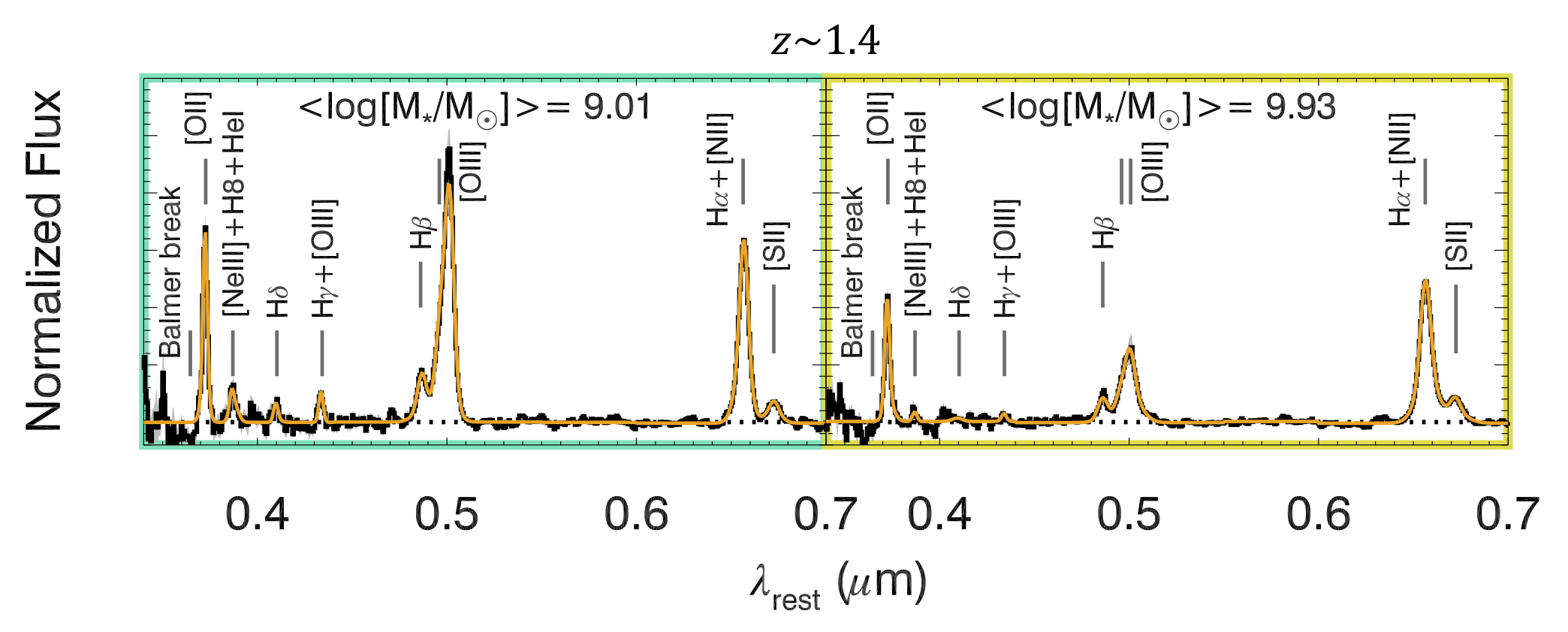}}\vspace{1mm} \\
\includegraphics[width=0.49\textwidth]{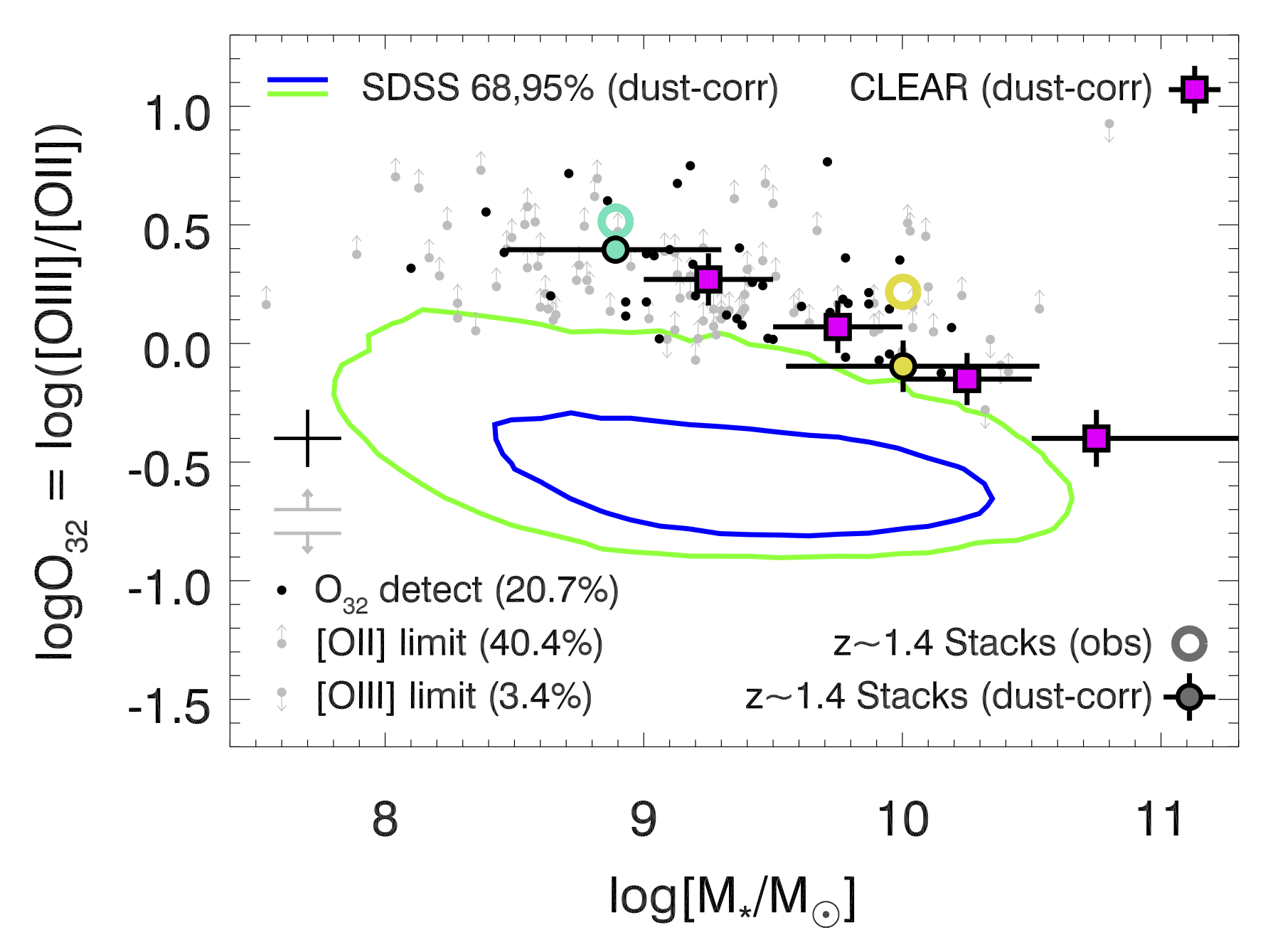} &
\includegraphics[width=0.49\textwidth]{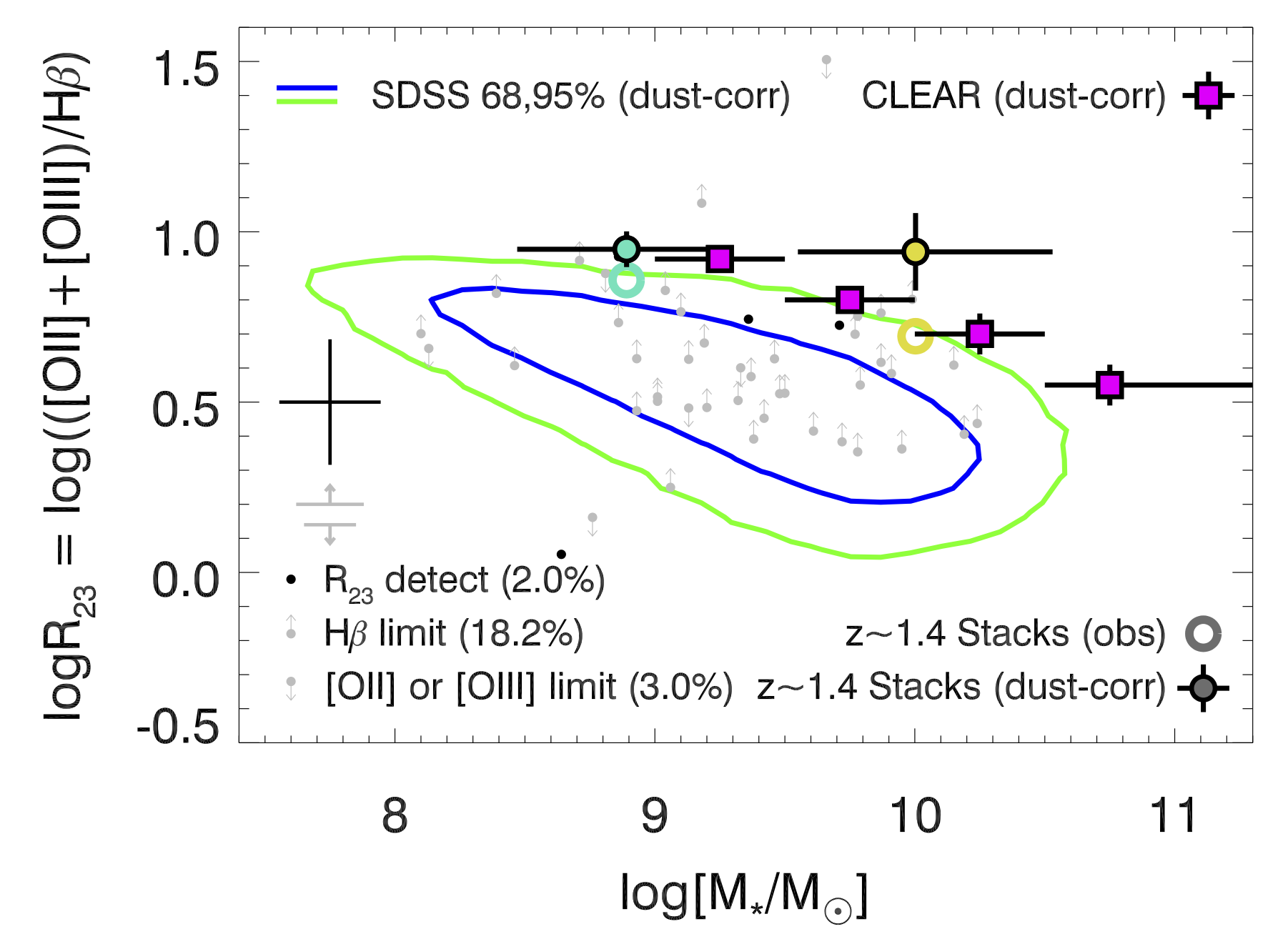}\\
\end{array}$
\vspace{-2mm}
\vspace{-2mm}
\caption{\textit{Top:} Similar to Figure~\ref{fig:oiii_hb_compare}, but now for the window covering [OII] to [SII] and the stacked spectra are normalised to \halpha+[NII] total flux. The numerous optical emission lines in this window allows several ISM diagnostics to be simultaneously available. The region around the Balmer break (0.3645$\mu$m) shows a discontinuity, which may result from the simplistic treatment of the stellar continuum (see Section~\ref{linefits}). \textit{Bottom:} We compare the $O_{32}$ and $R_{23}$ ratios vs stellar mass from WISP ($z\sim1.4$) to values from CLEAR \citep[$1.1<z<2.3$;][]{papovich22} and also show contours for SDSS  ($z<0.2$; coloured lines). 
Coloured open and closed symbols correspond to the stellar mass stacks before and after dust extinction correction, respectively, based on the measured Balmer decrement (see Section~\ref{full_window}). The WISP stacks are in rough agreement with CLEAR, lying above the SDSS sample, and showing a decreasing $O_{32}$ and $R_{23}$ with increasing $\log M_\star$. Larger $O_{32}$ ratios indicate of higher ionisation potential and/or harder ionizing spectrum. Larger $R_{23}$ indicate lower metallicities (for upper-branch, $12+\log(\mathrm{O}/\mathrm{H})>8.1$), however the $R_{23}$-metallicity relation has a turn-over at $\log R_{23}\sim1$ \citep[e.g.,][]{curti17}.
\label{fig:full_window_compare}}
\end{figure*}

We divide galaxies in this window into two equal-number stellar mass bins. The spectra of these stellar mass stacks are shown in Figure~\ref{fig:full_window_compare}.  The average galaxy properties and emission line values for these stacks are listed in Table~\ref{tab:stack_full}. Unlike previous stacks, we can perform dust corrections based on the measured Balmer decrement. This is based on the methods detailed in \citet{battisti22}, which require corrections for \halpha+[NII] blending and Balmer absorption based on empirical relations from \citet{faisst18} and \citet{kashino&inoue19}, respectively. We note that the average dust attenuation based on SED modelling is systematically lower than that inferred from the Balmer decrement, with $\log \mathrm{SFR(SED)}$ being $\sim$0.2~dex lower than $\log \mathrm{SFR}$(\halpha$_\mathrm{corr}$). This is consistent with the findings in \cite{battisti22} and is expected. In the absence of IR data, the age-dust degeneracy (older populations can produce redder colours) will result in older (lower SFR) templates in \magphys\ being able to reproduce the data.

We show a comparison of the $O_{32}$ and $R_{23}$ line ratios vs $\log M_\star$ for WISP relative to results from SDSS ($z\sim0$) and the CLEAR survey \citep[$1.1<z<2.3$;][]{papovich22}, which was an \hst\ grism program in GOODS-N and GOODS-S \citep{simons23}. We show both the observed and extinction corrected line ratios for WISP (only extinction corrected for CLEAR). For CLEAR, the extinction corrections are based on the attenuation measurements from the SED fitting that assume the \cite{calzetti00} attenuation curve and assuming the nebular and continuum reddening are the same \citep[see][for details]{papovich22}. For WISP, we use Balmer decrements and the MW extinction curve from \citet{fitzpatrick19}.  The WISP sample has similar values and behaviour to the CLEAR stacks for both line ratios, which are offset above local galaxies based on SDSS. The SDSS contours are based on $\sim$156000 galaxies at $z<0.2$ \citep[for selection criteria, see][]{battisti22}, noting that we use the stellar mass enclosed in the SDSS fiber and not the `total' stellar mass.
For $O_{32}$, the decreasing values in line ratios with increasing $\log M_\star$ suggests a lower ionisation parameter and/or softer ionizing spectrum with increasing stellar mass. For $R_{23}$, the decreasing values in line ratios with increasing $\log M_\star$ reflects an increase in metallicity with increasing stellar mass. However, we note that our WISP $R_{23}$ values are close to the $\log R_{23}\sim1$ turn-over in the $R_{23}$-metallicity relation\citep[e.g.,][]{curti17}, which introduces an ambiguity in relating $R_{23}$ to a metallicity. The CLEAR sample is representative of galaxies on the star-forming galaxy main-sequence \citep[see Figure~2 of][]{papovich22} and our agreement with their trends further supports the argument that the WISP sample also reflects main-sequence galaxies. For a more detailed analysis of the metallicity of WISP galaxies, we refer readers to \citet{henry21}.

\renewcommand{\arraystretch}{1.2} 
\begin{table*}
\caption{Average properties, emission line luminosities, and line ratios for stacked spectra in the [OII] to [SII] window ($1.27<z<1.45$) \label{tab:stack_full}} 
\begin{center}
\begin{tabular}{cccccccccccc}
 \hline \\[-1em]
\multirow{2}{*}{$N$} & 
$\log M_\star$ & 
\multirow{2}{*}{$\left\langle \log M_\star \right\rangle$} &  
[OII] & \hbeta & [OIII] &
\halpha+[NII] &
[SII] &
\halpha$_\mathrm{obs}$ & 
\multirow{2}{*}{$\tau_B^l$} & 
\multirow{2}{*}{$\log\mathrm{O}_{32}$} & 
\multirow{2}{*}{$\log\mathrm{R}_{23}$} \\ 
\vspace{0.04in} & range & & \multicolumn{6}{c}{($10^{41}$~erg s$^{-1}$)} \\
 \hline
     102 & [ 7.54, 9.39] &  9.01 &  5.03$\pm$0.24 &  2.46$\pm$0.14 & 16.41$\pm$0.31 & 10.11$\pm$0.09 &  1.45$\pm$0.09 &  9.28$\pm$0.26 &  0.27$\pm$0.07 &  0.40$\pm$0.06 &  0.95$\pm$0.05 \\ 
     101 & [ 9.40,11.26] &  9.93 &  6.56$\pm$0.35 &  2.39$\pm$0.21 & 10.83$\pm$0.32 & 18.29$\pm$0.16 &  3.29$\pm$0.11 & 14.50$\pm$1.72 &  0.71$\pm$0.19 & -0.10$\pm$0.11 &  0.94$\pm$0.11 \\ 
\hline
 \end{tabular}
\end{center}
\textbf{Notes.} \hbeta\ shown is not corrected for stellar absorption. [OIII] is the sum of [OIII]$\lambda\lambda$4959, 5007. [SII] is the sum of [SII]$\lambda\lambda$6716, 6731. The Balmer optical depth, $\tau_B^l=\ln((F(\mathrm{H}\alpha)/F(\mathrm{H}\beta)/2.86)$ is based on absorption-corrected lines (see Section~\ref{full_window}). Values listed for $\mathrm{O}_{32}$ and $\mathrm{R}_{23}$ are after extinction correction.
\end{table*}
     
\section{Conclusion}\label{conclusion}
Slitless spectroscopic surveys are an efficient method to perform large spectroscopic surveys of galaxies across a wide range of cosmic time. 
We present the public data release of reduced ancillary photometric images obtained for the WFC3 Infrared Spectroscopic Parallel (WISP) survey, a large pure-parallel \hst\ program, as well as a consistent photometric catalogue containing $\sim$230,000 sources. This catalogue is based on the \se\ and \texttt{TPHOT} codes, the latter of which uses the high spatial resolution \hst\ data to perform forced-photometry on the low spatial resolution ground-based and \spitzer\ data. 
We also present the WISP emission line catalogue containing $\sim$8,000 sources. This catalogue is based on a novel combination of an automated line detection algorithm and visual inspection.
These data can be used to study a broad range of topics in galaxy evolution over $\sim$60\% of cosmic time ($0.3\lesssim z\lesssim 3$) and will serve as a useful reference sample for future slitless surveys with \jwst, \euclid, and \rst.

We combine the WISP photometric and spectroscopic catalogues to examine the properties of WISP galaxies using stacked spectra in bins of stellar mass over four grism windows (redshift regimes) where specific emission line ratios are available to study their ISM properties (Section~\ref{stack_results}):
\begin{itemize}
    \item For \hbeta\ and [OIII] coverage ($0.74< z_\mathrm{grism} < 2.31$), we bin 1040 galaxies and examine their position on the Mass-Excitation diagram ([OIII]/\hbeta\ vs. $M_\star$; \citealt{juneau14}). We find that our stacks at $\log (M_\star /M_\odot)<10.2$ lie in the star-forming region of the diagram and $\log (M_\star /M_\odot)\gtrsim 10.2$ lie on/above the AGN region of the diagram. This suggests that there may be a non-negligible fraction of sources with AGN at higher masses and is qualitatively consistent with findings from other high-$z$ spectroscopic studies \citep[e.g.,][]{forsterSchreiber19}.
    
    \item For \halpha+[NII] and [SII] coverage ($0.30< z_\mathrm{grism} < 1.45$), we bin 1191 galaxies and examine their position on the galaxy main sequence \citep[SFR vs. $M_\star$;][]{leslie20}. Bins with $\log (M_\star/M_\odot) \gtrsim 9$ appear roughly consistent with the ``star-forming main sequence", indicating they are representative of `normal' star-forming galaxies. At $\log (M_\star/M_\odot) \lesssim 9$,  our bins reside above the main sequence.
    
    \item For \halpha+[NII], [SII], and [SIII] coverage ($0.30< z_\mathrm{grism} < 0.73$), we bin 318 galaxies to examine the [SIII]/[SII] ratio (ionisation parameter proxy). Our stacks are consistent with ratios found in MaNGA ($z\sim0$) and MOSDEF \citep[$z\sim1.5$;][]{sanders20} if WISP galaxies ($z\sim0.5$) contain a similar [SII] contribution from diffuse-ionised gas as MaNGA galaxies. 
    
    \item For [OII] to [SII] coverage ($1.27< z_\mathrm{grism} < 1.45$), we bin 203 galaxies to examine the [OIII]/[OII] ratio (ionisation parameter/spectral slope proxy) and ([OII]+[OIII])/\hbeta\ ratio (metallicity proxy). Our stacks are consistent with line ratios found in CLEAR \citep[$1.1<z<2.3$;][]{papovich22}. In particular, the relative strength of [OIII] emission is substantially higher than in local star-forming galaxies observed by SDSS.
\end{itemize}
These results indicate that the majority of WISP galaxies are representative of typical main-sequence galaxies. 

Finally, we note that several current large-area photometric surveys, such as the Dark Energy Survey \citep[DES, $grizY$-bands;][]{DES16}, the DECam Legacy Survey of the SDSS Equatorial Sky \citep[DECaLS, $grz$-bands;][]{dey19}, the Beijing–Arizona Sky Survey \citep[BASS, $gr$-bands;][]{zou17}, and the Mayall z-band Legacy Survey \citep[MzLS, $z$-band;][]{dey19}, provide shallow coverage ($\sim$22-24~AB mag) for some of the WISP fields that do not have existing coverage in this data release. In the future, the Legacy Survey of Space and Time \citep[LSST, $ugrizy$-bands;][]{ivezic19} with the Rubin Observatory will provide more uniform and deeper coverage (final depth $\sim$25-27~AB mag) that will supersede the depth of most existing optical data for WISP fields in the Southern Hemisphere ($\mathrm{Dec}<0$). Incorporating these datasets may be the subject of a future data release. 



\section*{Acknowledgments}
The authors thank the anonymous referee, whose suggestions helped to clarify and improve the content of this work. 
AJ Battisti thanks the staff at the WIYN and Magellan facilities for their assistance with technical aspects of the observations. 
AJ Battisti is also thankful for attending ASTRO 3D writing retreats that provided a helpful environment to complete portions of this manuscript. 
Parts of this research were supported by the Australian Research Council Centre of Excellence for All Sky Astrophysics in 3 Dimensions (ASTRO 3D), through project number CE170100013. 
We acknowledge the invaluable labor of the maintenance and clerical staff at our institutions, whose contributions make our scientific discoveries a reality. 
This research was conducted on Ngunnawal Indigenous land, as well as within the traditional homelands of the Dakota people. 
This work is sponsored by the National Key R\&D Program of China for grant No.\ 2022YFA1605300, the National Nature Science Foundation of China (NSFC) grants No. \ 12273051 and 11933003. HA is supported by CNES (Centre National d'Etudes Spatiales). AJ Bunker acknowledges funding from the "FirstGalaxies" Advanced Grant from the European Research Council (ERC) under the European Union’s Horizon 2020 research and innovation programme (Grant agreement No. 789056).
Support for WISP (HST programs GO-11696, 12283, 12568, 12902, 13517, 13352, and 14178) was provided by NASA through grants from the Space Telescope Science Institute, which is operated by the Association of Universities for Research in Astronomy, Inc., under NASA contract NAS5-26555.
This work uses data from WIYN.
The WIYN Observatory is a joint facility of the NSF's National Optical-Infrared Astronomy Research Laboratory, Indiana University, the University of Wisconsin-Madison, Pennsylvania State University, the University of Missouri, the University of California-Irvine, and Purdue University.
This paper includes data gathered with the 6.5 meter Magellan Telescopes located at Las Campanas Observatory, Chile.
This work uses data from Palomar.
The Palomar Observatory is a joint facility of the Caltech/Caltech Optical Observatories, Jet Propulsion Laboratory, Yale University, and National Astronomical Observatories of China.
This work uses data from SDSS.
Funding for SDSS-III has been provided by the Alfred P. Sloan Foundation, the Participating Institutions, the National Science Foundation, and the U.S. Department of Energy Office of Science. The SDSS-III web site is http://www.sdss3.org/.
SDSS-III is managed by the Astrophysical Research Consortium for the Participating Institutions of the SDSS-III Collaboration including the University of Arizona, the Brazilian Participation Group, Brookhaven National Laboratory, Carnegie Mellon University, University of Florida, the French Participation Group, the German Participation Group, Harvard University, the Instituto de Astrofisica de Canarias, the Michigan State/Notre Dame/JINA Participation Group, Johns Hopkins University, Lawrence Berkeley National Laboratory, Max Planck Institute for Astrophysics, Max Planck Institute for Extraterrestrial Physics, New Mexico State University, New York University, Ohio State University, Pennsylvania State University, University of Portsmouth, Princeton University, the Spanish Participation Group, University of Tokyo, University of Utah, Vanderbilt University, University of Virginia, University of Washington, and Yale University.
This work uses data from PAN-STARRS. The Pan-STARRS1 Surveys (PS1) and the PS1 public science archive have been made possible through contributions by the Institute for Astronomy, the University of Hawaii, the Pan-STARRS Project Office, the Max-Planck Society and its participating institutes, the Max Planck Institute for Astronomy, Heidelberg and the Max Planck Institute for Extraterrestrial Physics, Garching, The Johns Hopkins University, Durham University, the University of Edinburgh, the Queen's University Belfast, the Harvard-Smithsonian Center for Astrophysics, the Las Cumbres Observatory Global Telescope Network Incorporated, the National Central University of Taiwan, the Space Telescope Science Institute, the National Aeronautics and Space Administration under Grant No. NNX08AR22G issued through the Planetary Science Division of the NASA Science Mission Directorate, the National Science Foundation Grant No. AST–1238877, the University of Maryland, Eotvos Lorand University (ELTE), the Los Alamos National Laboratory, and the Gordon and Betty Moore Foundation.

\section*{Data Availability}
All of the WISP photometric and spectroscopic data used in this paper are publicly available through data releases on the WISP website\footnote{\url{https://archive.stsci.edu/prepds/wisp/}} on MAST. This release adds the following data products to the WISP website: (1) fully reduced $5\arcmin\times5\arcmin$ cutouts (centered on WISP field) of ground-based observations, (2) fully reduced \spitzer\ images, (3) the photometric catalogue (FITS binary table; described in Table~\ref{tab:FITS_photo_catalog}), (4) the emission line catalogue (FITS binary table; described in Table~\ref{tab:FITS_line_catalog}), (5) the full versions of Table~\ref{tab:line_depths} (grism depths for each field), Table~\ref{tab:wfc3ir_depth_completeness} (WFC/IR photometric depths/completeness for each field), Table~\ref{tab:phot_depth_catalog} (photometric depths for each field), and Table~\ref{tab:magphys_output} (MAGPHYS properties for subsample). 
Other data products can be made available upon reasonable request to the first author.

\bibliographystyle{mnras}
\bibliography{AJB_bib}

\appendix
\section{\texttt{TPHOT} Input and Configuration Details}\label{appendix:tphot_config}

In order for \texttt{TPHOT} to obtain reliable measurements, the geometric centers of the sources detected in different bands must be perfectly aligned. 
To obtain this result, the coordinates of the sources originally extracted from the low-resolution 
images are refined  by comparing the positions of the sources in the images at different wavelengths. This coordinate recentering is performed using an iterative procedure. Initially, 
source positions in the low-resolution data are determined by running \se\ on the original images. The most likely \hst\ counterparts are identified using an initial search radius of 2\arcsec. At this stage, each source 
is associated to only one potential counterpart. Then, an average shift correction for the 
coordinates is computed by comparing the (RA,Dec) positions of the counterparts in the two bands and the corresponding new set of world coordinate system (WCS) is applied to the low-resolution 
images. After this initial iteration, the entire process is repeated using, this time, a shorter searching radius (0.75\arcsec). Therefore, the WCS correction is refined by performed the second iteration using only sources with a more secure counterpart identification. At the end of the process described, the coordinates in the low-resolution 
images are precisely recentered to the \hst\ reference-frame.


\texttt{TPHOT} also requires input images with identical (or integer multiple) pixel-scale and identical pixel orientation. We use the \texttt{SWarp} software to obtain low-resolution 
images consistent with the \hst\ reference images. 
We note that these \texttt{SWarp} intermediate products (i.e., images with identical pixel scales and orientations) are not provided in the current data release.
\texttt{TPHOT} also requires both a catalogue of sources extracted in the high-resolution image (\hst), and a corresponding segmentation map. In particular, the IDs reported in the input catalogue must be identical to the pixel values of the corresponding sources in the segmentation map. Our catalogues were created by merging catalogues of sources detected in the $J$ and/or in the $H$ band (plus a catalogue obtained from a sum of the $J$+$H$ images). As a consequence, the IDs in our catalogues had to be reassigned (and ordered on a brightness basis). 
Finally, not all the sources extracted (and included in the segmentation maps) were included in our catalogues. For example, we removed all the sources located in the image borders, characterised by bad photometric measurements.

Due to the fact that not all sources uniquely correspond to a counterpart across the different \hst\ segmentation maps, we computed an appropriate unique segmentation image, 
from the original segmentation maps obtained by \se. In the new segmentation map, every source of the catalogue is uniquely related with a corresponding source (with identical ID) in the map. Moreover, all the sources removed from the catalogues are also removed from the final segmentation map.

In addition to the IDs, the \texttt{TPHOT} input catalogue must include, for each source, its (x,y) position in the \hst\ reference image, the values x$_{\mathrm{min}}$, x$_{\mathrm{max}}$, y$_{\mathrm{min}}$ and y$_{\mathrm{max}}$ defining the borders of the source in the segmentation map, a local background value (we set this value to 0 as the background was already subtracted from the images before running \texttt{TPHOT}), and the reference flux for each source in the high resolution band. 

One of the most important inputs required by \texttt{TPHOT} is the image of the kernel required to perform the inverse of the following convolution operation:
\begin{equation}
\mathrm{PSF(LRI)} = \mathrm{Kernel} * \mathrm{PSF(HRI)},
\end{equation}
where LRI and HRI stand for low- and high-resolution image, respectively.
We obtain the convolution kernel image by deconvolving  the low-resolution 
PSF using the \hst\ PSF as a deconvolution kernel. 
For this, we adopt the Richardson-Lucy approach \citep{richardson72, lucy74}. The average low-resolution 
and \hst\ PSFs are obtained averaging the PSFs of a selection of point-like, non-saturated sources detected in each image. In accordance with this choice, we set the \texttt{TPHOT} configuration parameter \texttt{usereal}= \texttt{True} (and \texttt{usemodels} and \texttt{useunresolved}=\texttt{False}). 

The \texttt{TPHOT} algorithm is organised in different ``stages'', each of which performs a specific task. The list and the order of the various stages in each iteration can be set by the user.
As described in the user manual, the best results are obtained when the various stages are run in two separate iterations (``passes''). We proceed in the default manner, setting the keyword \texttt{order} to \texttt{standard} and to \texttt{standard2} in the two iterations, respectively.

For each source, \texttt{TPHOT} computes (x,y) shifts during the first pass (``dance'' stage).
Using these corrections, in the second pass, local kernels are registered to each single source improving the accuracy of the outputs. This can be obtained by setting the keyword \texttt{multikernel}=\texttt{true} for the second iteration. We set the size of the region in which the PSF shift is computed (keyword \texttt{dzonesize}) to 100 pixels and the maximum shift allowed (keyword \texttt{maxshift}) to 20 pixels. We reduce these two parameters to 50 and 1, respectively, for the second iteration. The shifts computed are smoothed over 100 and 50 neighbors in the two passes respectively (keyword \texttt{nneighinterp}). Among the three different methods available for solving the linear systems during the fitting stage, (i.e., LU, Cholesky and the iterative biconjugate gradient), we selected the matrix inversion method of LU (default). Additionally, we select the option to clip out large negative fluxes before obtaining the final fit. We set all the remaining \texttt{TPHOT} parameters to the default options.

\section{WISP Photometric Catalogue Flag Description}\label{appendix:phot_flags}
There are two entries for quality flags in the photometric catalogue, each consisting of a sum of bit flags (i.e., sum of powers of 2). One is for the default \se\ `internal' flags\footnote{\url{https://astromatic.github.io/sextractor/Flagging.html}} 
(\texttt{FLAG\_[NIRFILTER]}) and the other is for the default \texttt{TPHOT} flags \citep[][\texttt{TPHOT\_FLAG\_[FILTER]}]{merlin15}. These are described below.

\texttt{FLAG\_[NIRFILTER]} contains a sum of 8 flag bits (i.e., sum of powers of 2): 
\begin{itemize}
\item 1 = photometry likely to be biased by neighboring sources or bad pixels
\item 2 = object has been deblended
\item 4 = at least one object pixel is saturated
\item 8 = object is close to image boundary
\item 16 = at least one photometric aperture is incomplete or corrupted
\item 32 = the isophotal footprint is incomplete or corrupted
\item 64 = a memory overflow occurred during deblending
\item 128 = a memory overflow occurred during extraction.
\end{itemize}
For example, a saturated detection close to an image boundary will have \texttt{FLAG\_[NIRFILTER]} = 4+8 = 12. \\

\texttt{TPHOT\_FLAG\_[FILTER]} contains a sum of 3 flag bits: 
\begin{itemize}
\item 1 = the prior has saturated or negative flux
\item 2 = the prior is blended
\item 4 = the source is at the border of the image.
\end{itemize}

\section{WISP Continuous Wavelet
Transform Algorithm}\label{appendix:cwt}

\begin{figure}
\begin{center}
\includegraphics[width=0.4\textwidth]{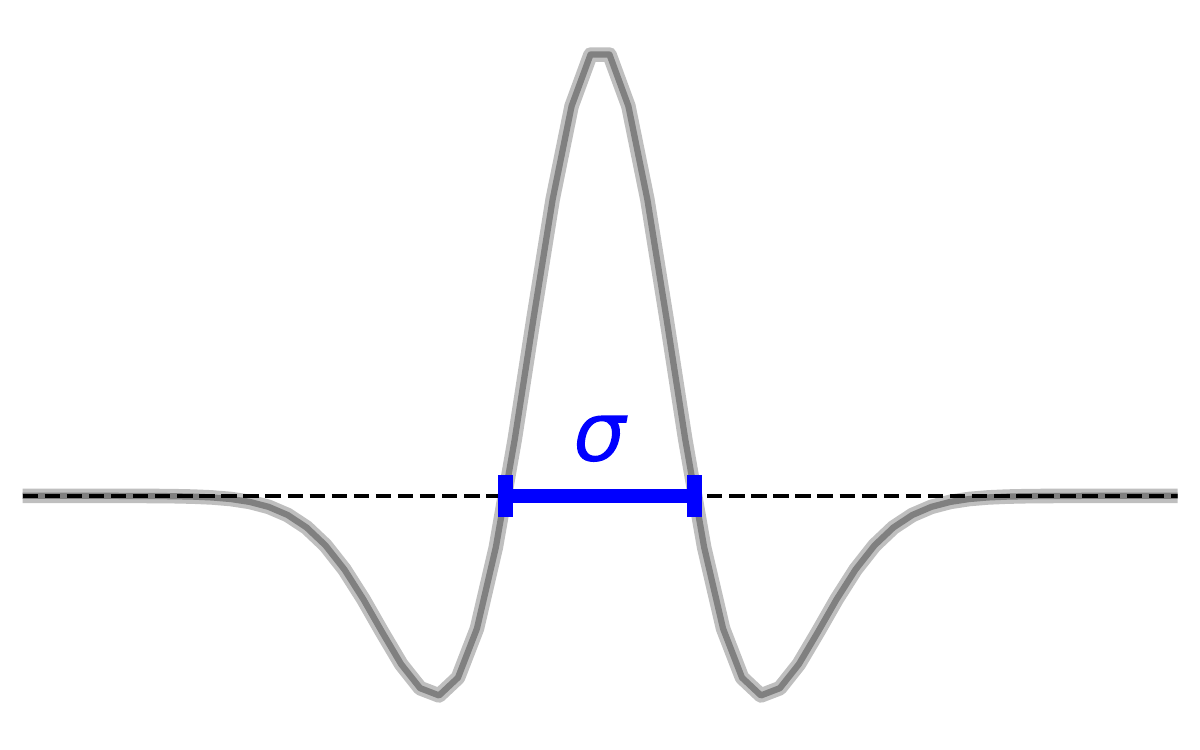}
\end{center}
\vspace{-5mm}
\caption{The Ricker wavelet used for emission line detection in the continuous wavelet transform. It is proportional to the second derivative of a Gaussian function and is defined by a width parameter, $\sigma$.
\label{fig:ricker}}
\end{figure}

Starting from the one-dimensional spectra extracted and calibrated by \texttt{aXe}, we perform a continuous wavelet transform on the spectrum using a Ricker wavelet, which is proportional to the second derivative of a Gaussian function. The Ricker wavelet models the function
\begin{equation}
f =  \frac{2}{\sqrt{3\sigma}\pi^{\frac{1}{4}}} (1-\frac{x^2}{\sigma^2}) e^{-\frac{x^2}{2 \sigma^2}} \,,
\end{equation}
and is illustrated in Figure~\ref{fig:ricker}.

The transform is performed using $N_{\sigma}=10$ kernels or wavelets of varying widths, with the minimum width set to 1.5 pixels, corresponding to $\sim$36\AA\ ($\sim$70\AA) in G102 (G141), and the maximum width equal to twice the full width at half maximum (FWHM) estimate for the object. Spectra obtained in slitless mode are essentially images of the source at each wavelength, and so the width of emission lines will be correlated with the source size. We approximate the FWHM as twice the dispersion measured along the semi-major axis (\texttt{A\_IMAGE} reported by \se), and use this FWHM$_{\mathrm{est}}$ to define upper bounds for both the wavelet widths and the FWHM of the emission line profile fits described in Section~\ref{linefits}.

\begin{figure}
\includegraphics[width=0.48\textwidth]{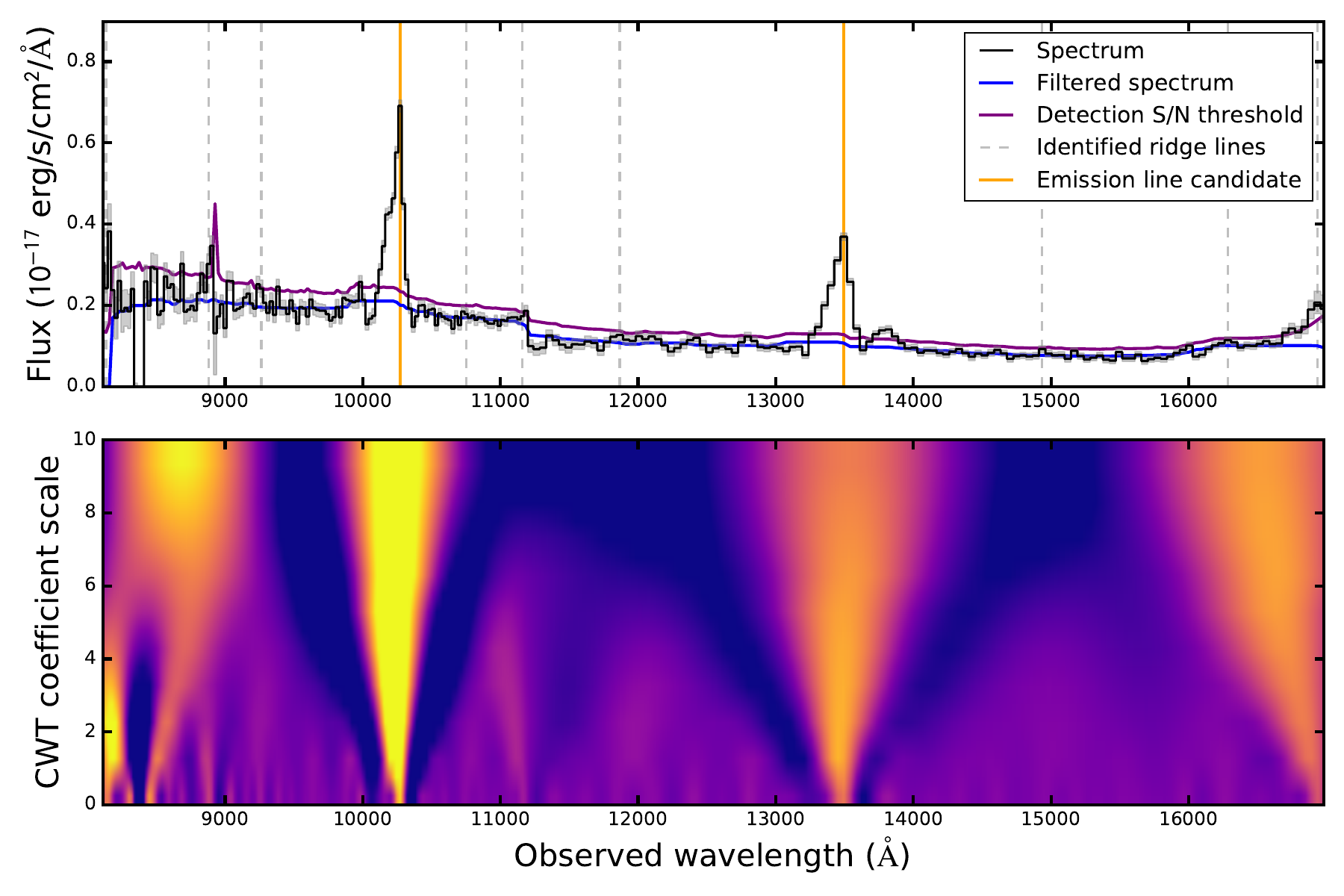}
\vspace{-6mm}
\caption{An example of the emission line detection algorithm. \textit{Top panel:} The input spectrum is plotted in black. The filtered spectrum (blue curve) is used as an estimate of the continuum. The purple curve is the detection threshold, defined as a minimum S/N$\geq$2.31 above the continuum.
\textit{Bottom panel}: The two-dimensional CWT coefficient matrix represents the correlation of the spectrum and wavelet at each scale and wavelength. Larger coefficients (yellow colours) identify regions of higher correlation. Emission lines are correlated with the wavelet at many scales, creating connected ridge lines in the CWT matrix. The dashed grey lines in the top panel show the positions of detected ridge lines. The yellow solid lines show the ridges that passed our additional selection criteria to be identified as true emission line candidates.
\label{fig:cwt}}
\end{figure}

The CWT transform compares the wavelet with the 1D spectrum, shifting the wavelet to cover all wavelengths and scaling or stretching to cover all input widths. The resulting array of CWT coefficients is a matrix of dimension $N_{\sigma} \times N_{\lambda}$ representing the correlation of the spectrum and wavelet at each scale and wavelength. Large CWT coefficients indicate regions of the spectrum with a strong correlation with the wavelet, and so the largest coefficients will occur where both the position and width of the wavelet best match a spectral feature. An example coefficient matrix is displayed in the bottom panel of Figure~\ref{fig:cwt}.

As can be seen in the bottom panel of Figure~\ref{fig:cwt}, strong emission peaks in the spectrum are strongly correlated with the Ricker wavelet at many scales. The resulting peaks in the two-dimensional CWT coefficient matrix extend to multiple scales and can be visualised as mountain ridges. Emission line features in the spectrum can now be identified using the ridges in the CWT coefficient matrix. For this step, we use the scipy program \texttt{find\_peaks\_cwt}, which is an implementation of the procedure presented in \cite{du06}. We briefly summarise the process here and refer the reader to \cite{du06} for a full description of the algorithm.

The first step in the peak identification process involves identifying ridge lines in the coefficient matrix. At each scale, a local maximum is matched with the nearest maximum calculated for the adjacent scale. The two local maxima constitute a ridge provided they are within a set distance, dist$_{\mathrm{max}}$, which we define as one pixel larger than the width of the wavelet at the given scale. A ridge line continues through the matrix as long as the local maximum at each scale is within $d_{\mathrm{max}}$ of the previously identified maximum. The ridge line is terminated if a local maximum is not matched at more than $g_{\mathrm{thresh}}$ consecutive kernels. We conservatively use $g_{\mathrm{thresh}}=1$~kernel.

\begin{table*}
\caption{Emission Line Detection Parameters \label{tab:cwt}}
\begin{center}
\begin{tabular}{lll}
 \hline \hline
Parameter & Value & Description \\
\multicolumn{3}{c}{\textbf{\underline{Continuous Wavelet Transform}}} \\[5pt]
$N_{\sigma}$         & 10 & number of CWT kernels (widths) used in transform \\[2pt]
$\sigma_{\mathrm{min}}$   & 1.5 [pixels] & minimum CWT kernel width \\[2pt]
$\sigma_{\mathrm{max}}$   & 2.0 & maximum CWT kernel width, fraction of FWHM$_{\mathrm{est}}$ \\[5pt]
\hline \\[-5pt]
\multicolumn{3}{c}{\textbf{\underline{Ridge and peak definitions}}} \\[5pt]
$d_{\mathrm{max}}$  & $\sigma +1$ [pixels] & maximum acceptable separation distance between local \\
 & & \hspace{5pt} maxima at each scale on the same ridge line \\[2pt]
$g_{\mathrm{max}}$   & 1 [kernels] & maximum acceptable gap between connected ridges \\[2pt]
$l_{\mathrm{min}}$ & 3 [kernels] & minimum acceptable \# of kernels in which peak is significant \\[2pt]
S/R$_{\mathrm{min,CWT}}$    & 1 & minimum acceptable ridge S/N at each scale \\[5pt]
\hline \\[-5pt]
\multicolumn{3}{c}{\textbf{\underline{Requirements for real lines}}} \\[5pt]
$C_{\mathrm{min}}$  & 0.15  & minimum acceptable contrast between peak and continuum fluxes \\[2pt]
$d_{\mathrm{edge}}$ & 5 [pixels] & minimum acceptable distance to edge of spectrum  \\[2pt]
$n_{\mathrm{min}}$ & 3 [pixels] & minimum number of pixels above the noise threshold \\[2pt] 
S/N$_{\mathrm{min,pix}}$ & 2.31 & minimum S/N per pixel \\[2pt] 
\hline
\end{tabular}
\end{center}
\end{table*}

The second step is to identify peaks from the detected ridge lines. We only consider ridge lines that cover at least $l_{\mathrm{min}}=3$ scales and have a ridge signal-to-noise of S/N$_{\mathrm{min,CWT}}\geq1$. The ridge S/N is estimated at each scale, and is taken to be the maximum CWT coefficient value on the ridge divided by the overall noise floor. The noise floor is calculated as the 10th percentile of the coefficients around the ridge line at the smallest scale ($\sigma_{\mathrm{min}}$). These identified peaks are represented as dashed grey lines in the top panel of Figure~\ref{fig:cwt}.

Following the peak finder, we additionally screen the selected peaks to remove sources that are likely to be spurious. The spectrum is filtered to remove noise using a median filter with a window size of 31~pixels, corresponding to $\sim$760\AA\ in G102 and $\sim$1440\AA\ in G141. 
The purpose of this filtering is to obtain a crude measure of the continuum, which is used in evaluating the strength of any identified peaks. The large window is chosen so that emission lines will not significantly affect the continuum measurement. From this continuum estimate, the contrast between the peak and the continuum is calculated 
$C = (f_{\lambda,\mathrm{peak}} - f_{\lambda,\mathrm{continuum}}) / f_{\lambda,\mathrm{continuum}}$, and any peak with $C < C_{\mathrm{min}} = 0.15$ is rejected. This criterion is approximately equivalent to a cut on very low equivalent width (EW) emission lines and is implemented in order to remove noise spikes. Next, all peaks that overlap with zeroth orders of nearby bright ($m<23.5$ in the corresponding direct image) sources are rejected. Recall that we also do not consider any spectra on the right edge of the detector where we cannot determine the position of zeroth orders. We additionally remove peaks that are within $d_{\mathrm{edge}} =5$~pixels of the edge of the spectrum to avoid convolution edge effects. Finally, we require that emission line candidates have an overall signal-to-noise for the emission line of S/N$_{\mathrm{line}}\geq4$.
For a line consisting of three contiguous pixels, this corresponds to S/N$_{\mathrm{min,pix}}\geq2.31$, where here S/N is defined traditionally as the flux in the continuum divided by the error. This resulting noise threshold is displayed as the purple curve in  Figure~\ref{fig:cwt}. A summary of the emission line detection parameters is provided in Table~\ref{tab:cwt}.

\section{WISP Survey Completeness Simulations}\label{appendix:sims}
The selection function in slitless spectroscopic data is complex, depending on line signal-to-noise (S/N), equivalent width (EW), galaxy size and concentration, and observed wavelength. Extensive simulations exploring the full parameter space are required. Moreover, the completeness strongly depends on the observing strategy and depth of the datasets. We therefore perform two sets of simulations, one each for the shallow and deeper WISP fields. We add synthetic sources to a selection of real WISP fields and reprocess the fields through the WISP pipeline and line finding procedures presented in Section~\ref{spec_pipeline}. We describe the full process below.

We create 10000 simulated galaxies and divide them equally between the shallow and deep fields. All sources are assigned a spectral template from the models of \cite{bruzual&charlot03} created with a \cite{chabrier03} initial mass function, a constant star formation history observed 100 Myr after the beginning of star formation, and one of three metallicities: $Z/Z_{\odot} = $0.02, 0.2, or 1. The template spectra are redshifted to the observed frame such that all emission line and continuum fluxes are normalised to the desired observed values. We add the following emission lines to the spectra:
[OII]$\lambda3727$, [OIII]$\lambda4363$, \hbeta, [OIII]$\lambda4959$, [OIII]$\lambda5007$, \halpha, [SII]$\lambda6716$, [SII]$\lambda6730$, [SIII]$\lambda9069$, and [SIII]$\lambda9531$. Each emission line is modeled as a Gaussian with $\sigma=3$ \AA.

The basic question addressed by completeness corrections is whether an input source or emission line is recovered by the reduction and processing performed on real data. As it is not a measure of the rates of source misclassification or redshift misidentification, the inputs need not represent the physical distributions observed in the universe. Unless otherwise noted, we therefore uniformly populate the input parameter space so we can determine the ranges most affected by incompleteness.

For each parameter, we choose input ranges that bracket the observed values. Source redshifts are pulled from a uniform distribution ranging from the redshift at which \halpha\ enters the wavelength coverage to that at which [OIII]$\lambda5007$ leaves the wavelength coverage\footnote{We do not add any emission lines to the spectrum in the two narrow wavelength ranges that \texttt{aXeSIM} uses for spectral normalisation: $10400-10600$ (G102) and $15400-15600$ \AA\ (G141). Emission lines in these ranges would artificially boost the normalisation factor thereby significantly reducing the simulated continuum of the spectrum. For more information, see the \texttt{aXeSIM} \citep{kummel07} manual available at \url{axe.stsci.edu/axesim/}. There are therefore several narrow redshift ranges that we do not populate with synthetic sources.}.
We adopt conservative wavelength cutoffs for each grism to avoid the wavelengths where the sensitivity drops rapidly: $8500 \leq \lambda_{\mathrm{G102}} \leq 11200$ \AA\ and $11200 \leq \lambda_{\mathrm{G141}} \leq 16500$ \AA\ for the G102 and G141 grisms, respectively. Real emission lines that lie outside of these wavelength ranges are flagged in the catalogue (see Appendix~\ref{appendix:line_flags} for details). These cutoffs correspond to redshift ranges of $0.3 \leq z \leq 2.3$ for the deep fields with coverage in both grisms and $0.7 \leq z \leq 2.3$ for the shallow fields. Input \halpha\ fluxes for sources in shallow fields are pulled uniformly from $5\times10^{-17} \leq f_{\mathrm{H}\alpha} \leq 1\times10^{-15}$ \esc, with a lower limit of $1\times10^{-17}$ \esc\ in the deep fields. We increase the number of faint line fluxes in the deep fields by separating the synthetic sources into two groups: half with fluxes pulled from a uniform distribution with a maximum at $1\times 10^{-16}$ \esc\ and half with fluxes extending up to $1\times 10^{-15}$. Using two upper flux limits allows us to populate the bright end where sources should be easily detected in the deep fields, while ensuring we have an adequate number of faint objects even if the recovered fraction is small. The input observed EW distribution is uniform across the range $20 \leq \mathrm{EW}_{\mathrm{H}\alpha\mathrm{,obs}} \leq 700$ \AA. The flux density in the continuum at the observed wavelength of \halpha\ --- i.e., the ratio of input \halpha\ flux and EW --- is used to normalise the spectral template to the desired observed units and brightness.

There is a range of observed emission line ratios in the WISP catalog, which we account for in the simulations by varying the input \halpha/[OIII]$\lambda5007$ ratios in the synthetic spectra. Although we are not attempting to quantify the redshift misidentification in the catalog, simulating emission lines with a variety of flux ratios is necessary to include any biases related to single versus multi-line emitters. For example, reviewers are more likely to identify low S/N lines if there are additional emission lines visible in the spectrum to confirm the source's redshift. Input log$_{10}$(\halpha/[OIII]$\lambda5007$) ratios are drawn from a Gaussian distribution centered at $\mu=0$ with $\sigma=0.2$, matching the observed distribution in the catalogue but with a slightly larger full width at half maximum. The intrinsic \halpha/\hbeta\ ratio for case B recombination is adopted for all sources, 2.86 \citep{osterbrock89}. For the remaining emission lines, we adopt the ratios from \cite{anders03} with respect to [OIII]$\lambda5007$ for each metallicity assuming an electron density of $n_e=100$ cm$^{-3}$ and electron temperature of $T_e = 10000$ K. The input $H$-band magnitudes are not assigned to the sources, but instead are the result of the flux normalisation of the template spectral continua and depend on the \halpha\ fluxes and EWs of each source. We do not add the effects of dust to the spectra, but instead rely on the range of line ratios to cover observed values. We note that the adopted case B Balmer decrement, \halpha/\hbeta, is the only line ratio that remains unchanged and therefore always exhibits the value expected for dust-free galaxies. However, we remind the reader that as we are not trying to replicate reality, this choice will not affect the resulting completeness calculations.

The completeness of sources in the WISP catalogue depends strongly on object size and shape. Object size first affects the completeness in imaging, where the low surface brightness of faint, extended objects may fall below the adopted \se\ detection thresholds while the higher surface brightnesses of more compact sources are detected. In addition, the pipeline removes the most extremely elongated detected sources from the catalogue in an attempt to remove artifacts such as diffraction spikes and persistence from bright first orders. Since emission lines observed via slitless spectroscopy are essentially images of the sources at the given wavelengths, the source shape and size will also affect the completeness of the line finding procedure. Extended sources with a low EW are missed by the peak finder, and reviewers are more consistent with their treatment of compact, high-S/N emission lines. We therefore assign each object a profile RMS along the major (minor) axis pulled from a uniform distribution in the range $0.05\arcsec \leq a \leq 1.2\arcsec$ ($0.05\arcsec \leq b \leq a$), again matching the range but not the shape of the observed distribution in the catalog. The input parameter distributions are summarised in Table~\ref{tab:sim_inputs}.

\begin{table*}
\caption{Input parameters for simulated sources \label{tab:sim_inputs}}
\begin{tabular}{c|cc}
\hline \hline
Parameter & Deep Fields & Shallow Fields \\
\hline
Redshift & $0.3 \leq z \leq 2.3$ & $0.7 \leq z \leq 2.3$ \\[4pt]
Semi-major axis$^a$ & $0.05\arcsec \leq a \leq 1.2\arcsec$ & $0.05\arcsec \leq a \leq 1.2\arcsec$ \\[4pt]
Elongation$^b$ & $1 \leq a/b \leq 24$ & $1 \leq a/b \leq 24$ \\[4pt]
Observed \halpha\ flux (\esc) & $1\times10^{-17} \leq f \leq 1\times10^{-15}$ & $5\times10^{-17} \leq f \leq 1\times10^{-15}$ \\[4pt]
Observed \halpha\ EW (\AA)& $20 \leq EW_{\mathrm{obs}} \leq 700$ & $20 \leq EW_{\mathrm{obs}} \leq 700$ \\[4pt]
\halpha/[OIII]$\lambda5007$ ratio & $\mathrm{Gaussian:} \ \mu=0, \ \sigma=0.2$ & $\mathrm{Gaussian:} \ \mu=0, \ \sigma=0.2$  \\[4pt]
$H$-band magnitude$^c$ & $16.8 \leq m_H \leq 27.6$ & $16.8 \leq m_H \leq 26.2$ \\
\hline
\end{tabular} \\
\vspace{1mm}\textbf{Notes:} Parameters indicated with a range are uniformly populated between that range. \\
$^a$The semi-major axis $a$ is treated as the profile RMS along
the major axis.\\
$^b$The semi-minor axis $b$ values are pulled from the range $0.05\arcsec \leq b \leq a$, and so these elongations $a/b$ list the minimum and maximum possible values.\\
$^c$The $H$-band magnitude is computed from the input template spectra normalised according to the \halpha\ flux and EW. The $m_H$ here list the minimum and maximum possible values.
\end{table*}

We add 25 simulated sources at random locations to the raw images for a set of WISP fields, using the \texttt{aXeSIM} \citep{kummel07} software package to create the synthetic direct and grism images of each source. These fields are then fully processed as real data, including the visual inspection by two reviewers. As mentioned in Section~\ref{completeness}, in order to save on the time and effort required for this step, the reviewers only inspect the spectra of simulated sources that were identified by the line finding algorithm. This is not to say that all emission line candidates were real. Some were noise spikes, contamination, or the result of poorly fit continua. However, it does mean we cannot use the simulations to measure the rates of contamination or redshift mis-identification in the catalog.

The WISP completeness calculations, described in
Section~\ref{completeness}, involve determining the fraction of sources  that have been recovered in bins of source size and emission line flux and EW. After the simulated sources are fully processed through the WISP software, we compare all properties of the recovered sources with their input values. This step is necessary to confirm that the recovered sources are counted in the proper bins, i.e., that the parameter values are not systematically different due to the simulation and reduction processes. The input and output fluxes for \halpha\ and [OIII] show a clear correlation down to $f\sim7\times10^{-17}$ \esc. At fainter fluxes, there is a slight trend toward brighter output fluxes, which may be caused by spectral contamination. Overlapping spectra can boost a source's measured emission line flux, and we expect the severity of this contamination to increase with decreasing line flux. Yet there are far too few recovered \halpha\ or [OIII] lines with fluxes $<7\times10^{-17}$ \esc\ to properly evaluate the trend. The other emission lines are almost always fit as secondary lines and will therefore be fainter and at a lower S/N than the primary lines. There is a similarly good agreement between input and output \halpha\ and [OIII] EWs.

A comparison of the source sizes, however, shows that the inputs are systematically larger than the outputs. This effect is not surprising, since the flux in the wings of the simulated Gaussian sources can fall below the \se\ detection limit. The extracted ``footprints'' of the sources are then smaller than what was simulated. We must understand the relationship  between the input and output sizes in order to properly determine the  number of sources that are recovered as a function of size. We model this relationship with a combination of a fourth order polynomial for semi-major and minor axes $\lesssim 0.6\arcsec$ and a linear fit for larger sizes. The WISP survey completeness is applied to sources according to their observed fluxes, EWs, and sizes, as their intrinsic values are unknown. The completeness is therefore calculated as a function of the output values measured for the simulated sources rather than the input values. We use the models to scale the input $a$ and $b$ to their measured values. This step is necessary to ensure that input sources that are not recovered are counted in the correct bins.

\section{WISP Emission Line Catalogue Flag Description}\label{appendix:line_flags}
There are nine entries for quality flags in the emission line catalogue, all consisting of a sum of bit flags (i.e., sum of powers of 2). These are described below.

\texttt{FILTER\_FLAG} indicates the filter coverage as a sum of 10 flag bits: 
\begin{itemize}
\item 1 = F110 coverage
\item 2 = F140 coverage
\item 4 = F160 coverage
\item 8 = UVIS1 coverage (F475X or F606W)
\item 16 = UVIS2 coverage (F600LP or F814W)
\item 32 = IRAC coverage (Ch1 and/or Ch2)
\item 64  = $u$ coverage
\item 128 = $g$ coverage
\item 256 = $r$ coverage
\item 512 = $i$ coverage.
\end{itemize}

\texttt{GRISM\_FLAG} contains a sum of 2 flag bits: 
\begin{itemize}
\item 1 = G102 coverage
\item 2 = G141 coverage.
\end{itemize}

\texttt{[GRISM\_FILTER]\_FLAG} contains a sum of 5 flag bits: 
\begin{itemize}
\item 1 = Artifact, a satellite trail, strange features, significant persistence, etc.
\item 2 = Sky subtraction problem, residual sky, or structure remaining in the background
\item 4 = One or more bright sources present that significantly contaminate the field
\item 8 = A very crowded field, usually indicating some type of star cluster
\item 16 = Scattered light, leading to sensitivity depths that vary significantly depending on source position.
\end{itemize}

\texttt{EDGE\_FLAG} contains a sum of 5 flag bits: 
\begin{itemize}
\item 0 = Object is not near an image edge in direct image
\item 1 = Object is within $\sim$20 pixels of bottom edge of direct image
\item 2 = Object is within $\sim$20 pixels of top edge of direct image
\item 4 = Object is within $\sim$20 pixels of left edge of direct image
\item 8 = Object is potentially within the region along the right edge in which the position of zeroth orders is unknown. The exact x-position is different for both grisms and the wavelength of the emission line must also be considered. The line-finding algorithm takes these details into account.
\end{itemize}

\texttt{REDSHIFT\_FLAG} contains a sum of 8 flag bits: 
\begin{itemize}
\item 0 = Redshift agreement within 1$\sigma$ errors
\item 1 = Redshift disagreement within 1$\sigma$ errors; adopted redshift taken as the case with three or more lines with $S/N>3$
\item 2 = Redshift disagreement within 1$\sigma$ errors; adopted redshift taken as the case with line identified as \halpha. If both reviewers measured \halpha\ (i.e., they identified two different lines as
\halpha), the one with the best $\chi^2$ is taken.
\item 4 = Redshift disagreement within 1$\sigma$ errors and neither reviewer identified multiple high S/N lines nor \halpha; adopted redshift taken as the case with the best $\chi^2$.
\item 8 = Redshift based on a single line
\item 16 = Only one reviewer identified object.
\end{itemize}
In rare cases where both redshift errors are zero, the redshifts are considered to be in agreement if the percent difference in redshift is <1\%.\\

\texttt{FWHM\_FLAG} contains a sum of 5 flag bits: 
\begin{itemize}
\item 0 = FWHM agreement within 1$\sigma$ errors
\item 1 = Reported FWHM is larger than 2$\times$\texttt{A\_IMAGE}
\item 2 = One or more measurements had $\sigma_\mathrm{FWHM} = 0$. This occurs when the FWHM bumps up against either the upper or lower bounding constraints set on the model fit.
\item 4 = FWHM disagreement within 1$\sigma$ errors; reported FWHM is from best $\chi^2$ fit to full spectrum
\item 8 = Reported FWHM is based on measurement from only one reviewer. This either indicates there was redshift disagreement, in which case the reported FWHM is from the best $\chi^2$ fit to full spectrum, or only one reviewer accepted the object.
\end{itemize}

\texttt{[LINE]\_FLAG} contains a sum of 6 flag bits: 
\begin{itemize}
\item 0 = Line flux and EW agreement within 1$\sigma$ errors
\item 1 = EW disagreement within 1$\sigma$ errors; reported EW is from best $\chi^2$ fit to full spectrum
\item 2 = Flux disagreement within 1$\sigma$ errors; reported flux is from best $\chi^2$ fit to full spectrum
\item 4 = Only one reviewer measured flux. The other reviewer has line masked or outside coverage. Alternatively, due to redshift disagreement, all line measurements are from one reviewer.
\item 8 = Flux limits reported for line
\item 16 = No measurement for line; masked or outside of wavelength coverage
\end{itemize}

\texttt{[LINE]\_CONTAM} contains a sum of 4 flag bits: 
\begin{itemize}
\item 0 = Uncontaminated
\item 1 = Reviewer marked general contamination
\item 2 = Reviewer marked continuum contaminated. May affect emission line fit and therefore flux measurement.
\item 4 = Reviewer marked emission line contaminated.
\end{itemize}

\texttt{[LINE]\_EDGE\_FLAG} contains a sum of 4 flag bits: 
\begin{itemize}
\item 0 = Emission line is away from grism edges
\item 1 = Emission line is at $\lambda_\mathrm{obs}\leq 8500$\AA.
\item 2 = Emission line is at $\lambda_\mathrm{obs}\geq 16750$\AA.
\item 4 = Emission line is in grism overlap region, $10900 \leq \lambda_\mathrm{obs}\leq 11500$\AA.
\end{itemize}

\bsp	
\label{lastpage}
\end{document}